\documentclass{mn2e}

\usepackage{amssymb}
\usepackage{graphicx}
\usepackage{psfrag}

\makeatletter

\def\r14{$r^{1/4}$}

\def\R14{$R^{1/4}$}
\def\Rn{$R^{1/n}$}
\def\rms{{\it rms}}
\def\Re{$R_{\rm e}$}
\def\Rd{$R_{\rm d}$}

\def\Ie{$I_{\rm e}$}
\begin{document}

\input epsf


\title[Elliptical galaxies from mergers of discs]{Elliptical galaxies from mergers of discs}
\author[A. C. Gonz\'alez-Garc\'{\i}a \& M. Balcells]{Antonio C\'esar Gonz\'alez-Garc\'{\i}a$^{1,2}$ \& Marc Balcells$^2$\\
$^1$ Kapteyn Astronomical Institute, P.O. BOX 800, 9700 AV Groningen, The Netherlands\\
$^2$ Instituto de Astrof\'{\i}sica de Canarias, V\'{\i}a L\'actea s/n, La Laguna, 38200, Spain}

\maketitle

\begin{abstract} 
We analyse N-body galaxy merger experiments involving disc galaxies. 
Mergers of disc-bulge-halo models are compared to those of bulge-less,
disc-halo models to quantify the effects of the central bulge on the merger
dynamics and on the structure of the remnant.  Our models explore galaxy
mass ratios 1:1 through 3:1, and use higher bulge mass fraction than
previous studies.  A full comparison of
structural and dynamical properties to observations is carried out.  The
presence of central bulges results in longer tidal tails, oblate final
intrinsic shapes, surface brightness profiles with  higher S\'ersic index,
steeper rotation curves, and oblate-rotator internal dynamics.  Mergers of
bulge-less galaxies do not generate long-lasting tidal tails, and their
strong triaxiality seems inconsistent with observations; these remnants show
shells, which we do not find in models including central bulges. 
Giant ellipticals with boxy isophotes and anisotropic
dynamics cannot be produced by the mergers modeled here; they could be
the result of mergers between lower luminosity ellipticals, themselves plausibly formed
in disc-disc mergers.   
\end{abstract}

\begin{keywords}
galaxies:interactions-- kinematics and dynamics-- structure-- elliptical -- numerical simulation
\end{keywords}

\section{Introduction}
\label{Sec:Introduction}

Hierarchical galaxy formation models assume a merger origin for spheroids (e.g.\ Cole et al.\ 2000).  Such merger hypothesis was initially put forward by Toomre (1977), and Toomre's formulation is still useful for detailed studies of the links between merger dynamics and elliptical galaxy structure. 
Observations of merger sequences (e.g.\ Hibbard \& Mihos 1995) generally support the merger hypothesis, as mergers can explain most of the characteristics of elliptical galaxies deemed 'peculiar', such as shells, ripples, kinematically distinct cores, misaligned rotation axes and disky-boxy isophotal shapes (see Schweizer 1998 for a review).  N-body simulation techniques are ideal to check whether the collisionless evolution of galaxies in mergers leads to objects showing the structural and kinematic properties of ellipticals, and many papers have studied the merger hypothesis using these techniques.    
An underlying question, which remains open today, is whether the solution to the divergences between models and real ellipticals requires the refinement of the N-body models; the modeling of new stars formed out of the gas component of the precursor galaxies; or the ruling out of  the merger hypothesis in favor of classical collapse models.

Since the late eighties, improvements in the models and the computing power have led to a progressive increase in the understanding of the merger process.  
Barnes (1988, 1992) used equal-mass model galaxies composed of a non-rotating bulge, a disc and a dark matter halo. 
The remnants from his experiments show de Vaucoulers profiles compatible with real E's and they are supported by random motion rather than by rotation. Hernquist and collaborators (1992, 1993; Hernquist et al.\ 1993; Heyl et al.\ 1994, 1995, 1996) studied equal-mass mergers of both bulge-less systems and systems with a bulge. Bulge-less systems were deemed to be unlikely progenitors of today's elliptical galaxies due to the low density in the inner regions of discs. This difficulty was first put forward by Ostriker (1980) in a short but famous paper, and later by Carlberg (1986). Two ingredients may in principle alleviate this problem: the introduction of gas, or of a third component, namely a bulge. Hernquist et al. found that, as expected, the presence of bulges in the progenitors boosts the central density of the remnant, thus providing a plausible solution to the central density problem.  

Barnes (1999) extended the exploration of N-body mergers to unequal-mass progenitors.  His mergers with 3:1 mass ratio generally lead to flattened systems, irrespective of the orbital configuration, in contrast with 1:1 merger remnants, whose shapes strongly depend on the initial orbital geometry.  This led to the proposal that S0 galaxies may be the result of mergers between unequal-mass discs.  

The link between unequal-mass mergers and remnants harboring discs was furthered by Naab, Burkert \& Hernquist (1999), who analysed the isophotal properties of 1:1 and 3:1 mass-ratio mergers. Isophotes from 1:1 remnants tend to have boxy deviations from elliptical shape, while 3:1 mergers lead to disky objects instead. They thus showed that collisionless mergers can account for the range of isophotal shapes of elliptical galaxies, without recourse to gas dynamics to explain disky objects.  More recently Naab \& Burkert  in a series of papers (2001, 2003) extended the study with a large number of disc-disc mergers similar to the ones found in previous papers. They find that mass ratio and initial orbital configurations are important in the shaping of a number of characteristics of the remnants.

To what degree kinematic diagnostics support the merger hypothesis is unclear at present. 
Reproducing the kinematic peculiarities in ellipticals, such as misaligned rotation and counter-rotation, with collisionless mergers is not difficult;   
we (Balcells \& Gonz\'alez 1998) modeled collisions with mass ratios 1:1, 2:1 and 3:1, using galaxy models which included bulge components in the progenitors.  Unequal-mass mergers in retrograde orbits lead to remnants with counterrotating cores, without recourse to dissipative gas dynamics (Hernquist \& Barnes 1991) to concentrate the counterspinning material in the nuclear regions of the remnant; also, Bendo \& Barnes (2000), using merger simulations with mass ratios 1:1 and 3:1, found that the merger models can explain kinematic properties of real ellipticals such as misaligned rotation and counter-rotation.  However, whether such mergers can explain the more fundamental question of the amount of rotation of ellipticals remains to be answered.  Part of the difficulty rests in whether the goal is to reproduce the slow rotation of giant ellipticals or the rapid rotation of intermediate- and low-luminosity ellipticals, and whether the comparison is to be performed in the inner parts of the galaxies or beyond the effective radius, where observational data is scarce but with some evidence for rapid rotation (Rix, Carollo, \& Freeman 1999).  Bendo \& Barnes (2000) claim that their models reproduce the rapid rotation found in the sample of Rix et al.\ (1999).  However, Cretton et al.\ (2001), using similar simulations, conclude that their models are unable to yield the rapid rotation in the Rix et al.\ data.  

The study of the asymmetry term $h_3$ in the line-of-sight velocity distributions (LOSVD) provides additional clues.  Observationally, besides the well-known trend that $h_3$ has opposite sign to the mean velocity (e.g.\ Bender et al.\ 1994), elliptical galaxies show high $h_3$ values near the galaxy centres, where $v/\sigma$ is low (Bender et al.\ 1994; their Figure 15).  Naab \& Burkert (2001) tried to reproduce this result using mergers with 1:1 to 3:1 mass ratios.  They concluded that their merger remnants do not follow the trends unless they added a disc with at least 10 to 20$\%$ of the spheroid mass. They proposed gas as the means to produce this disc.

Thus, although mergers of disc systems account for a great deal of both global and fine-structure features of ellipticals, several issues remain open. We highlight four questions: (i) can the central densities of ellipticals be explained through collisionless merger dynamics of the precursors' bulges, or are they the traces of gas infall with ensuing star formation, (ii) do the diskiness properties of ellipticals require gas, (iii) what conclusions can be drawn from the relatively unexplored 3:1 mergers, and (iv) do the rotation and LOSVD properties of collisionless merger remnants match those of real galaxies.

In the present paper we address these questions through the analysis of 18 simulations of mergers between disc galaxies.  We run each merger with initial  models comprising disc, bulge and dark halo, as well as bulge-less models comprising a disc and a halo only.  
We have explored three different mass ratios of the progenitor systems (1:1, 2:1, and 3:1) and three different orbital orientations, keeping the initial orbits for the 18 runs the same. The survey is clearly not comprehensive, but it expands on previous works in several ways.  The studies by Barnes (1992), Hernquist (1992), Hernquist (1993), Hernquist et al.\ (1993a), Hernquist et al.\ (1993b), Heyl et al.\ (1994), Heyl et al.\ (1995), Heyl et al.\ (1996) involve equal-mass galaxies, hence do not take into account the effects of different mass ratios on the structure of the final system. Barnes (1999), Balcells \& Gonz\'alez (1998), Naab et al.\ (1999), Bendo \& Barnes (2000), Cretton et al.\ (2001) and Naab \& Burkert (2001, 2003) do explore mergers between progenitors with mass ratios of 1:1 and 3:1 including a bulge. Naab \& Burkert (2003) extend their range up to mass ratio 4:1. Our models are complementary to those above in that our galaxy models have more massive bulges, with disc-to-bulge ratio D/B=2:1, i.e.\ we model 'earlier type' galaxy mergers.  Finally, we analyse bulge-less, non-equal mass mergers, which are not addressed in previous papers.  Together, our 'early-type' and bulge-less models provide a wide base-line in bulge-to-disc ratio (B/D), which is useful to address the central density problem as well as the controversy on the rotation properties of the merger remnants.  

The initial models and the setup of the simulations are described in Section 2. Section 3 presents a comprehensive analysis of the merger remnants, including tidal tails, shells, intrinsic shapes, surface density profiles, isodensity contour shapes, rotation curves, pressure support, and kinematic anisotropy.  In section 4 we study the effect of particle resolution. We discuss our results in Section 5.  

\section{Initial models \label{kkdb}}
\label{Sec:InitialModels}

\begin{table}
\begin{center}
\caption{Initial input parameters to generate dbh and dh models with the GalactICS code. \label{tabdisc}}
\begin{tabular}{cccccc}
\hline
{\bf model} & {\# halo} & { $\psi_{\rm o}$} & { $v_{\rm o}$} & &  \\
\hline
$dbh$ & 30000 & -4.6 & 1.42 & & 
\\
$dh$ & 30000 & -3.0 & 1.5 & & 
\\
\hline
{\bf model} &{\#disc} & {$M_{\rm d}$} & {\Rd}& {$R_{\rm outer}$} & {$z_{\rm d}$} \\
\hline
$dbh$ & 12000 & 0.867 & 1 & 5 & 0.1  \\
$dh$ &15000 & 1.5 & 1 & 5 & 0.1 \\
\hline
{\bf model} &{\# bulge} & {$\rho_{\rm b}$}& {$\sigma_{\rm b}$}&& \\
\hline
$dbh$ &6000 & 14.45 & 0.714 & &\\
$dh$ & 0 & 0 & 0 & &  \\
\hline
\end{tabular}
\end{center}
\end{table}

\begin{table}
\begin{center}
\caption{Initial mass ratios and disc central radial velocity dispersion for dbh and dh models.\label{tabdisc2}}
\begin{tabular}{cccc}
\hline
{\bf model} & {$(M_{\rm lum}/M_{\rm tot})_{R_{\rm out}}$}& {$(M_{\rm lum}/M_{\rm tot})_{\rm total}$} & {$\sigma_{\rm R,0}$}\\
\hline
$dbh$ & 0.459 & 0.241&0.410\\
$dh$ & 0.463 & 0.129 &0.320\\
\hline
\end{tabular}
\end{center}
\end{table}

The initial models were built using the Kuijken-Dubinski GalactICS code (Kuijken \& Dubinski 1995), made freely available by the authors. 
The halo is built from an Evans model (Kuijken \& Dubinski 1994); the bulge is a King (1966) model, thus having a nearly-exponential surface density profile; and the disc is a Shu (1969) model generalized to three dimensions; its surface density is nearly exponential, and it follows a sech$^2(z)$ profile vertically.  Parameters for the initial realizations are given in Table~\ref{tabdisc}.  Haloes are given by the global central potential $\psi_0$ and the asymptotic circular velocity $v_0$.  Discs are given by their mass $M_{\rm d}$, the exponential scale length \Rd, the outer truncation radius $R_{\rm outer}$ and the vertical scale height $z_{\rm d}$.  Bulges are given by their central density $\rho_{\rm b}$ and central velocity dispersion $\sigma_{\rm b}$.  The number of particles and the mass of each component are given in columns (2) and (3), respectively (units are described in section~\ref{Sec:Units}).  Neither the bulges or the haloes are given net rotation. It should be noted that the models generated by the code where scaled to the desired masses as presented below.
 Table~\ref{tabdisc2} lists the ratios of luminous-to-dark masses in each of the models, and the central radial velocity dispersion for the models with a $M_{\rm tot}=1$. 

\begin{figure}
\centering
\includegraphics[width=4cm]{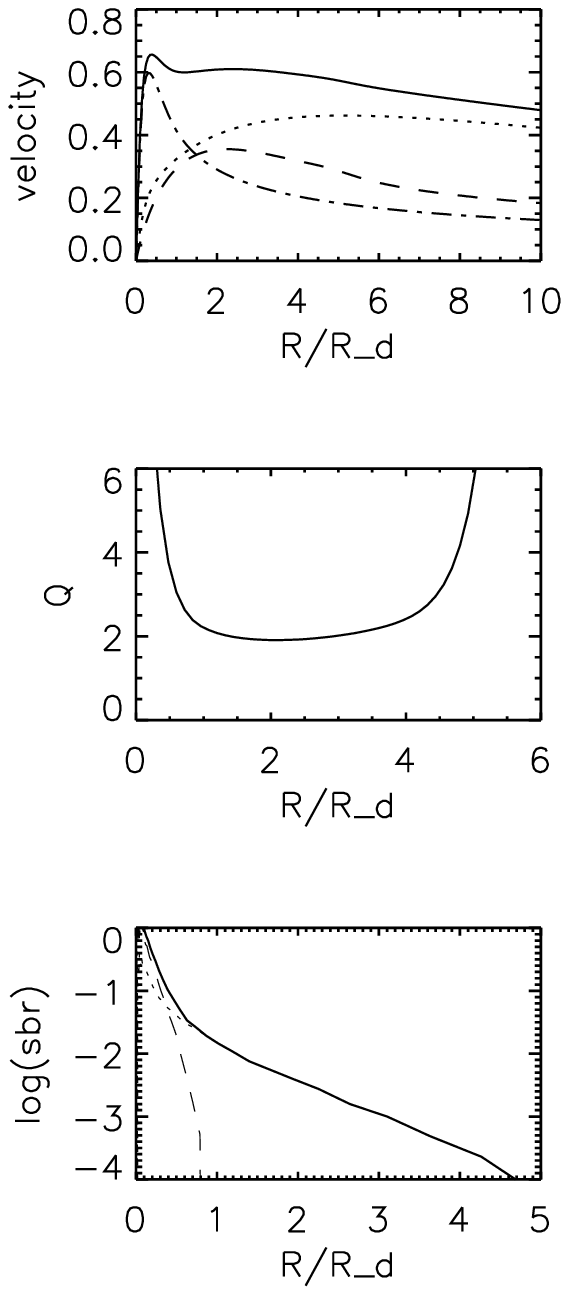}
\hspace{0.cm}
\includegraphics[width=4cm]{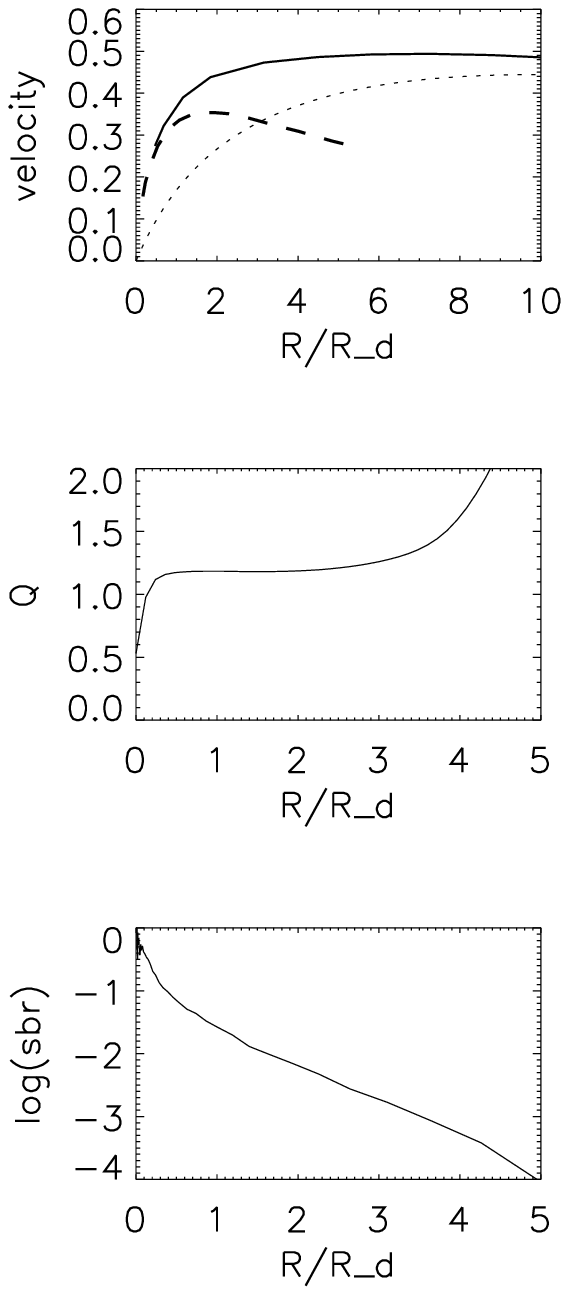}
\caption{{\it Left panels:}  the initial dbh models. {\it Right panels:} the initial dh models.  
{\it Top panels:} Circular velocity curves ({\it solid lines}), and contributions from the disc ({\it dashed}), the bulge ({\it dot-dashed}), and halo ({\it dotted}).  
{\it Middle panels:} Toomre Q parameter.
{\it Bottom panels:} face-on surface density profiles for the luminous matter ({\it solid lines}), for the disc ({\it dotted}) and for the bulge ({\it dashed}).  \Rd~ is the exponential scale length of the disc. \label{initial}}
\end{figure}

Our disc-bulge-halo model (hereafter dbh models), identical to model A from Kuijken \& Dubinski's (1995) models for the Milky Way, has a bulge-to-disc mass ratio \mbox{B/D=1:2}.  Basic properties of this model are shown in Figure~\ref{initial}, left panels.  
The circular velocity curve is flat out to the disc truncation radius ($R_{\rm outer}=5$) and gently declines further out.  Note that the model does not have a maximum disc, rather it is bulge-dominated out to $r=1.5$\Rd~ and halo-dominated elsewhere.  The surface density profile shows an exponential profile for the bulge component, and a characteristic exponential profile in the disc region.  Note however that the disc is not exponential all the way to the galaxy nucleus.  

The bulge-less, disc-halo models (hereafter dh), were also built using the Kuijken \& Dubinski (1995) algorithm.  A bulge-less disc is quite unstable to bar formation (Hernquist 1992), and building a stable model proved to be difficult.  We obtained a solution by lowering the concentration of the halo and rising that of the disc.  The model has nearly maximum disc, a low Toomre parameter Q=1.4 at $r \simeq 3.5$\Rd, and a surface density profile that deviates from a pure exponential in the inner regions (Fig.~\ref{initial}, right panels).

\subsection{Units}
\label{Sec:Units}

We use model units so that the constant of gravity is $G=1$. The total mass of the smallest system is taken to be $M=1$. The scale length of the smallest disc is \Rd~$\simeq 0.2$. A set of physical units that match the dbh models to the Milky Way are:

\vspace{0.5cm}
\begin{equation}
	        [M] = 3.24 \times 10^{11}  \; \rm{M_{\odot}},
\end{equation}
\begin{equation}
		[L] = 14.0  \rm  \; {kpc} ,
\end{equation}
\begin{equation}
		[T] =  4.71\times10^7 \; \rm{yr} ,
\end{equation}
with

\begin{equation}
		[v] =  315  \rm  \; {km/s} .
\end{equation}

\begin{figure}
\centering
\includegraphics[width=6cm]{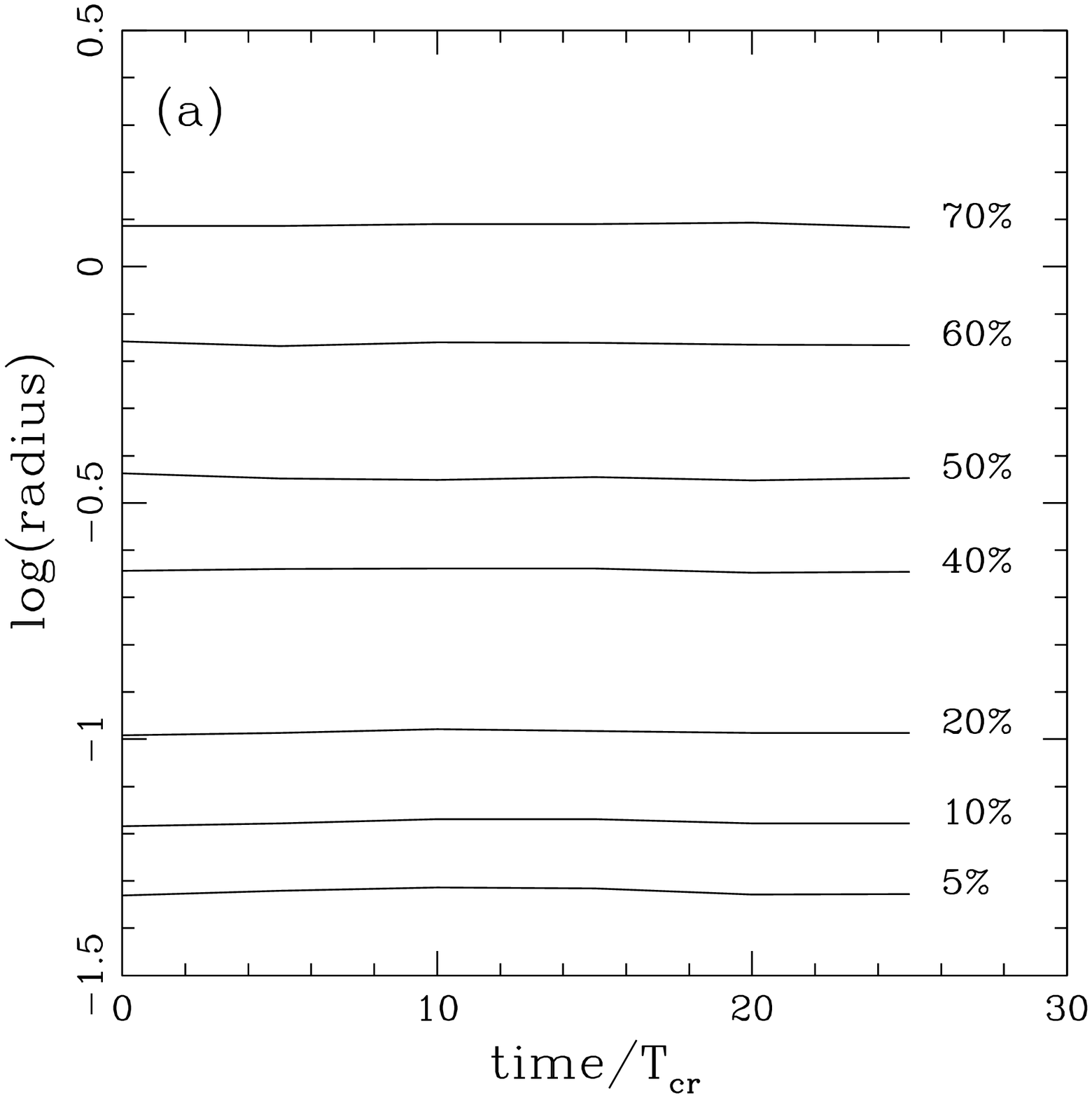}
\vspace{0.cm}
\includegraphics[width=6cm]{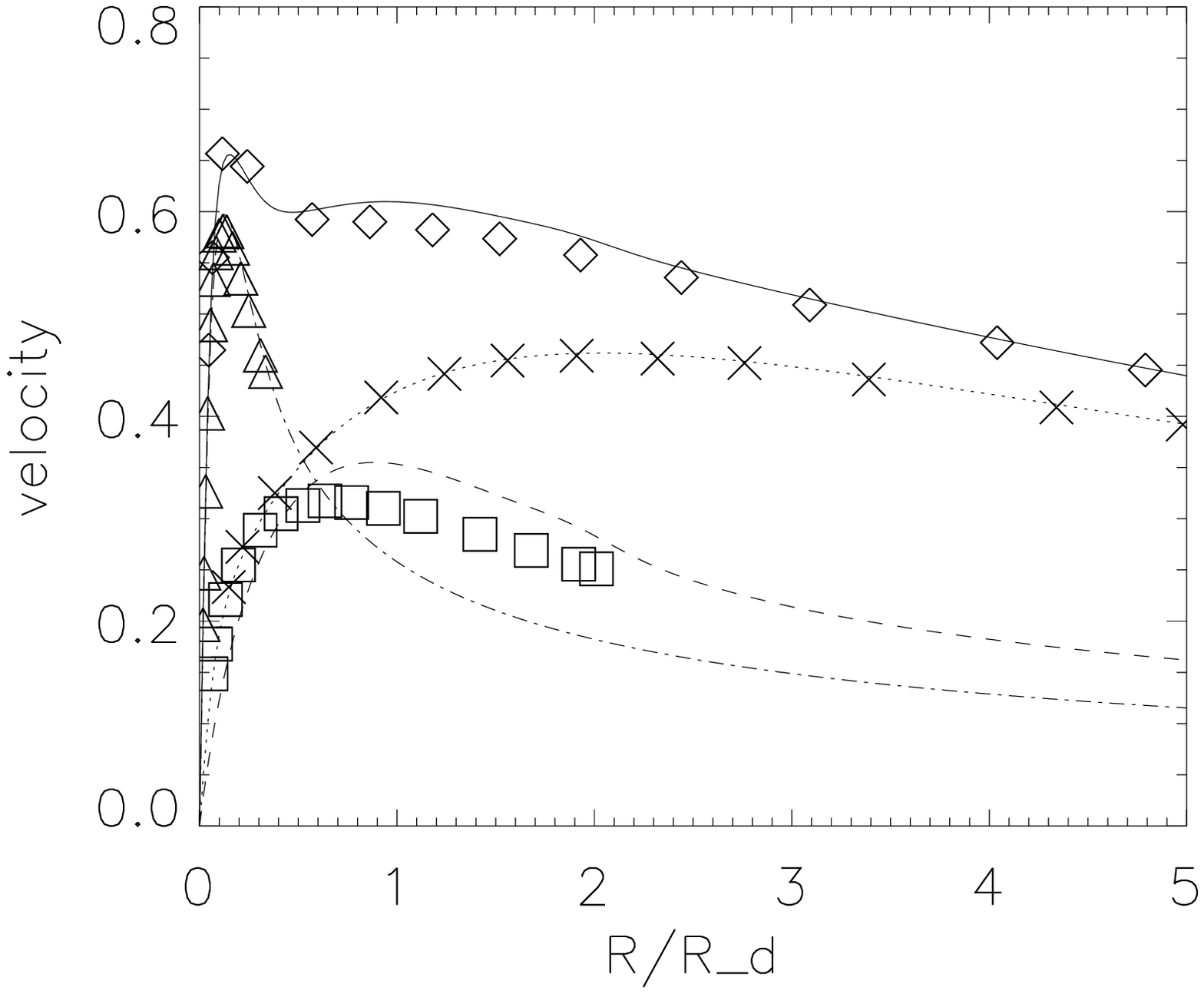}
\vspace{0.cm}
\includegraphics[width=6cm]{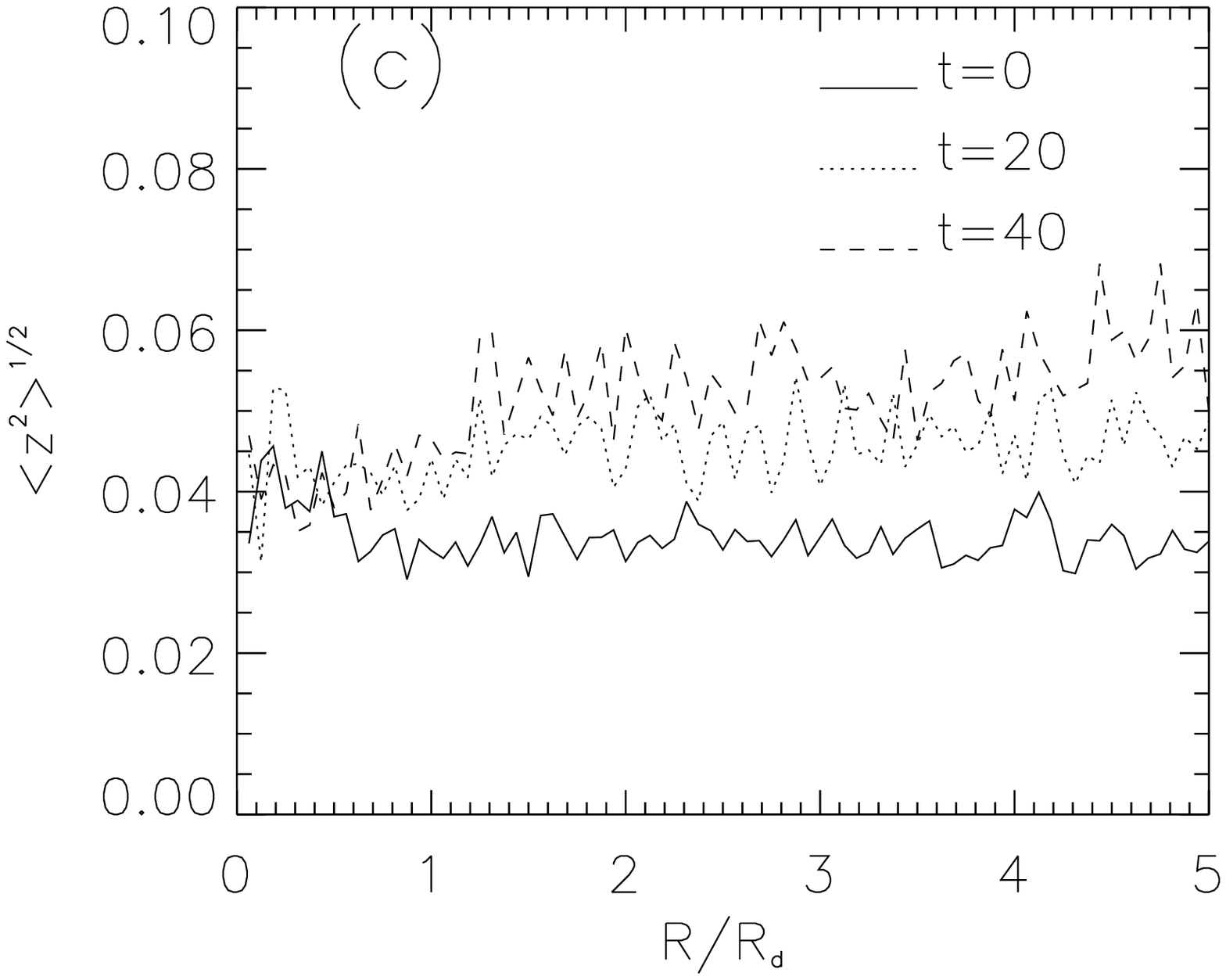}
\caption{(a) Evolution of the radial luminous mass distribution for the initial dbh model. (b) The final circular velocity for the total mass (diamonds), the disc (squares), bulge (triangles) and halo (crosses). The stability can be check comparing with the initial curves for the total mass (solid line), the disc (dashed line) the bulge (dot-dashed line) and for the halo (dotted line). (c) Evolution of the thickness of the disc during the stability run. \label{stabdbh}}
\end{figure}

\begin{figure}
\centering
\includegraphics[width=6cm]{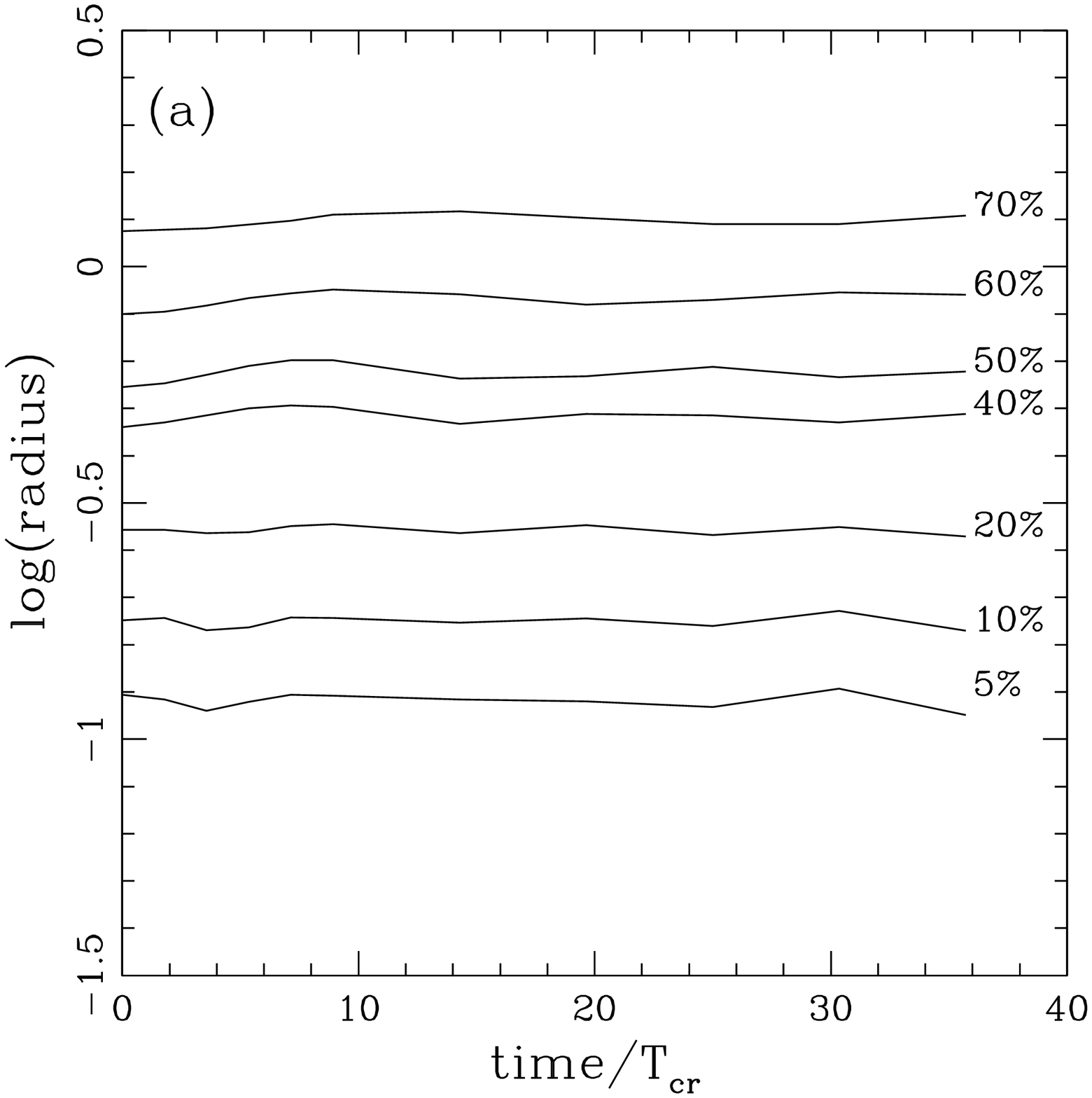}
\vspace{0.cm}
\includegraphics[width=6cm]{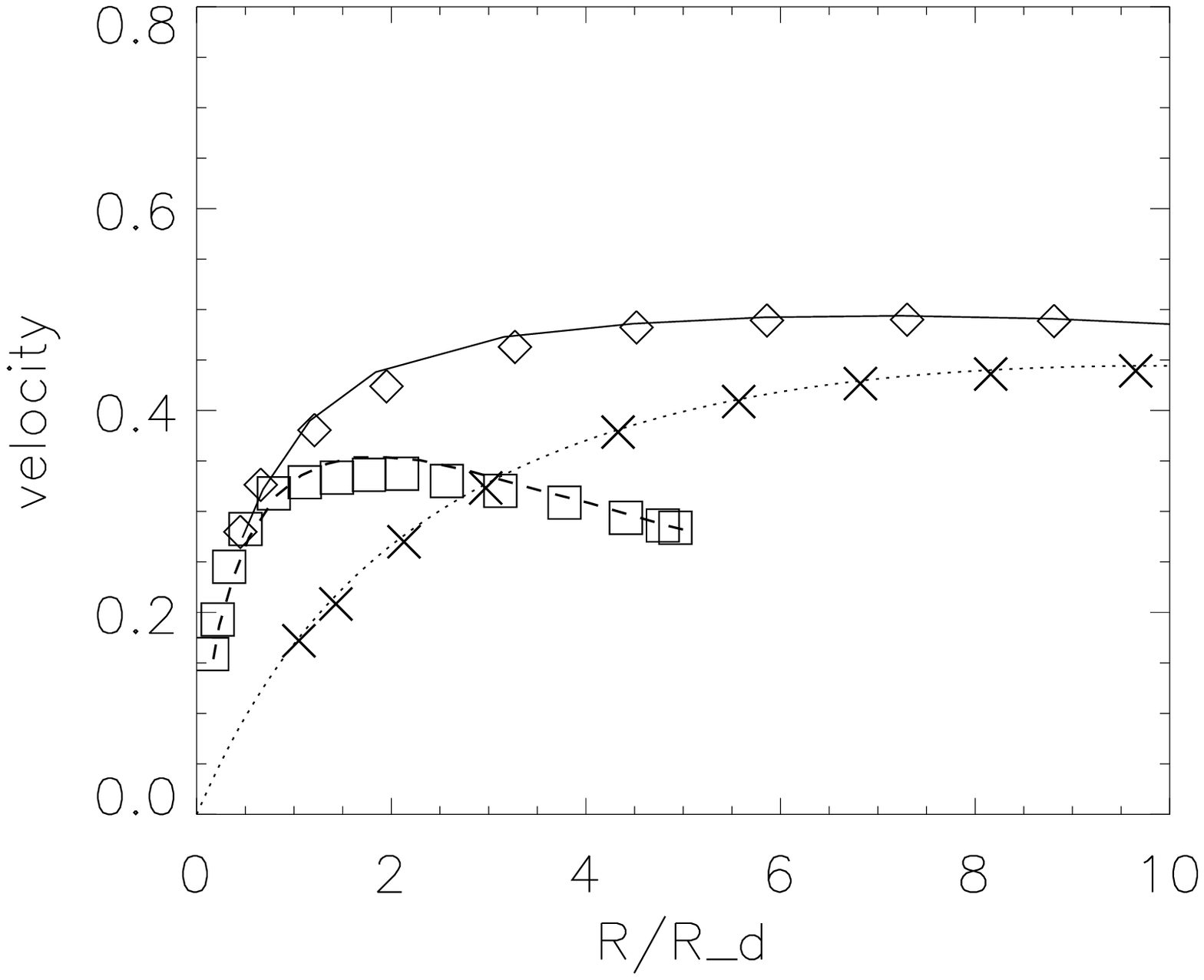}
\vspace{0.cm}
\includegraphics[width=6cm]{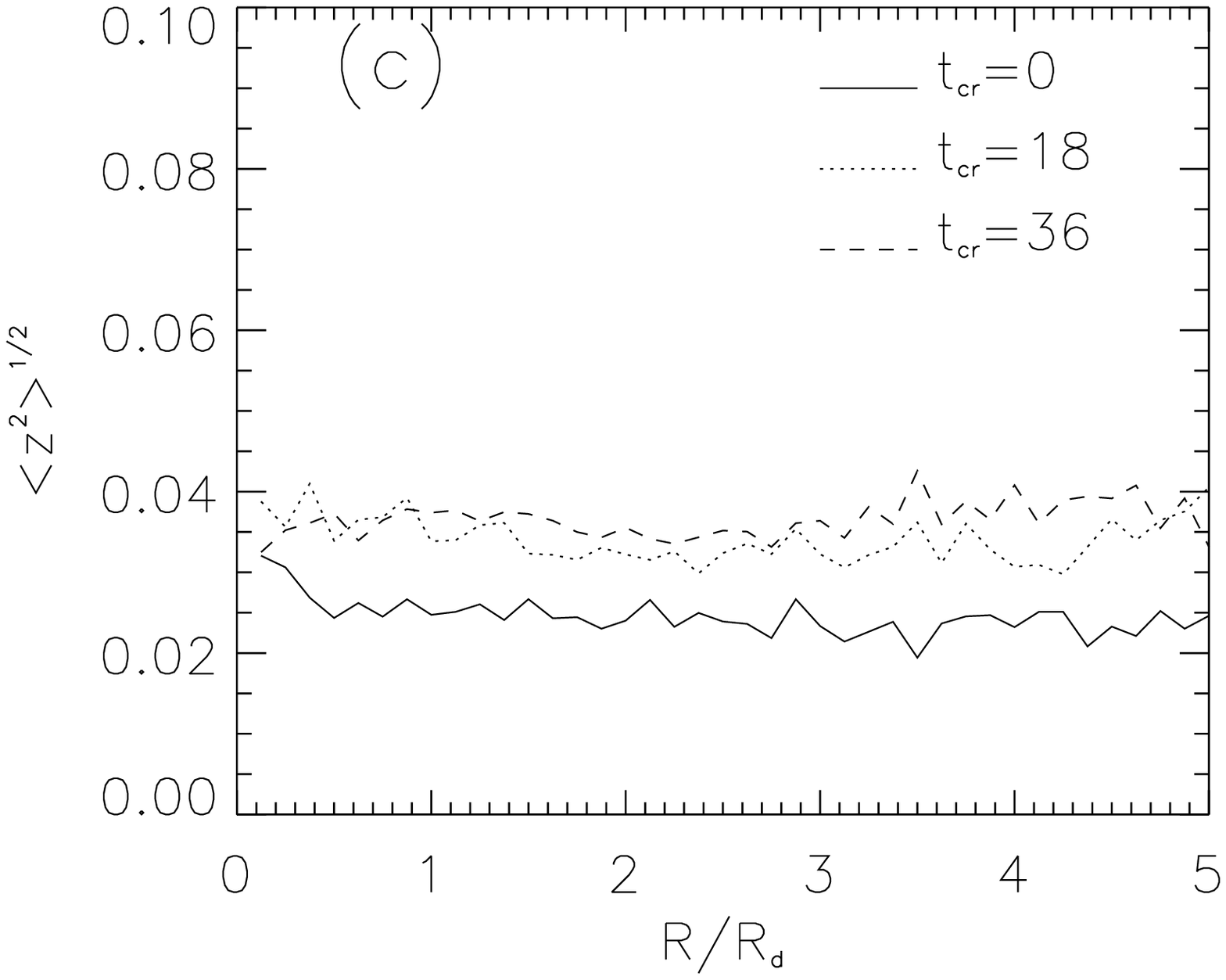}
\caption{(a) Evolution of the radial luminous mass distribution for the initial dh model.  (b) The final circular velocity for the total mass (diamonds) the disc (squares) and the halo (crosses). This can be compared with the curves for the initial model for the total mass (solid line), for the disc (dashed line) and for the halo (dotted line). (c) Evolution of the thickness of the disc during the stability run.\label{rdh}}
\end{figure}

\subsection{Stability of initial models}
\label{Sec:Stability}

The stability properties of the initial dbh and dh models are analysed by evolving the model galaxies in isolation and studying the time evolution of the radial mass distribution, the disc thickness and the circular velocity curve. 
These runs last for about 25 half-light crossing times of the disc galaxy. 

The results for dbh models are summarized in Figure \ref{stabdbh}.  The radial mass distribution (Fig.\,\ref{stabdbh}a) and the rotation curve $V_{\rm c}=\sqrt{GM(r)/r}$  (Fig.\,\ref{stabdbh}b) remain stable.  Figure \ref{stabdbh}b  shows the final circular velocity curve, to be compared with the one in Figure \ref{initial}. The disc vertical scale height increases with time due to bombardment by the halo particles (Fig.\,\ref{stabdbh}c).  The increase of $\sim$70\% over the duration of the experiment is comparable to that measured by Kuijken \& Dubinski (1995).

Figure \ref{rdh} shows the stability results for model dh. Over 35 half-light crossing times the system suffers a mild reorganization of the mass distribution. The disc shows some transient spiral features but is stable against bar formation. Figure \ref{rdh}a shows the evolution with time of fractional mass radii. We find that after relaxation the curves have not changed much. Figure  \ref{rdh}c shows that the disc thickness increases by 60\%.

\subsection{Scaling rules}
\label{Sec:Scaling}

We scale the sizes of model galaxies of different masses using the Tully-Fisher (TF) relation, assuming constant mass-to-light ratios as detailed below.  According to the TF relation, the maximum of the velocity curve scales as a power $\alpha$ of the luminosity of the galaxy:

\begin{equation}
	L \propto (V_{\rm max})^\alpha,
\end{equation}
where $\alpha$ ranges from 2.5 (Ziegler et al.\ 2002) for the B-band for distant galaxies, to 3.2 for the B band, 3.5 for R and 4 for the I band (Sakai et al.\ 2000) for nearby galaxies.

Now, assuming virial equilibrium, $2E_{\rm T}=-E_{\rm W}$, we have

\begin{equation}
	V^2 \propto \frac{GM_{\rm tot}}{r_{\rm G}},
\end{equation}
where $V^2$ is twice the kinetic energy per unit mass of the galaxy and $r_{\rm G}$ is the gravitational radius, defined by the equation $E_{\rm W} = -G\,M_{\rm tot}/r_{\rm G}$.  

Assuming that the luminous mass scales with the luminosity ($M_{\rm lum}/L =$constant),

\begin{equation}
	M_{\rm lum} \propto (V_{\rm max})^{\alpha},
\end{equation}
with $V \propto V_{\rm max}$ we obtain:

\begin{equation}
	\frac{M_{\rm tot}}{r_{\rm G}}\propto M_{\rm lum}^{2/\alpha}.
\end{equation}

Finally, we assume that $M_{\rm total}/M_{\rm lum}=$constant, and 
we obtain the scaling law:

\begin{equation}
	\frac{r_1}{r_2}=\left[ \frac{M_1}{M_2} \right]^{1-2/\alpha}.
	\label{Eqn:Rscaling}
\end{equation}

We have chosen $\alpha=3.5$, matching the TF exponent in the R-band (Sakai et al.\ 2000).

\begin{figure}
\centering
\includegraphics[width=8.cm]{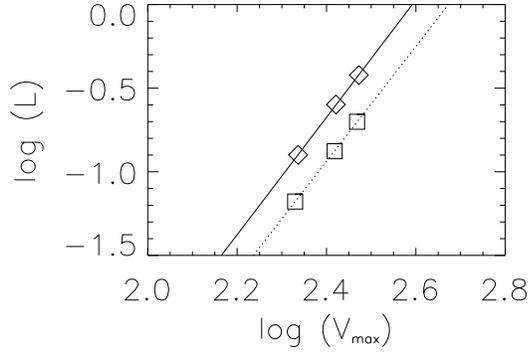}
\caption{Luminosity versus maximum rotation velocity for the initial disc models. The luminosity is taken equal to the luminous mass (bulge+disc in dbh models and disc in dh models). Diamonds are dbh models. Solid line is a fit to those points. The slope of this line is 3.51. Squares are dh models and the dotted line is a fit to these points; it has a slope of 3.45.\label{tf}}
\end{figure}

Figure\,\ref{tf} shows that the models scaled with this procedure indeed follow a Tully-Fisher relation. The observed shift between the two relations reflects a different zero-point due to the different masses of the luminous components of dh and dbh models.

\subsection{Merger experiments}
\label{Sec:MergerExperiments}

\begin{table}
\begin{center}
\caption{Input configurations for disc-bulge-halo models (first 9 entries) and for disc-halo models (maked with *).\label{tabdbh}}
\begin{tabular}{@{}ccccccc}
\hline

{\bf Mod.} &{\bf $\frac{M_2}{M_1}$}& {\bf $(\theta_1,\phi_1)$,$(\theta_2,\phi_2)$} & {\bf $r_{\rm i}=a$} & {\bf $e$}& {\bf $r_{\rm peri}$}& {\bf $T_{\rm peri}$}\\
(1) & (2) & (3) & (4) & (5) & (6) & (7) \\
\hline
\hline
$1p$ & 1:1 & (10,-10);(70,30) & 11.25 & 0.7 & 3.38 & 23.23 \\
$1a$ & 1:1 & (130,60);(70,30) & 11.25 & 0.7 & 3.38 & 23.23 \\
$1r$ & 1:1 & (170,-10);(110,30) & 11.25 & 0.7 & 3.38 & 23.23 \\
\hline
$2p$ & 2:1 & (10,-10);(70,30) & 24.76 & 0.7 & 7.43 & 61.94 \\
$2a$ & 2:1 & (130,60);(70,30) & 24.76 & 0.7 & 7.43 & 61.94 \\
$2r$ & 2:1 & (170,-10);(110,30) & 24.76 & 0.7 & 7.43 & 61.94\\
\hline
$3p$ & 3:1 & (10,-10);(70,30) & 26.98 & 0.7 & 8.07 & 60.99 \\
$3a$ & 3:1 & (130,60);(70,30) & 26.98 & 0.7 & 8.07 & 60.99 \\
$3r$ & 3:1 & (170,-10);(110,30) & 26.98 & 0.7 & 8.07 & 60.99\\
\hline  
\hline
$1p*$ & 1:1 & (10,-10);(70,30) & 12 & 0.7 & 3.60 & 25.60 \\
$1a*$ & 1:1 & (130,60);(70,30) & 12 & 0.7 & 3.60 & 25.60 \\
$1r*$ & 1:1 & (170,-10);(110,30) & 12 & 0.7 & 3.60 & 25.60 \\
\hline
$2p*$ & 2:1 & (10,-10);(70,30) & 20 & 0.7 & 6.00 & 44.97 \\
$2a*$ & 2:1 & (130,60);(70,30) & 20 & 0.7 & 6.00 & 44.97 \\
$2r*$ & 2:1 & (170,-10);(110,30) & 20 & 0.7 & 6.00 & 44.97\\
\hline
$3p*$ & 3:1 & (10,-10);(70,30) & 22 & 0.7 & 6.60 & 44.93 \\
$3a*$ & 3:1 & (130,60);(70,30) & 22 & 0.7 & 6.60 & 44.93 \\
$3r*$ & 3:1& (170,-10);(110,30) & 22 & 0.7 & 6.60 & 44.93\\
\hline  

\end{tabular}
\end{center}
\end{table}

We run mergers involving either two dbh models or two dh models.  Mass ratios of 1:1, 2:1 and 3:1 are done in each case.  For each combination of initial model and mass ratios, three relative orientations of the galaxy spins with the orbital angular momentum are run, yielding a total of 18 independent merger models.  Merger orbital configurations are not varied for each combination of initial model and mass ratio.  Hence, our experiments are suited for an investigation of the effects of mass ratio, of spin orientation and of the presence of a bulge on the structure of the final remnant. However, we do not intend to map the complete parameter space. 

Merger parameters are listed in Table~\ref{tabdbh}. A right-handed coordinate system has been adopted in which the initial models are placed on the $x$-axis, positive toward the less massive model and the orbital angular momentum defines the positive $z$-axis.  Column (1) gives the model name. The number denotes the mass ratio, as given in column (2); note that masses reflect total, luminous plus dark components, that $M_2$ denotes the most massive system, and that $M_1=1$ in all cases.  Model names without an asterisk denote dbh mergers, names with an asterisk denote dh mergers.  The letter describes the spin orientations: '$p$' models have a prograde coupling of both galaxy spins with the orbit; '$a$' ("antiparallel") models have a prograde coupling of the spin of the most massive galaxy with the orbit, and a retrograde coupling of the least massive galaxy;  and '$r$' ("retrograde") models have both spins retrograde with respect to the orbit. The exact orientations of the initial spins are given in column (3), with $\theta$ and $\phi$ being the usual polar and azimuthal spherical coordinates.  We avoided exact parallel or anti-parallel spin orientations since these are not representative of a general encounter between galaxies.  Columns (4)-(7) list the initial separation, ellipticity, pericentre separation and time to pericentre for the Keplerian orbit with the same initial parameters, respectively.   The orbits are slightly sub-parabolic, and ensure that the mass distributions overlap at pericentre, leading the systems to a fast orbital decay.  

After the merger is complete, systems are let to relax for at least 40 half-light radius crossing times before analysis.

\subsection{Merger simulation details}
\label{Sec:SimulationDetails}

Merger simulations are run using Hernquist's version 3 of the {\small TREECODE} (Hernquist 1990) on an Ultra Sparc Station. Softening is set to one-fifth of the bulge half-mass radius for the smallest galaxy in each simulation, $\varepsilon = 0.02$ for dbh models, and to one-fifth of the disc half-mass radius for dh models, $\varepsilon = 0.09$. The tolerance parameter is set to 0.8, and quadrupole terms are included in the force calculation. Individual time step calculation is included.  Energy conservation errors are below $1\%$.

We have modeled the bulge with 6000 particles, the disc with 12000, and the halo with 30000 particles for dbh models. For dh models the disc has 15000 particles and the halo has 30000 particles. In the choice of particle numbers we took care that the particle masses for the various components were not too dissimilar to minimize particle-particle heating. The maximum ratio of particle masses is 7.5 for dbh models  and 11.6 for dh models. The total particle numbers are rather low but they are comparable to Barnes (1992) and Hernquist (1992, 1993) and not so different from later work by these authors and others (Heyl et al.\ 1996, Naab et al.\ 1999).  But our choice of particle number is clearly lower than more recent work (e.g.\ Naab \& Burkert 2003).  As a check of the discreteness and resolution consequences of a low particle number, we have replicated one of our experiments (model $2a$), increasing the number of particles by a factor of 5.  This experiment, named $2aG$, was run with the GADGET code (Springel et al 2001) on a Beowulf cluster.  The properties of this model are represented with a filled diamond in the relevant figures throughout the paper.  We discuss it in Section~\ref{sec:cluster}.   We find that model $2aG$ shows only minor differences with respect to its low-$N$ counterpart, which suggests that the diagnostics reported in the paper for the merger models do not suffer from important discreteness or resolution effects.  

\section{Results}
\label{Sec:Results}

The 18 merger remnants of our experiments display a wide variety of morphological, kinematic and 'photometric' characteristics.  These must reflect the effects of varying mass ratio, of the spin orientations and of the presence or absence of a bulge. 

\begin{figure*}
\centering
\includegraphics[width=14cm]{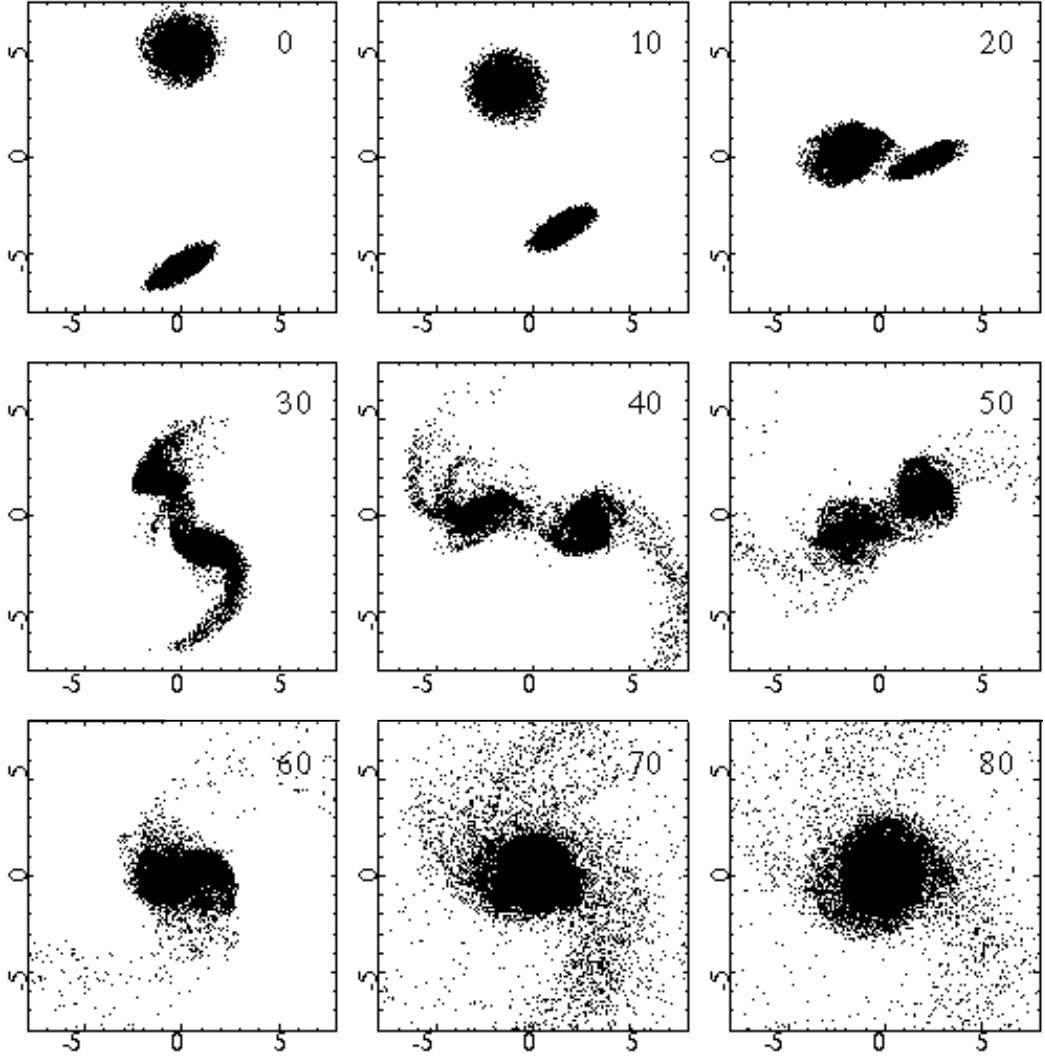}
\caption{Time evolution of the luminous matter in run $1p$ of the disc-bulge-halo series. Time is indicated in the upper right corner. The half mass radius of the final merged system is 0.52. \label{ev1a}}
\end{figure*}

\begin{figure*}
\centering
\includegraphics[width=14cm]{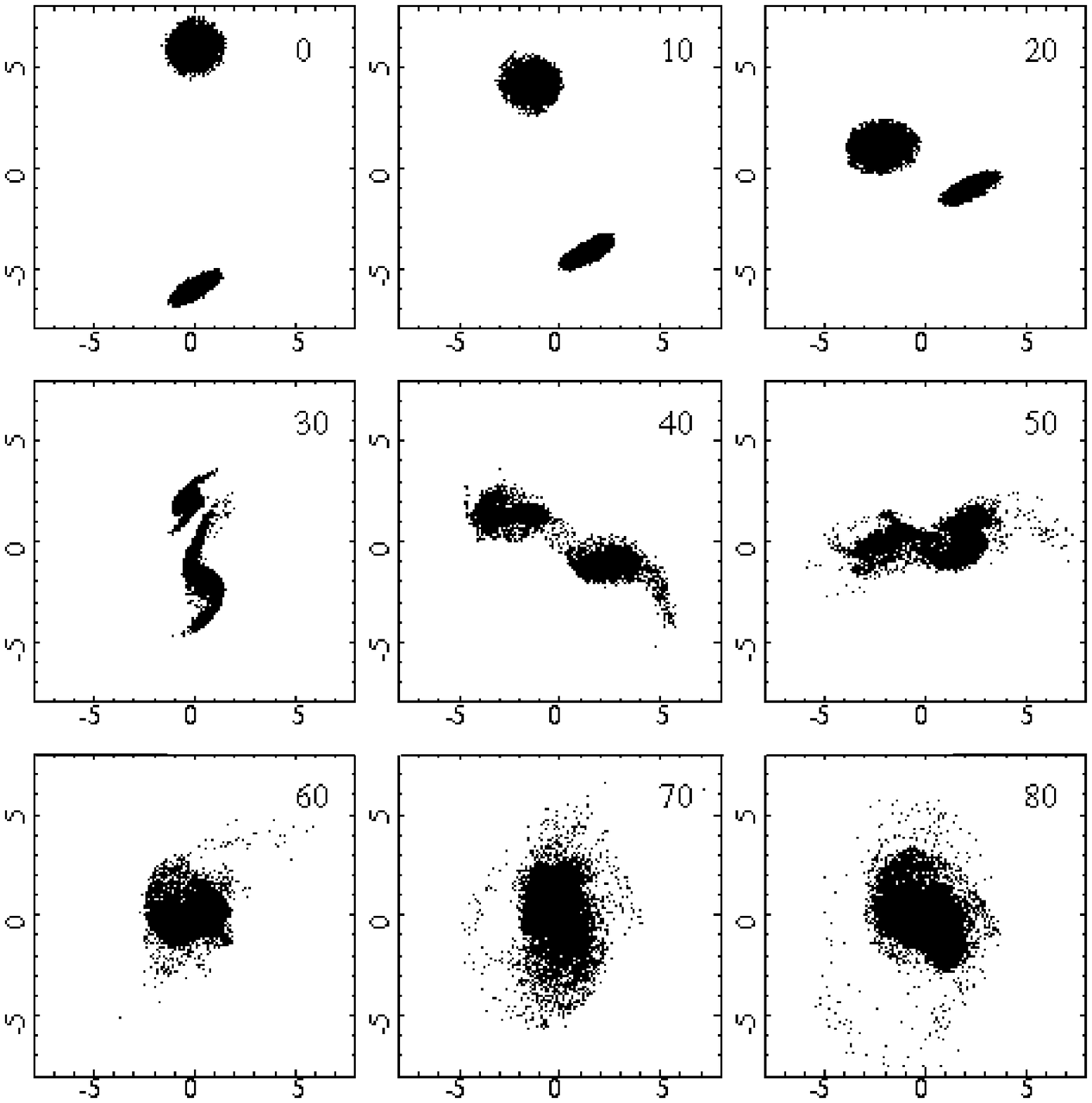}
\caption{Time evolution of the luminous matter in run $1p*$ of the disc-halo series. Time is indicated in the upper right corner. The half mass radius of the final merged system is 0.57. \label{ev1a*}}
\end{figure*}

Figures \ref{ev1a} and \ref{ev1a*} show the time evolution of the luminous matter in two representative runs. Figure\,\ref{ev1a} gives the evolution of run $1p$.  Two disc-bulge-halo models collide with their spins partially aligned with the orbital angular momentum. Prominent tails are created after the first pass through the pericentre. These tails are still prominent well after the inner parts are merged. Figure\,\ref{ev1a*} shows the time evolution of model $1p*$. Here two dh models merge from the same orbit as in Figure\,\ref{ev1a}. The tails formed after the first passage through pericentre are less prominent, and, by the time the merger has completed, the tails have disappeared.

\subsection{Tails}
\label{Sec:Tails}

\begin{figure*}
\centering
\includegraphics[width=14cm]{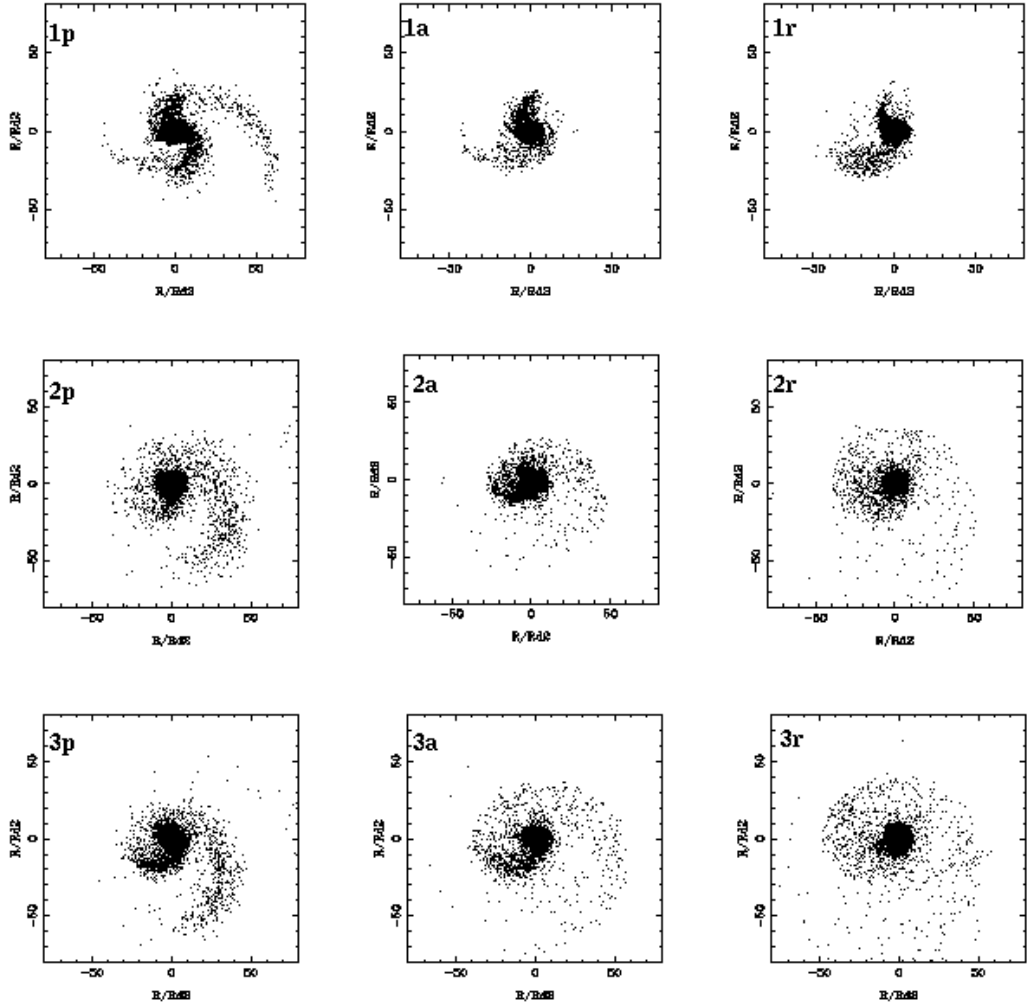}
\caption{Final distribution of the luminous matter in each of the dbh simulations. Each box measures 1.4 Mpc on a side, or approx.\ 200 effective radii \Re of the final models.  The frame sizes are chosen to comprise out to the outermost tidal material.  The systems are viewed from the positive $z$-axis; the merger orbits run counter-clockwise on the plane of the figure.  
\label{Fig:Tailsdbh}}
\end{figure*}

\begin{figure*}
\centering
\includegraphics[width=14cm]{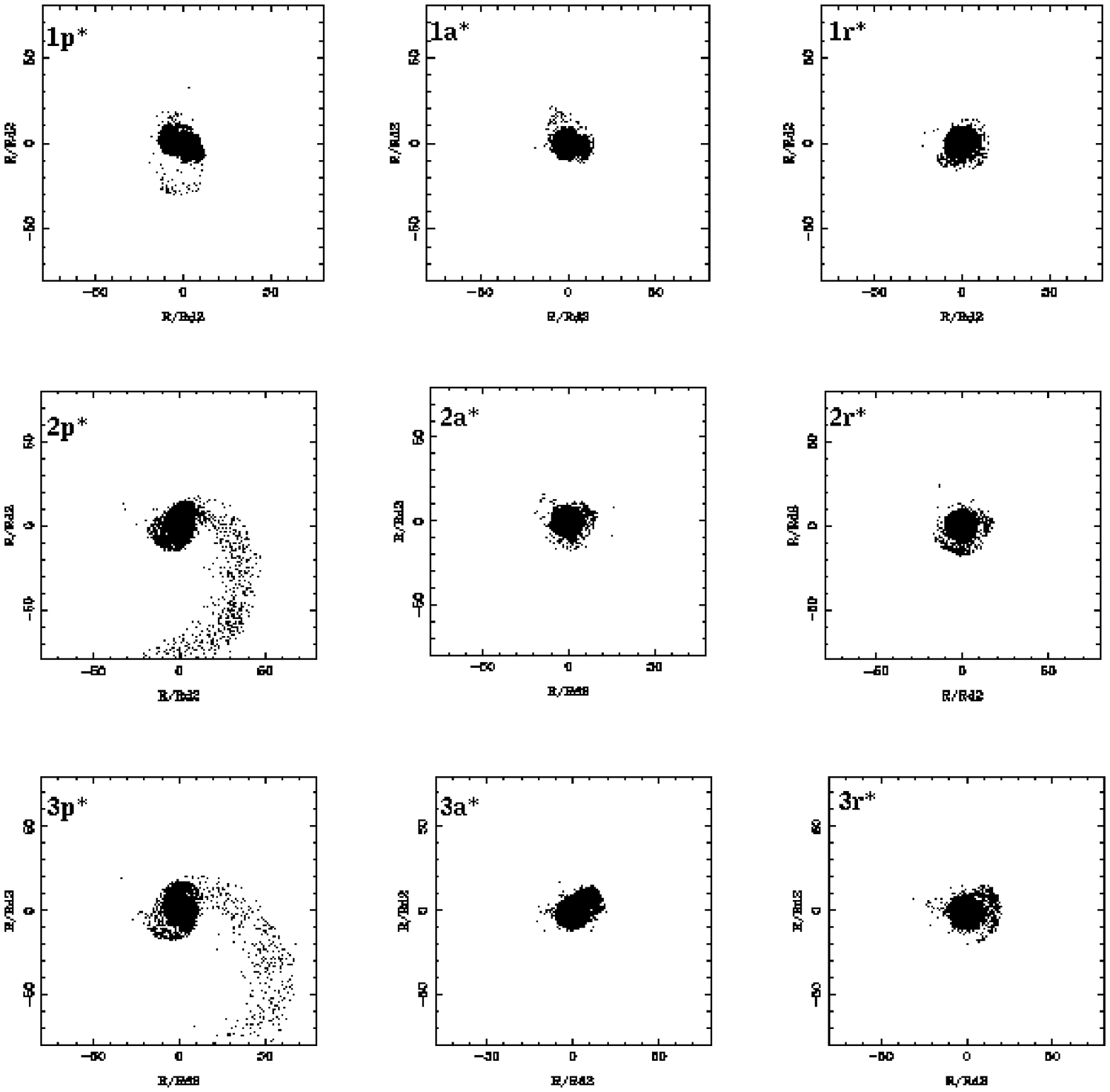}
\caption{Final distribution of the luminous matter in each of the bulge-less simulations.  Each box measures 1.4 Mpc on a side, or approx.\ 200 effective radii \Re of the final models; this is the same spatial scale as used in Figure~\ref{Fig:Tailsdbh}.  The systems are viewed from the positive $z$-axis; the merger orbits run counter-clockwise on the plane of the figure.  
\label{Fig:Tailsdh}}
\end{figure*}

Tail morphology in the nine dbh merger remnants is depicted in Figure~\ref{Fig:Tailsdbh}, which shows the large-scale distributions of the luminous matter roughly 1 Gyr after the mergers are complete. The size of each system has been scaled by the initial scale length of the largest disc, $R_{\rm d2}$ (see eqn.\,\ref{Eqn:Rscaling}), taken here as a reference yardstick for comparing models with varying masses and sizes.

\begin{table}
\begin{center}
\caption{Comparison of the tidal radius of each dbh system at pericentre passage for the different components.  Column (1) gives the model name, (2) production (Y) or not (N) of a tidal tail.  Column (3) gives the ratio between the tidal radius ($r_{\rm T1}$) versus the radius enclosing $99\%$ of the luminous mass ($r_{\rm peri1}$) of the smallest system  at pericentre.  Column (4) lists the same information for the largest system. The tidal radius is computed with the impulsive approximation. \label{tabrjr}}
\begin{tabular}{cccc}
\hline
{\bf model} & {Tail}& {$r_{\rm T1}/r_{\rm peri1}$}&  {$r_{\rm T2}/r_{\rm peri2}$}\\
(1) & (2) & (3) &(4) \\
\hline
1p& Y &     0.912 &      1.117 \\
1a& Y &   1.142   &      1.107 \\
1r & N &      1.096   &      1.134  \\
   & Y &     0.236   &   0.292  \\
\hline
2p & N &      2.010   &      2.264 \\
   & Y? &      1.258   &     1.695 \\
   & Y &     0.341   &    0.804  \\
\hline
2a & N &      2.096   &     2.219 \\
   & N &      1.557   &       1.700 \\
   & Y &     0.504   &    0.865  \\
\hline
2r & N &      2.105   &     2.383  \\
   & N &      1.369  &       1.733  \\
   & Y &     0.439  &       1.075  \\
\hline
3p & N &      1.990  &     2.302  \\
   & Y? &      1.231  &       1.757  \\
   & Y &     0.300  &      0.936 \\
\hline
3a & N &      1.980  &      2.251 \\
   & N &      1.427  &      1.809 \\
   & Y &     0.500  &     1.057  \\
\hline
3r & N &      2.020  &      2.330  \\
   & N &      1.153  &      1.693 \\
   & Y &     0.257  &   0.997 \\

\hline
\end{tabular}
\end{center}
\end{table}

Our models confirm previous work that prograde equal-mass mergers yield two tails, one coming from each of the precursor galaxies (Toomre \& Toomre 1972, hereafter TT72; Barnes 1992). However, model $1p$ shows that tails from prograde, equal-mass mergers can be quite asymmetric, when, as is the case in our models, spins are not exactly aligned with the orbit.  Also, tails become highly asymmetric, approaching one-tail configurations as soon as the mass ratio departs from unity (models $2p$, $3p$).  Hence, galaxies exhibiting one single, prominent tidal tail are not necessarily the result of a minor merger.

All dbh remnants, from both prograde and retrograde merger orbits, show tails.  From spin-orbit coupling (TT72; Barnes 1992; Hernquist 1993), tails were expected in $p$ models (both spins aligned with ${\bf J}_{\rm orb}$) and in $a$ models (primary aligned with ${\bf J}_{\rm orb}$), those of the $p$ model being longer and more sharply defined.  But extended tidal material is also present in $r$ models, in which both spins are retrograde with respect to the orbit.   

Clues to this behavior come from the analysis of which precursor galaxy hosted the stars now in each tidal tail.  Such information is also useful for the interpretation of observed merger remnants.  We find that all the tails in our non-equal mass merger remnants originate from the less massive system.  In particular, this is true for the $a$ models, in which the less massive galaxy's spin is antiparallel to the orbit, and the primary spin is aligned with the orbit (cf.\ Table\,\ref{tabdbh}).  Clearly, the tails  in the $a$ systems do not arise from spin-orbit coupling.  
Rather, tails in our strongly-interpenetrating merger orbits arise from tidal impulses.  Wherever the tidal radius is smaller than the radius of the disc, material must be expelled from the galaxy.  We estimate the strength of the tidal impulse using the impulsive approximation (e.g.\ Binney \& Tremaine 1987) to compute tidal radii.  Table~\ref{tabrjr} lists the ratios between tidal radius and disc radius, measured at each pericentre passage, for all the dbh models.  We also list whether tails appear at each given pericentre passage.  (In equal-mass encounters, long tails form in the first passage through the pericentre, with subsequent pericentre passages yielding additional, shorter tails that sometimes project on to the longer ones; in non-equal mass models, tails form in the second or third pass though pericentre).  Table~\ref{tabrjr} shows that tails appear whenever $r_{\rm T}/r_{\rm peri} < 1$, showing that the tidal impulse is indeed responsible for the formation of tails.  This explains that tails appear in models lacking spin-orbit coupling.  Only in equal-mass mergers does the recipe fail, possibly because the impulsive approximation fails in this case. 
Tidal impulse tails are also formed from the large system, for example in models $3p$ and $3r$. This tails are weak and do not extend much.

The fraction of mass expelled into tidal tails varies with the mass ratio and the spin-orbit coupling.  It ranges from a mere $2\%$ for model $3r$ to up to $10\%$ for $1p$ .   Of this, a large fraction has negative binding energy, and is falling, or will fall back, on to the remnant (Hibbard \& Mihos 1995).  Only in model $1r$ do we find the tail material with positive energy (all particles in the tail at radius larger than 20 units:  about $1\%$ of the luminous mass of the total system).  This small escape fraction is a result of the sub-parabolic nature of our merger orbits;  the escape fraction is 6\% in the models of Barnes (1992). 

Rather different results hold for dh (bulge-less) systems. Figure~\ref{Fig:Tailsdh} shows snapshots of the large-scale distribution of luminous matter at the end of the nine dh simulations, as viewed from the $z$-axis.  Tails appear only in models $2p*$ and $3p*$ (prograde, unequal mass mergers).  These tails are composed of secondary material, indicating that spin-orbit coupling is the dominant mechanism for tail formation in these models.  In general though, comparison with Figure~\ref{Fig:Tailsdbh} reveals a general lack of extended tidal material in db mergers as compared with dbh models; this is surprising if we recall that mass ratios, merger orbits and spin orientations are identical in both sets of models.   By inspection of the luminous mass distributions at intermediate stages during the merger orbit, we find that weak tails did actually develop in most of the models;  however, this tail material has already fallen back to the main body by the time the merger is complete.  

The strength of tidal features therefore depends not only on tidal strength and spin-orbit coupling but also on the presence of a bulge in the precursor galaxies.  This is due to bar formation in the merging discs.  As highlighted by Hernquist (1992), the central bulge in dbh models helps stabilize the disc against bar distortion.  Bulge-less models develop strong bars due to the perturber's tidal field.  The interaction of these bars with the haloes drives angular momentum from the discs to the haloes, depriving the discs from the angular momentum needed for the development of large tails.  Bar formation occurs in the bulge-less models of Hernquist (1992); however, some of his initial configurations have a perfect alignment of the discs with the orbital plane, creating a maximal spin-orbit coupling; this might lead to the formation of more prominent tails than we find here.

The strong disparity in tail strength, geometry and life-times between our dbh and db models, (cf. Figs.~\ref{Fig:Tailsdbh} and \ref{Fig:Tailsdh}) highlights the difficulties in constraining the masses and potential depths of the dark matter haloes of galaxies from the lengths of the tidal tails, as attempted by Dubinski et al.\ (1996), Mihos et al.\ (1998) and Dubinski et al.\ (1999). In this last work the authors perform a survey to check the influence of halo mass profiles on the development of tails. They give the quantitative relation $v_{\rm e}/v_{\rm c} \lesssim 2.5$ to produce tidal tails, with $v_{\rm e}^2= 2|\phi(r)|$ the escape velocity and $v_{\rm c}^2 = G M(r) / r$ the circular velocity at a radius of approximately two times \Rd. Our initial models do show different concentrations for the halo potential so we could expect differences in tail production. For our initial disk models the values for the $v_{\rm e}/v_{\rm c}$ ratio are 2.12 for the dbh model and 2.19 for the dh case, which are both in agreement with the previous relation, and therefore we would not expect the differences reported earlier. Still, different tidal tails are produced, thus highlighting the role played by the bulge in stabilizing the disk against bar formation.

\subsection{Shells}
\label{Sec:Shells}

\begin{figure*}
\centering
\includegraphics[width=14cm]{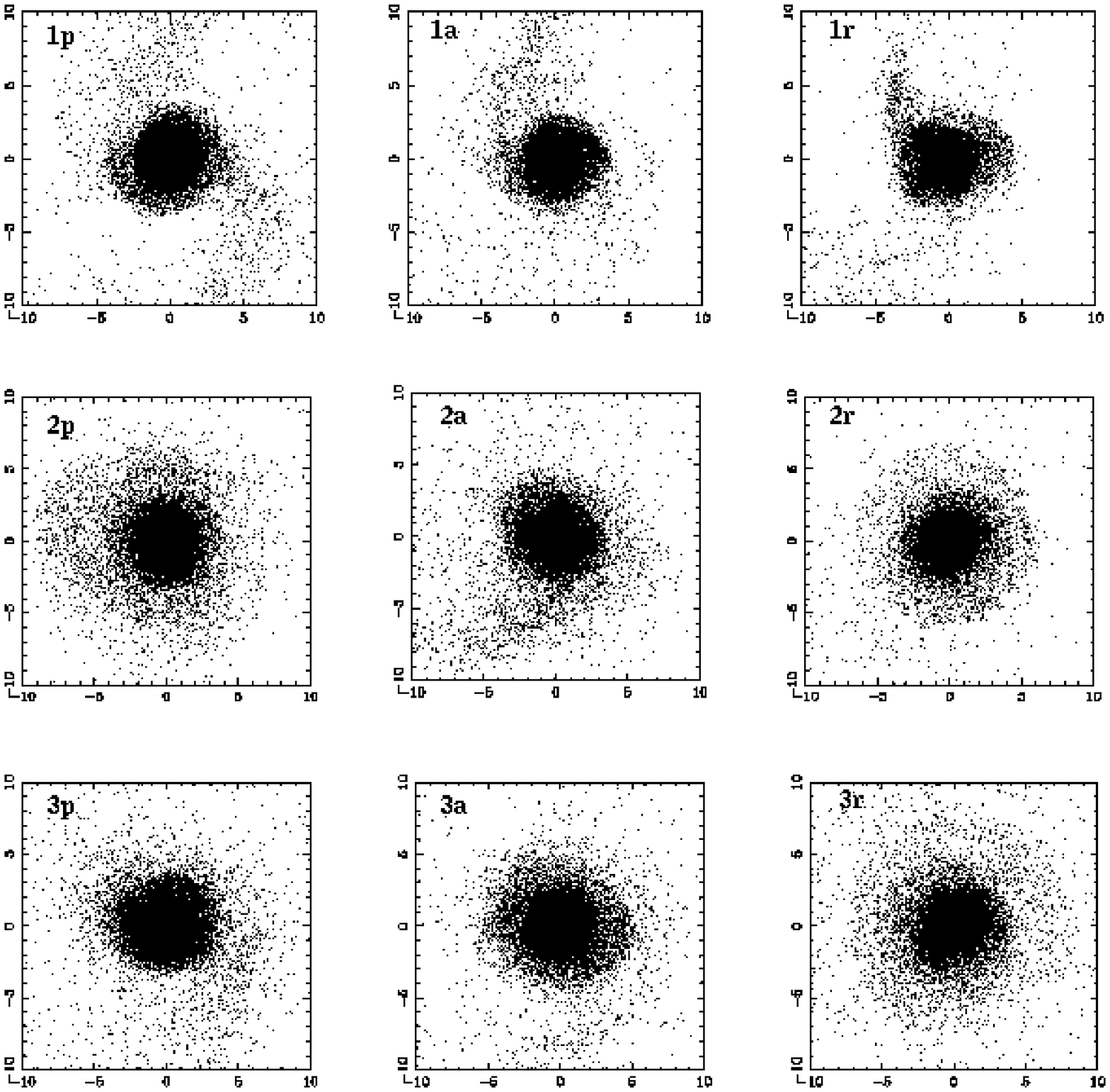}
\caption{Close-up views of the final distributions of luminous matter for the nine dbh simulations.  The size of each box is $\sim$40 \Re (see Table~\ref{tabr14}). The systems are viewed from the positive $z$-axis.\label{Fig:Shellsdbh}}
\end{figure*}

\begin{figure*}
\centering
\includegraphics[width=14cm]{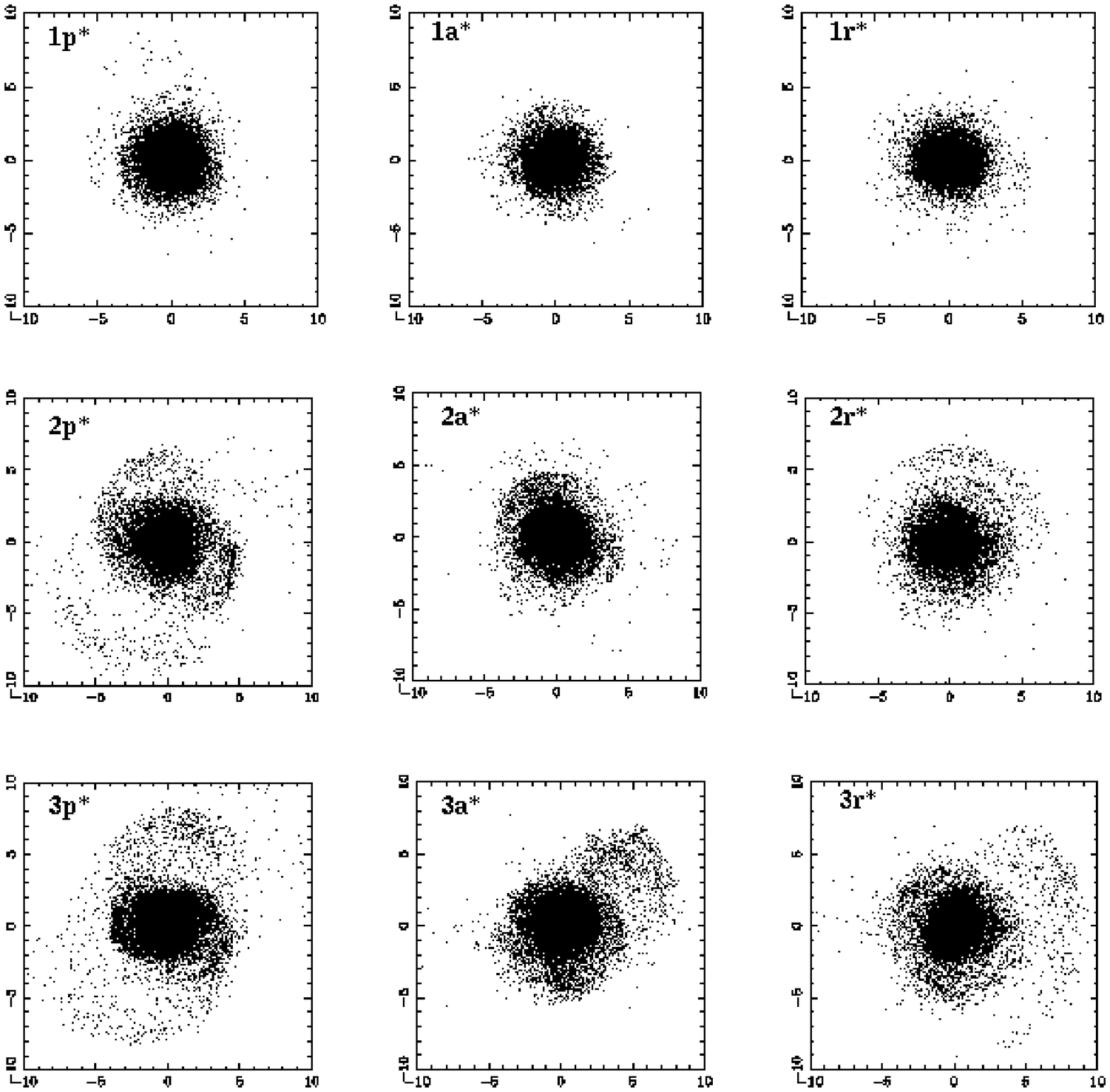}
\caption{Close-up views of the final distributions of luminous matter for the nine dh simulations.  The size of each box is $\sim$40 \Re (see Table~\ref{tabr14}). The systems are viewed from the positive $z$-axis. \label{Fig:Shellsdh}}
\end{figure*}

\begin{figure*}
\centering
\includegraphics[width=14cm]{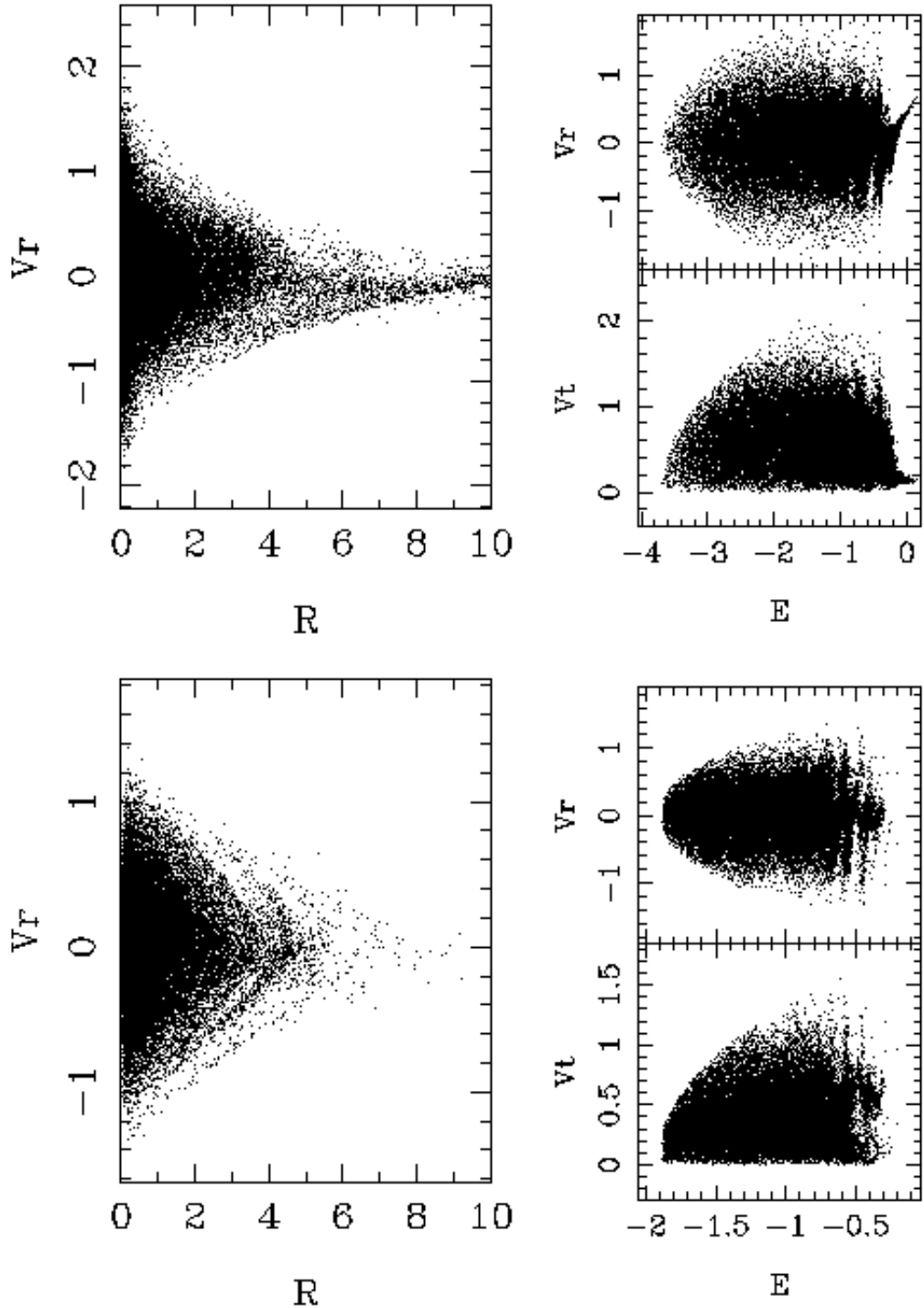}
\caption{Left panels: $V_{\rm r}$ vs. $R$ diagram for (top) simulation $2a$, and (bottom) simulation $2a*$. Right panels: radial and tangential velocities vs.\ energy for (top) simulation $2a$ and (bottom) simulation $2a*$.  \label{Fig:shellsvre}}
\end{figure*}

\begin{figure}
\centering
\includegraphics[width=8.5cm]{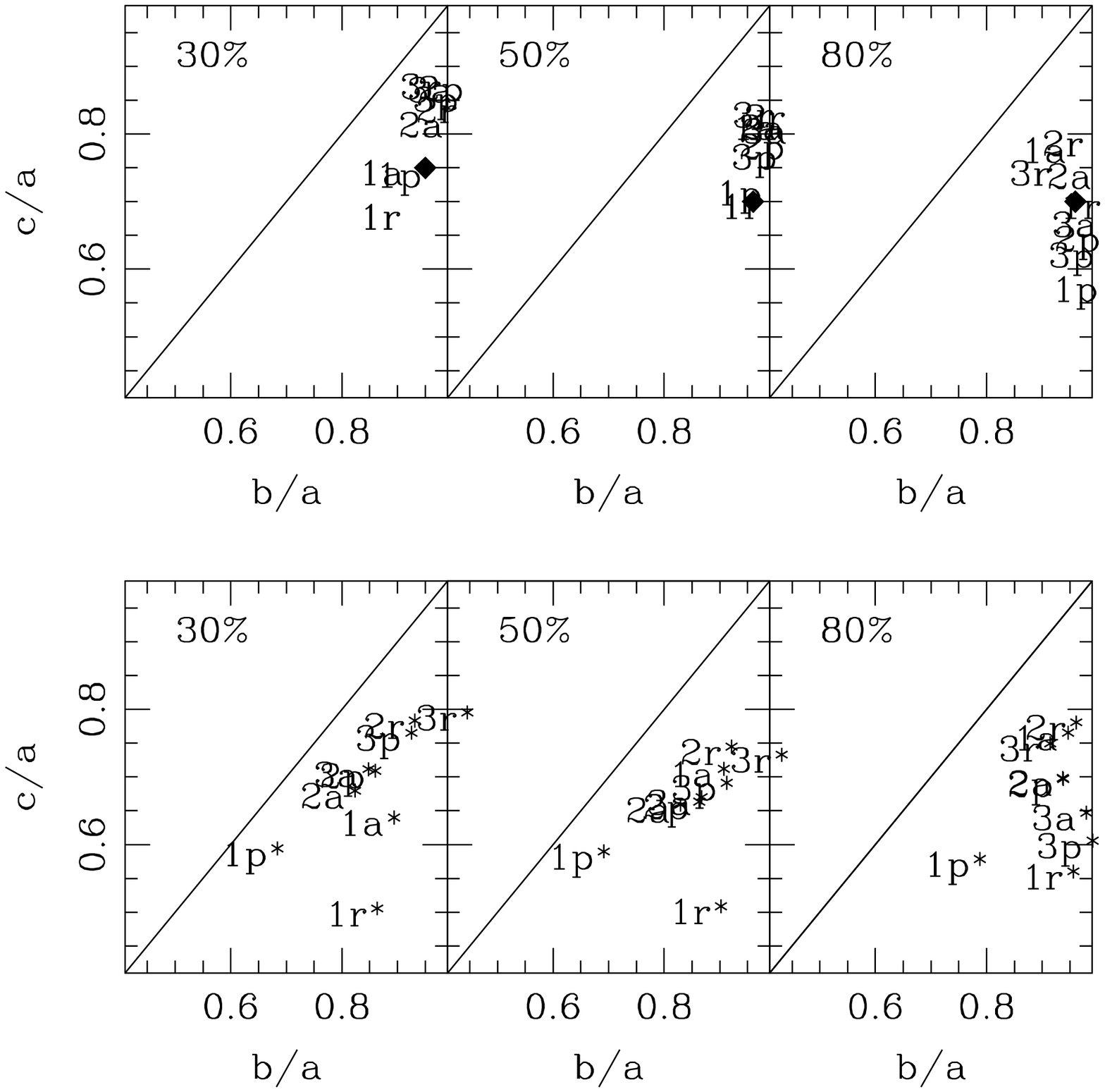}
\caption{Axis ratios of the final models, measured within radii enclosing (left) 30\%, (middle) 50\% and (right) 80\% of the luminous matter.  
Top panels are dbh models, bottom panels are dh models. The diamond indicates model $2aG$. \label{axidi}}
\end{figure}

Hernquist \& Spergel (1992), first noted that major mergers of disc galaxies may lead to the formation of shells, and we find that this is the case in our models.  We extend their results by analysing mass ratios different than unity and model galaxies with and without bulges.  On the other hand, our lower number of particles results in a lower definition for outlining existing shells.  

The presence of shells in our merger remnants may be seen in Figures~\ref{Fig:Shellsdbh} and \ref{Fig:Shellsdh}, which provide close-up views of the luminous mass distribution for dbh and dh merger models, respectively.  These Figures confirm that shells do form in S-S mergers, without recourse to the technique employed by Hernquist \& Spergel of "filtering" the particle distribution to plot only particles with near-zero radial velocity, i.e.\ near their apocentres.  From the particle distributions in Figures~\ref{Fig:Shellsdbh} and \ref{Fig:Shellsdh}, the numbers and sharpness of shells vary with  the mass ratio and with the presence or absence of a central bulge.  Shells are much less prominent in dbh models (Fig.\,\ref{Fig:Shellsdbh}), where we find shell-like structures in models $1r$, $2r$ and $3a$.  Sharp features are present in models $1p$ and $1a$, but those link with tails further out and may be described as well as inner, wound-up tails.

Models without bulges lead to more prominent shells (Fig.\,\ref{Fig:Shellsdh}).  Shells, of type II (all-round) or type III (irregular) following the definitions by Prieur (1990), are seen in models $2a*$, $2r*$, $3a*$, $3r*$.  We note the lack of shells in remnants of equal-mass mergers and on all prograde mergers.  This contrasts with the shell system presented by Hernquist \& Spergel (1992), a prograde merger of two equal-mass, bulge-less discs.  The perfect alignment of the disc spins with the orbital angular momentum may have favored the formation of shells in their model.  

The lack of particle resolution prevents further analysis of the shell distribution, radial range or interleaving.  We can nevertheless enquire whether the shells are phase-wrapped or space-wrapped, by switching from coordinate space to phase space.  Figure~\ref{Fig:shellsvre} shows four representations of phase space for models $2a$ and $2a*$.  In $V_{\rm r}$ -- $r$ space, the V-shaped features provide the classical signature of phase-wrapped shells from material on radial orbits (Quinn 1984; Merrifield \& Kuijken 1998).  Model $2a$ shows very faint (if any) shells, while these are more prominent for $2a*$. Whether these shells are indeed on radial orbits may be investigated by plotting $V_{\rm r}$ and $V_{\rm t}$ vs.\ the energy $E$ (Fig.~\ref{Fig:shellsvre}, right panels).  In $V_{\rm r}$ vs.\ $E$, vertical bands trace phase-wrapping, i.e.\ material with a range of radial velocities but common energy and apocentre distance.  For model $2a*$, at the energies of the three most prominent bands, the $V_{\rm t}$ vs.\ $E$ diagram shows clumps near $V_{\rm t}=0$.  Therefore, the strongest shells in this model are radial, phase-wrapped shells.  On the other hand, a clump of particles with $V_{\rm t} \approx 0.5$ and $V_{\rm r} = 0$ is seen ($E\approx -0.4$).  This configuration may correspond to a space-wrapped shell.  

Sharp features in phase-space distributions are less prominent for the dbh models, thus confirming the seemingly inabiltiy of galaxies with prominent bulges to form shells through mergers. One could argue that the material in the tails could produce more prominent shells for dbh models when that material falls back to the main body. We have followed model $2a$ futher in time and found that there is no major difference with the picture given here, confirming the result given above. 

Gonz\'alez-Garc\'{\i}a \& van Albada (2005a, 2005b) find that major mergers between elliptical galaxies also produce faint shells when there is a contrast in the potential wells of the progenitors.

\subsection{Axis ratios}
\label{Sec:AxisRatios}

We study the intrinsic shapes of the luminous mass distributions by computing axis ratios $b/a$, $c/a$, as was done in de Zeeuw \& Franx (1991)'s study of the intrinsic shapes of elliptical galaxies. The quantities $a > b > c$ are the axes of the inertia ellipsoid.  For a homogeneous ellipsoid with axes $2a$, $2b$ and $2c$, the eigenvalues of the inertia tensor are:

\begin{equation}
 E_1=(b^2+c^2)/5, \\\;\;
 E_2=(a^2+c^2)/5, \\\;\;
 E_3=(a^2+b^2)/5. \\\;\;
\end{equation} 
Defining the axes such that $E_1\leq E_2\leq E_3$, it follows that the axis 
ratios are:

\begin{equation}
        \frac{a}{b}=\sqrt{\frac{E_3+E_2-E_1}{E_1+E_3-E_2}},
\end{equation}
\begin{equation}
        \frac{a}{c}=\sqrt{\frac{E_3+E_2-E_1}{E_1+E_2-E_3}}.
\end{equation}
We compute these ratios for radii  including $30 \%$, $50\%$ and $80\%$ of the luminous mass.

The axis ratios for dbh and dh models are plotted in Figure \ref{axidi}, top and bottom, respectively.  In such diagrams, spherical objects ($a=b=c$) lie at the upper right corner, oblate objects lie on the $b/a = 1$ line, and prolate objects lie on the diagonal ($b/a=c/a$). 

Figure \ref{axidi} shows a strong dichotomy of shapes between dbh and dh models.  Mergers with bulges yield remnants that are oblate at all radii, whereas dh models are strongly triaxial, approaching prolate shapes, out to the inner half-mass radii.  Further out, dh models approach the oblate line, a trend toward outer oblate shape already noted by Gerhard (1983), Barnes (1992) and Hernquist (1992, 1993) for equal-mass mergers.  

The triaxiality of dh merger remnants may be understood from the susceptibility of bulge-less galaxies to become bar-unstable during orbital decay, due to the tidal field: by the time the two cores are about to merge, their shapes are nearly prolate.  In contrast, thanks to the stabilizing effect of the bulge potential, dbh models are robust against bar formation, and the discs as well as the bulges remain nearly oblate throughout orbital decay.  Thus, bar instability of bulge-less discs, which explained tail morphology differences for dh and dbh merger remnants (\S\,\ref{Sec:Tails}), lies also at the origin of the differences in final shapes between dh and dbh models.  

Franx et al. (1991), and Statler (1995) showed that real ellipticals have nearly oblate shapes.  In the light of Figure~\ref{axidi}, such intrinsic shapes are hard to explain within the collisionless merger hypothesis for ellipticals, unless the progenitor discs had prominent central bulges before the merger: bulge-less galaxies  generate remnants that are too triaxial.  

For dbh models, the most significant trend is an outward increase of $c/a$ flattening (e.g.\ from $c/a = 0.7$ at 30\% radius to $c/a = 0.55$ at 80\% radius, for model $1p$).  This is most likely due to the inner axis ratios reflecting the shapes of the merged spheroids (the bulges), while the outer distributions reflect the scattering of disc material over the merger orbital plane.  Equal-mass dbh models also show mild triaxiality in their inner regions ($1p$, $1r$, $1a$, 30\% mass diagram).  This may be explained by the general rule that equal-mass galaxies embedded in extended haloes merge on nearly-radial orbits, due to absorption of orbital angular momentum by the halo (e.g.\ Barnes 1992, Gonz\'alez-Garc\'{\i}a \& van Albada 2005b).  In unequal mass mergers, in contrast, the smaller galaxy spirals in around the dominant galaxy, and the final shape is more nearly oblate.  

We are unable to reproduce Barnes' (1999) claim that 3:1 mergers consistently 
lead to more flattened ellipticals than equal-mass mergers.  Naab \& Burkert (2003) have shown that 3:1 and 4:1 remnants have higher ellipticities than 1:1 remnants (see their figure 4). Figure~\ref{axidi} shows that final flattening 
$c/a$ does not scale with the mass ratio of the merging galaxies.  And, in the cores (30\% enclosed mass), equal-mass mergers actually lead to more flattened structures, as a result of the head-on final merger discussed above to explain the mild triaxiality of these cores.  Barnes' and our results may differ due to differences in the surveyed merger orbits, in the relative orientations of the initial spins, and in the bulge-to-disc ratios ($B/D=1:3$ in Barnes', $B/D=1:2$ in our models).  We get similar axis ratios for 1:1 mergers (e.g., models 2-3-5 in Barnes 1992), arguing against systematic differences in the derivation of axis ratios.  Both Naab \& Burkert and  Barnes have simulated larger samples than presented here for less massive bulges. Therefore the initial orientation of the disks is unlikely to result in the differences reported here. In either event, our models show that the flattening of 3:1 merger remnants is similar to that of the 1:1 remnants;  this occurs whether or not the parent galaxies harbor central bulges.  As a result, stating from axis ratio considerations that 3:1 mergers of disc galaxies lead to flattened, S0-type galaxies is not supported by our data; the statement might be valid for higher mass ratios, 5:1 or 10:1.  

The above analysis indicates that figure shapes and axis ratios of merger remnants strongly depend on the presence of a central spheroidal bulge in the precursor galaxies.  Disc galaxies with central bulges yield oblate remnants similar to real ellipticals, while galaxies without bulges yield objects that are triaxial within the effective radius.  That these objects are rare among elliptical galaxies may be a new problem for the merger hypothesis for the formation of ellipticals from collisionless mergers of bulge-less disc galaxies.  

\subsection{Surface density profiles}
\label{Sec:Sden}

\begin{figure*}
\centering
\includegraphics[width=4.5cm]{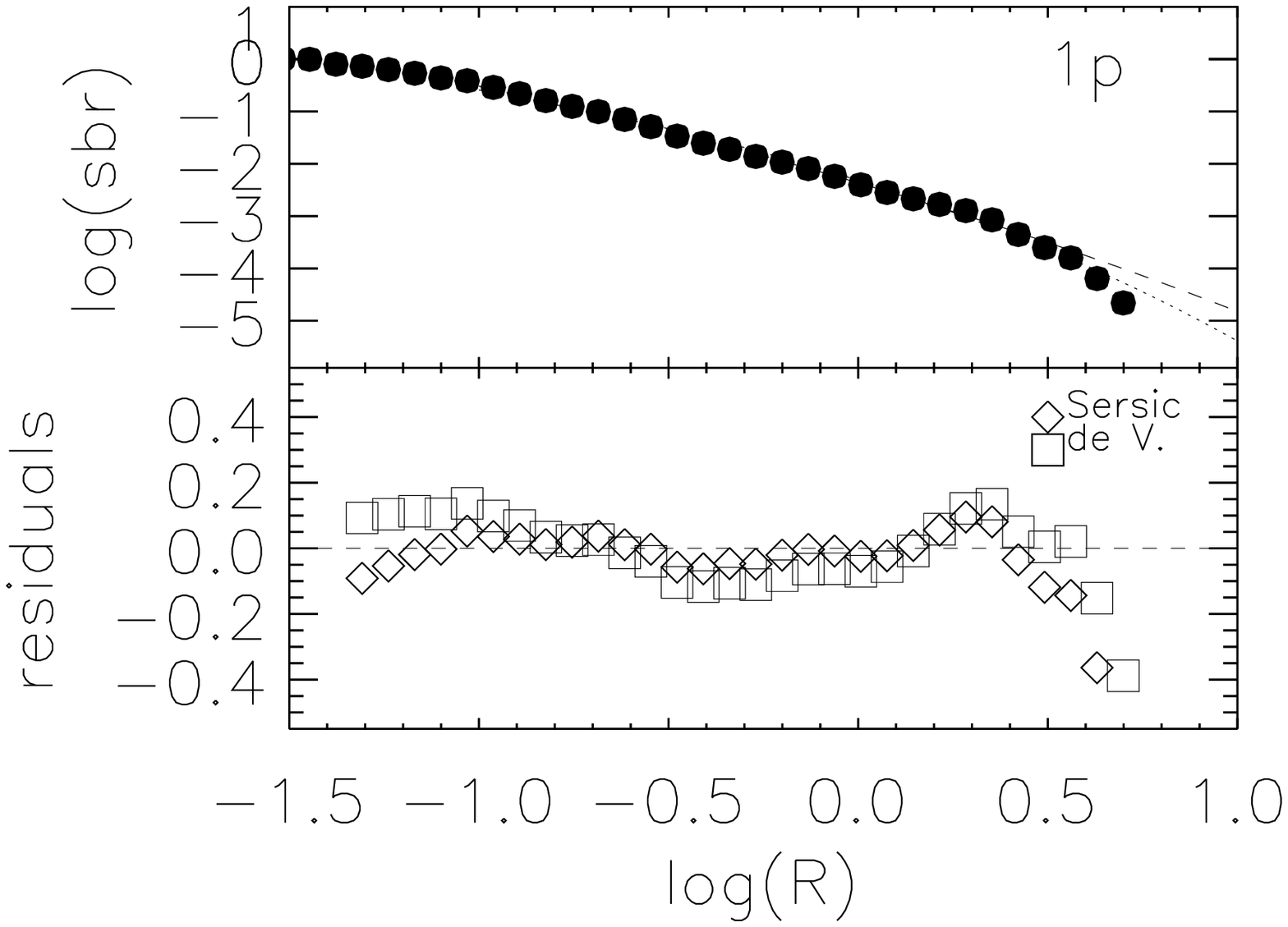}
\hspace{0.cm}
\includegraphics[width=4.5cm]{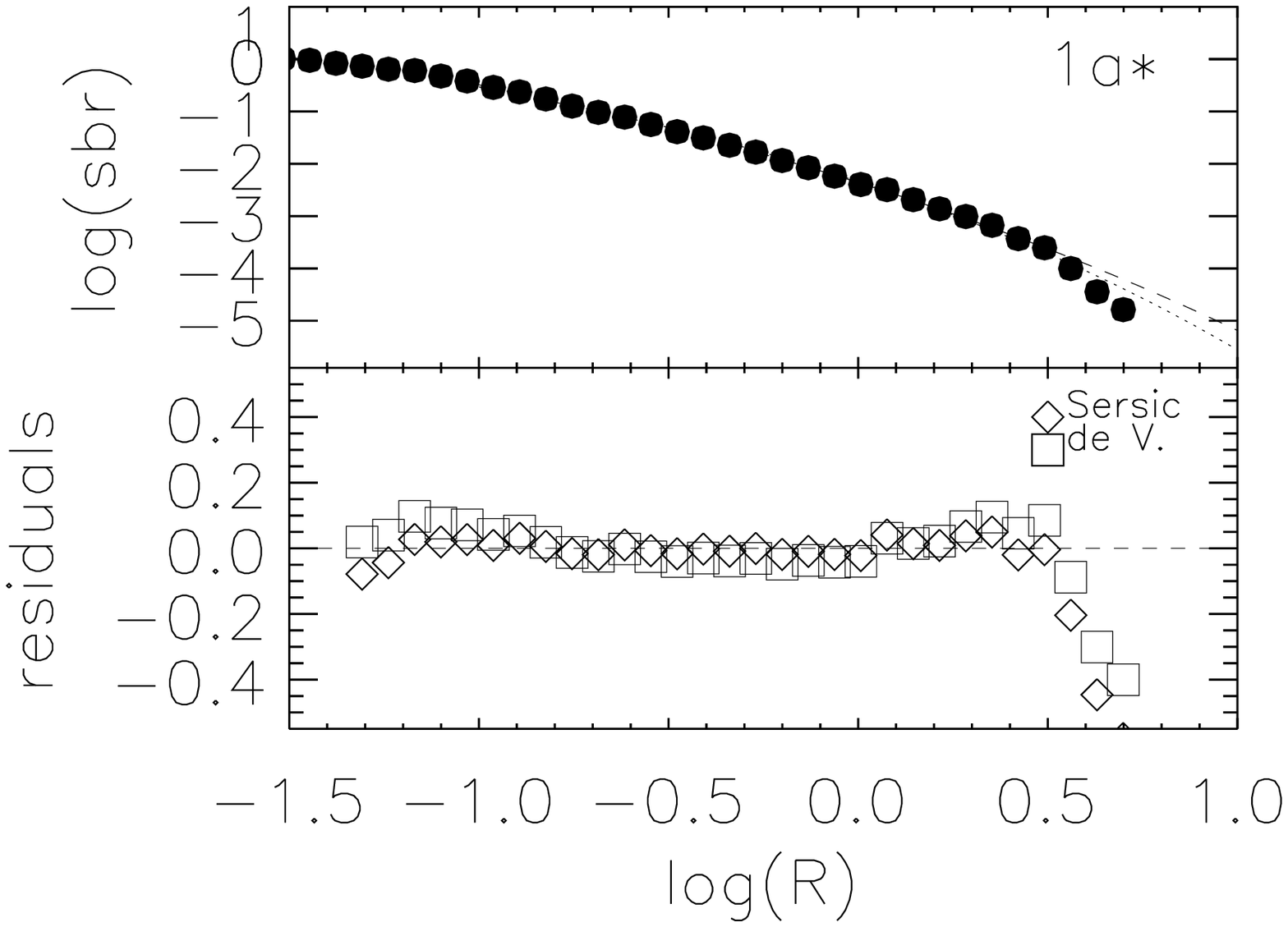}
\hspace{0.cm}
\includegraphics[width=4.5cm]{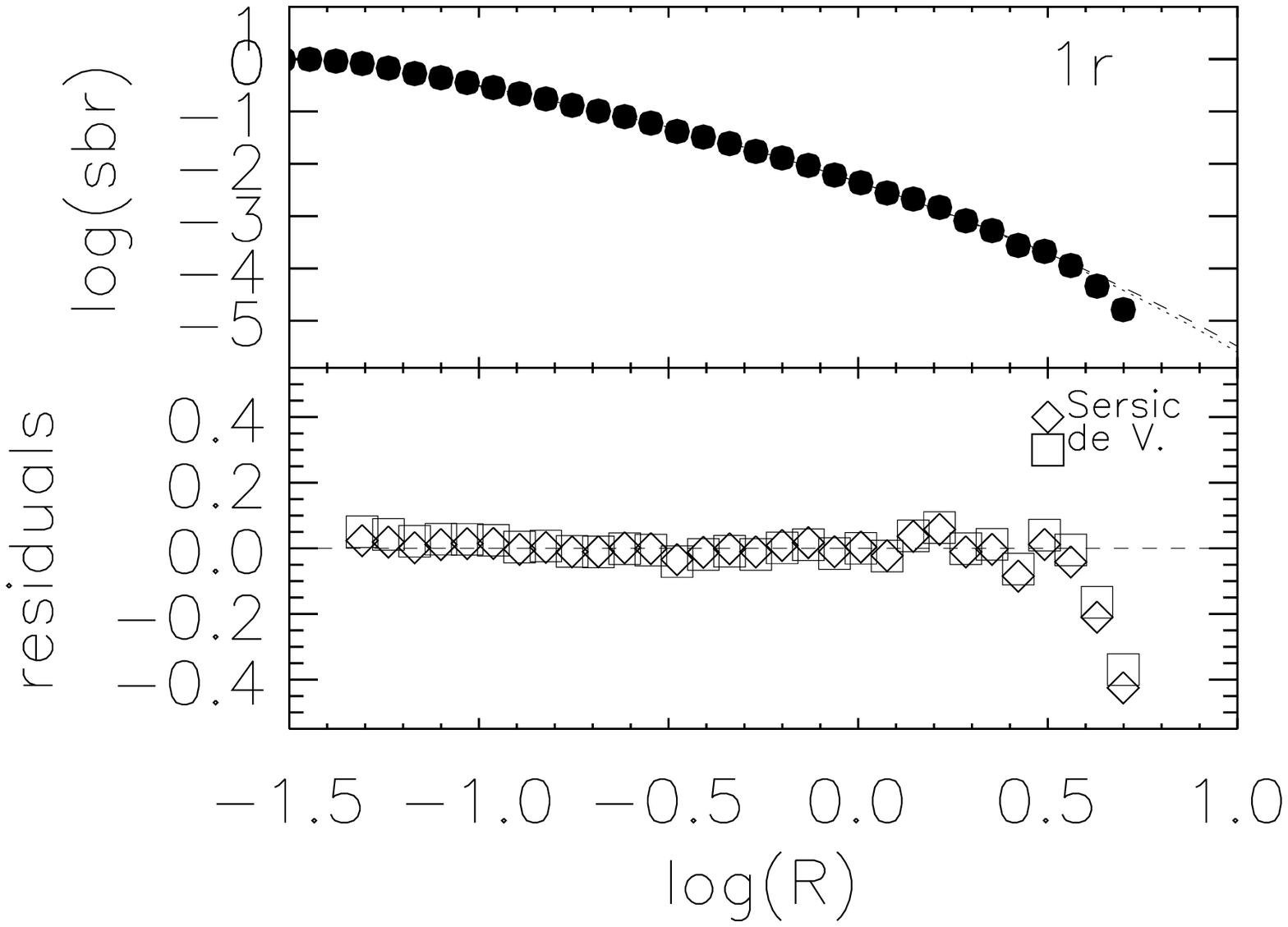}
\vspace{0.cm}
\includegraphics[width=4.5cm]{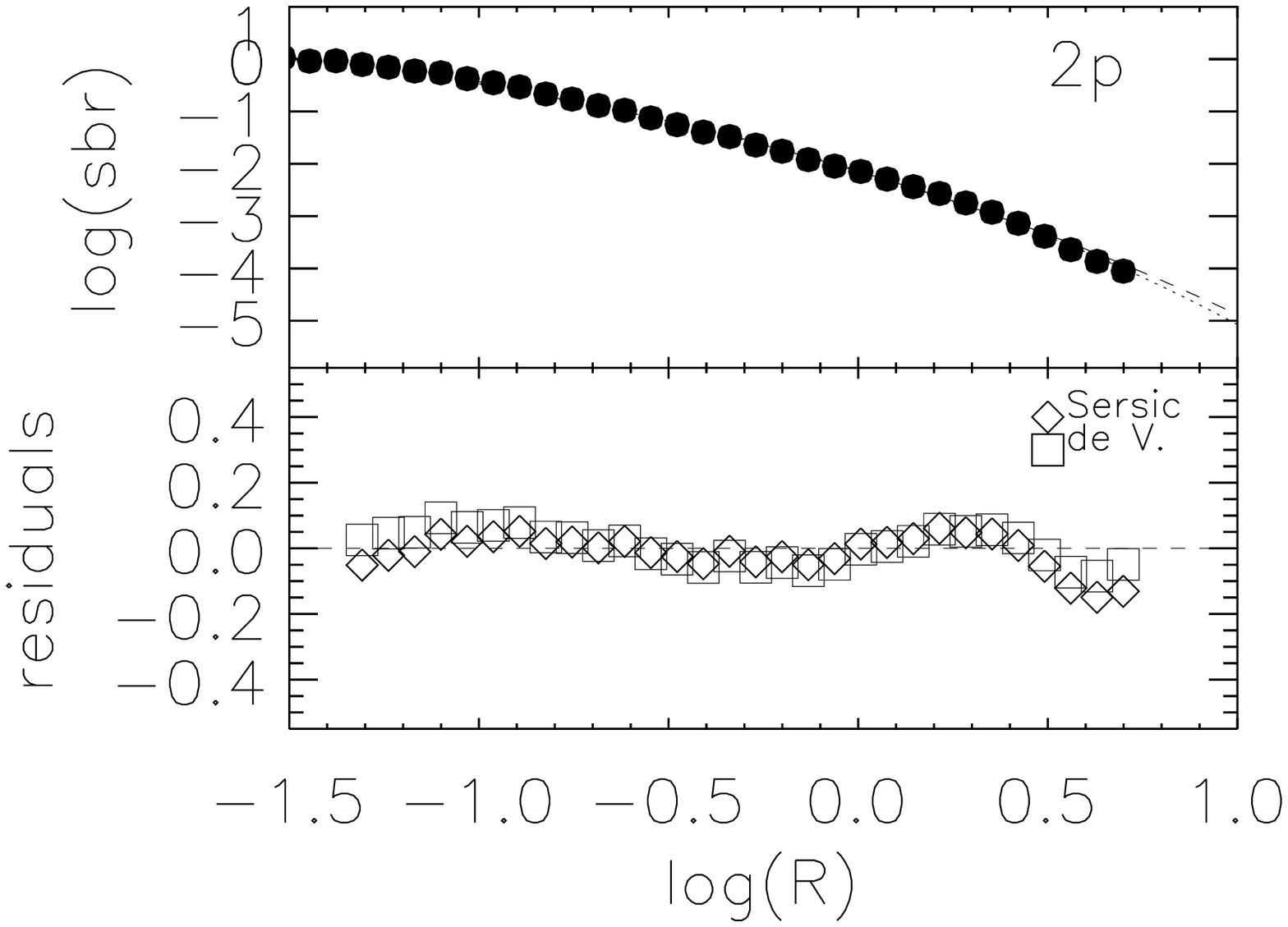}
\hspace{0.cm}
\includegraphics[width=4.5cm]{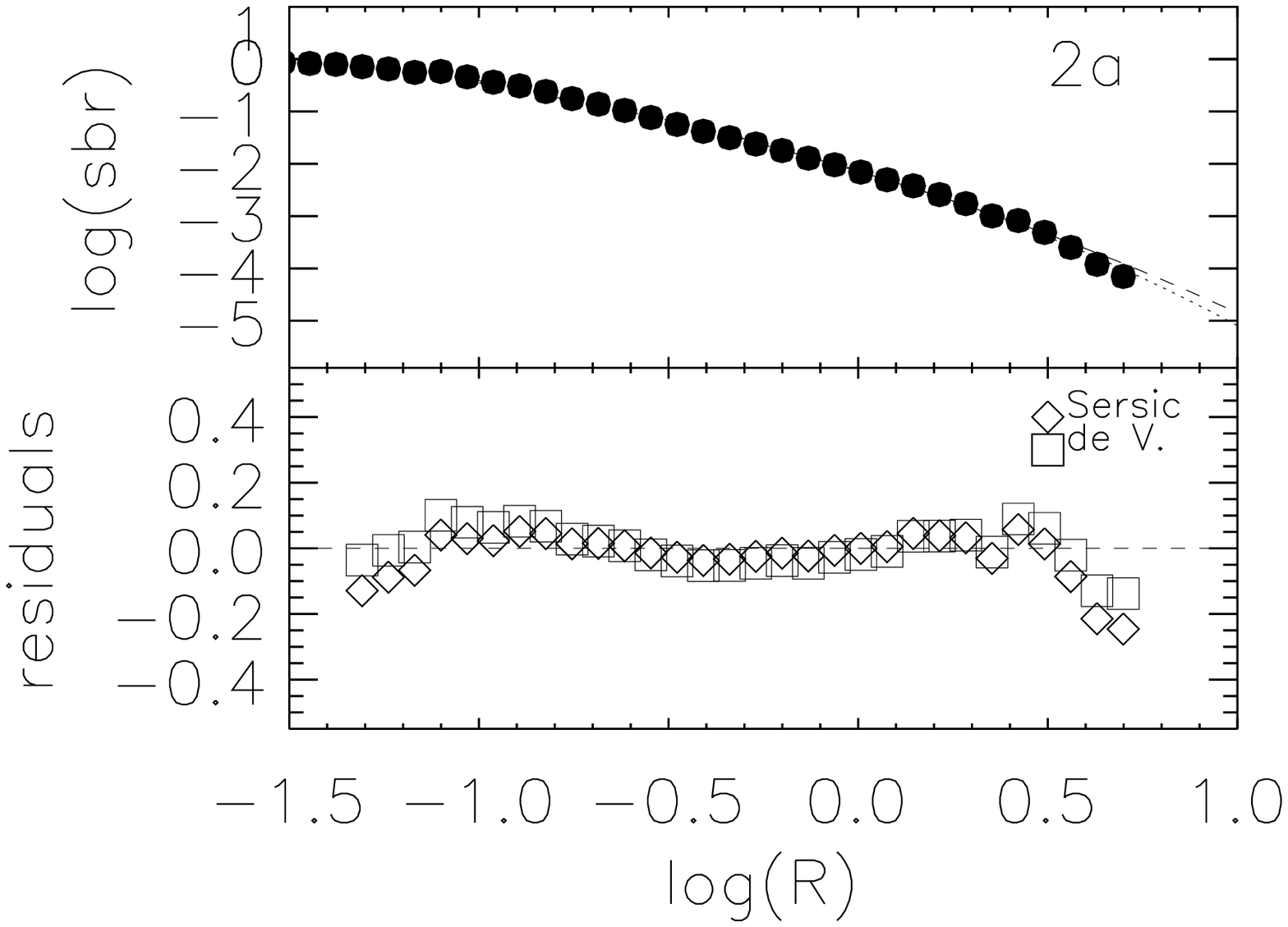}
\hspace{0.cm}
\includegraphics[width=4.5cm]{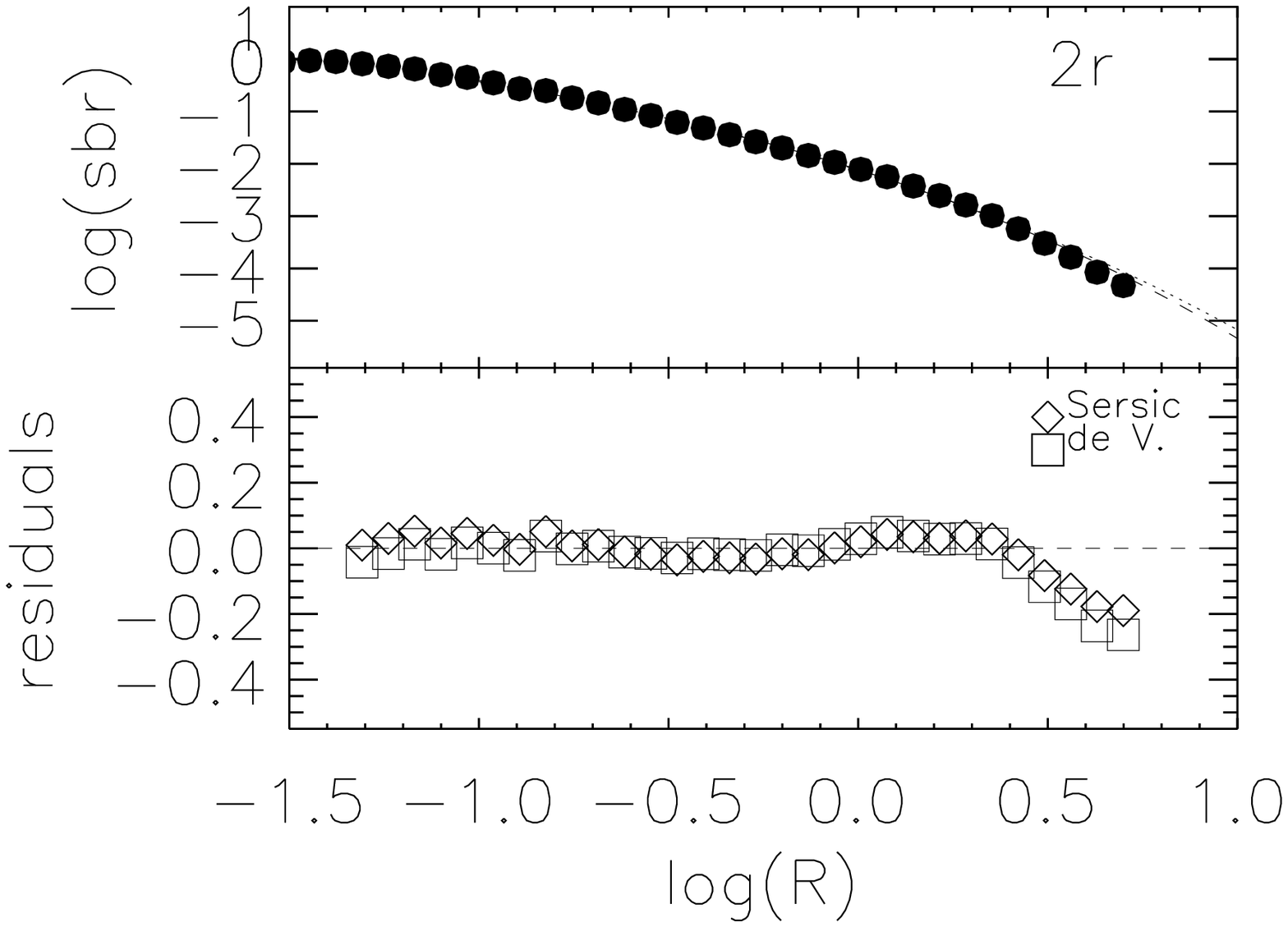}
\vspace{0.cm}
\includegraphics[width=4.5cm]{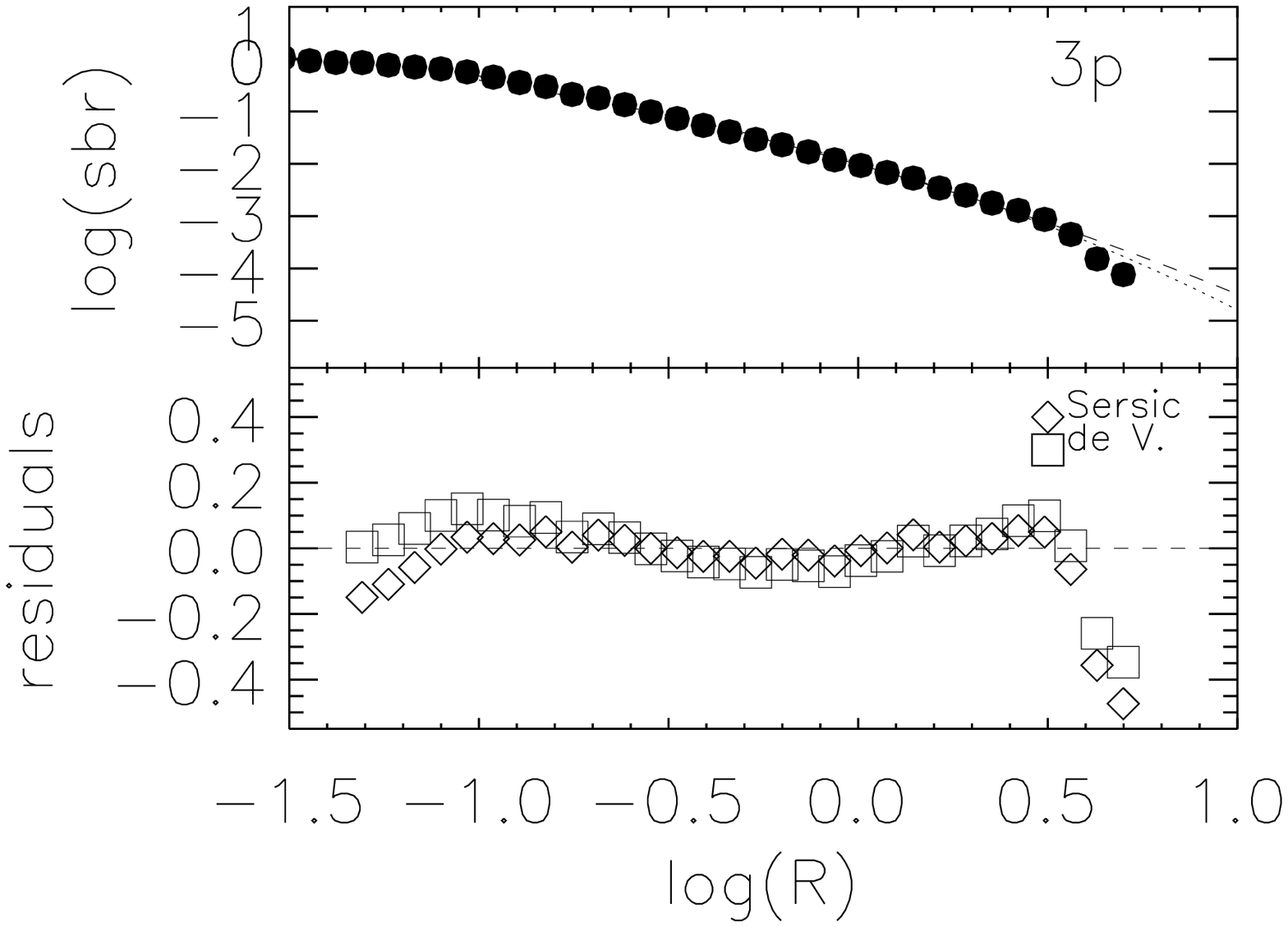}
\hspace{0.cm}
\includegraphics[width=4.5cm]{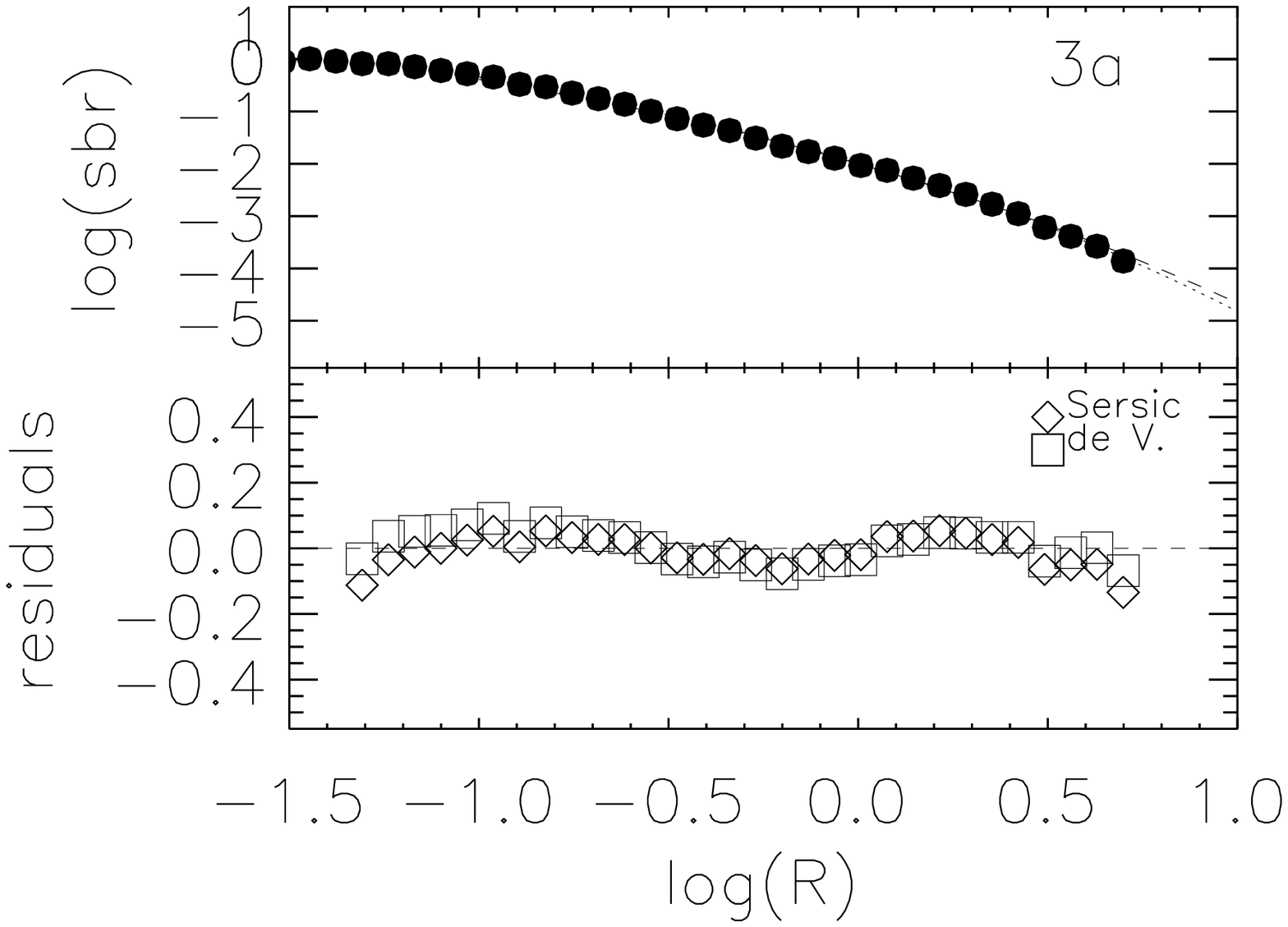}
\hspace{0.cm}
\includegraphics[width=4.5cm]{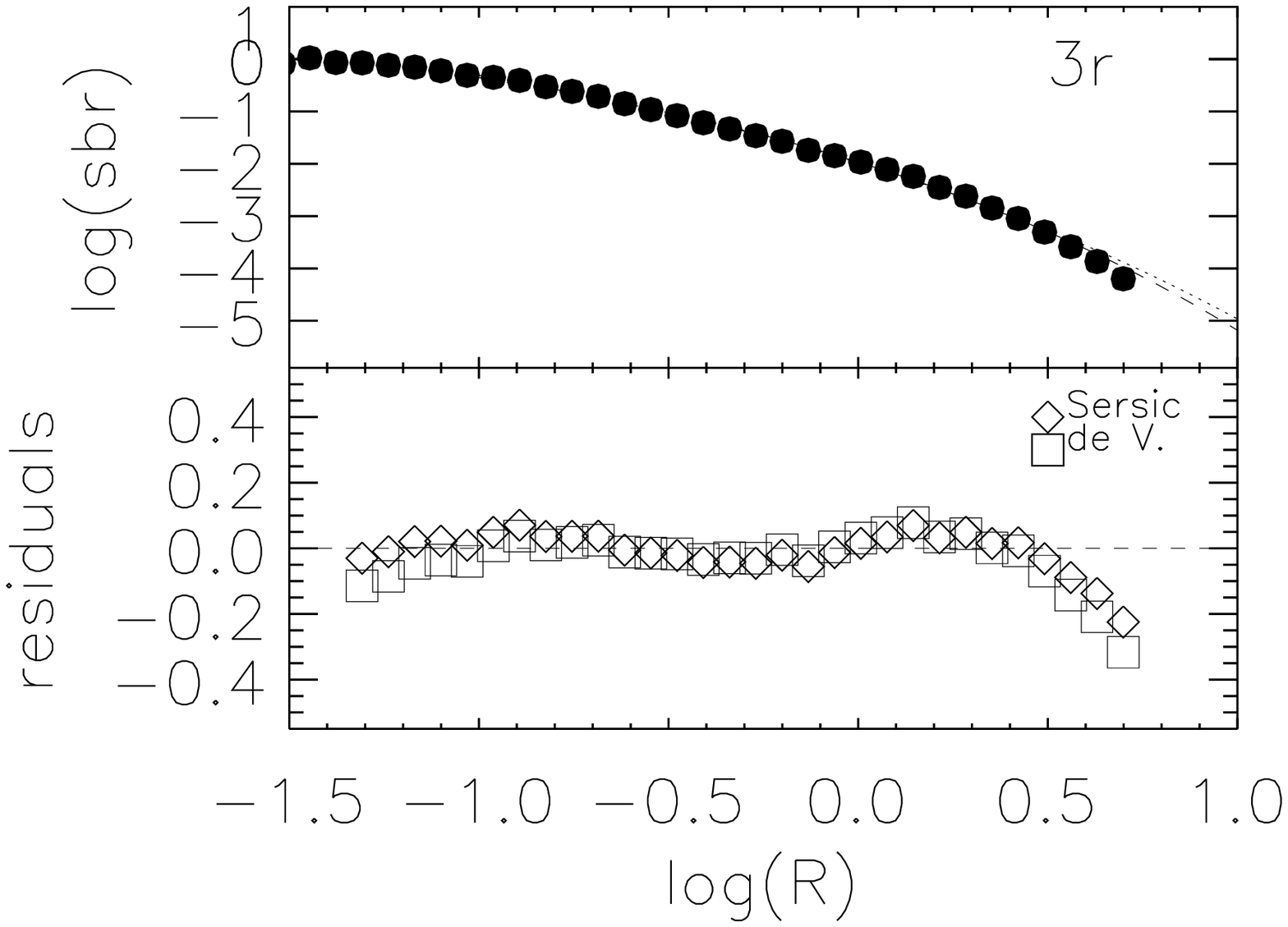}
\vspace{0.cm}
\caption{ Surface density profiles for the nine dbh models. For each model, the top panel shows the surface density profile (dots), the best-fitting \R14\ model (dotted line), and the best-fitting S\'ersic model (dashed line). The bottom panel shows the residuals from the two fits.\label{sbrdbh}}
\end{figure*}

\begin{figure*}
\centering
\includegraphics[width=4.5cm]{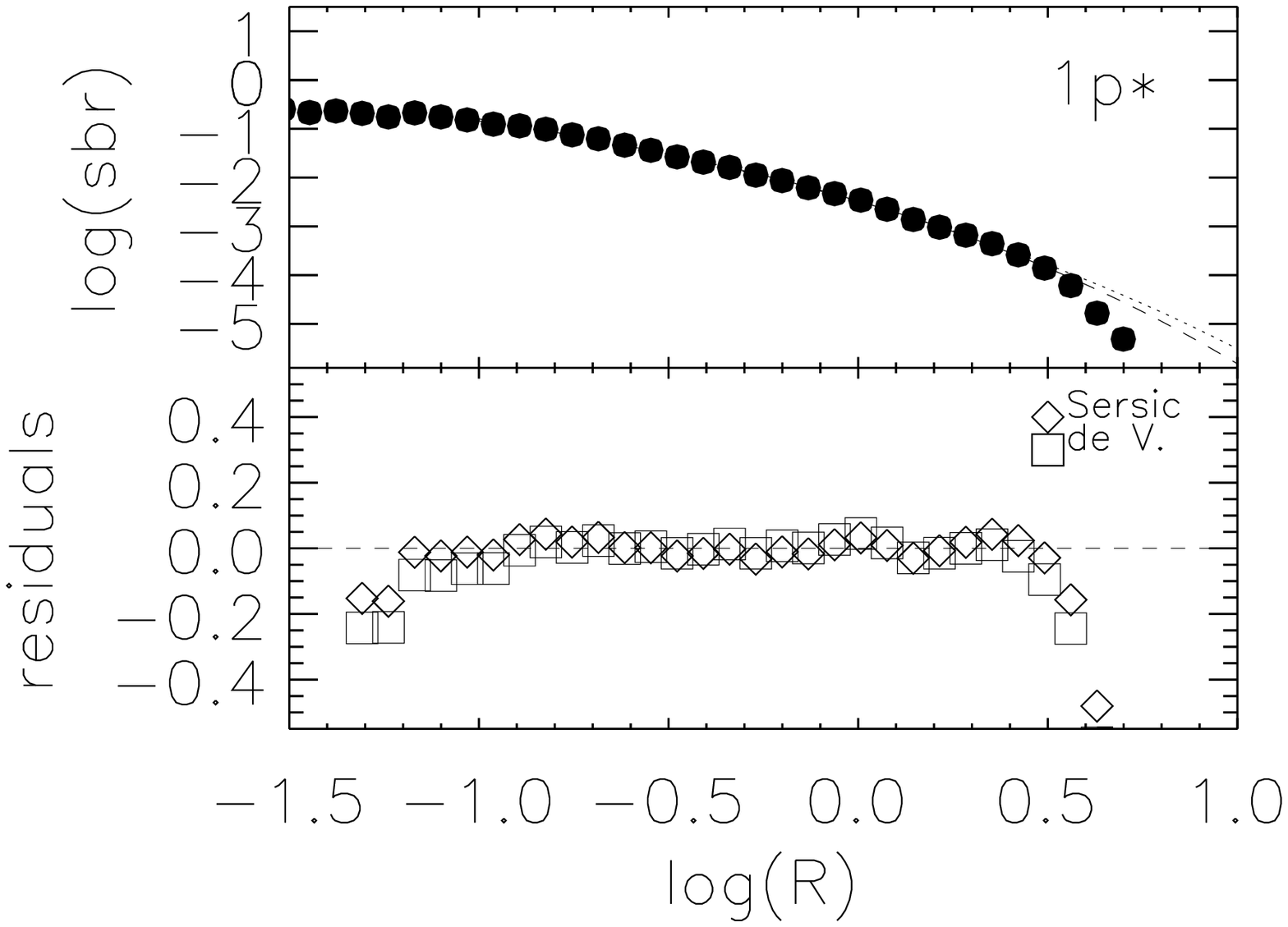}
\hspace{0.cm}
\includegraphics[width=4.5cm]{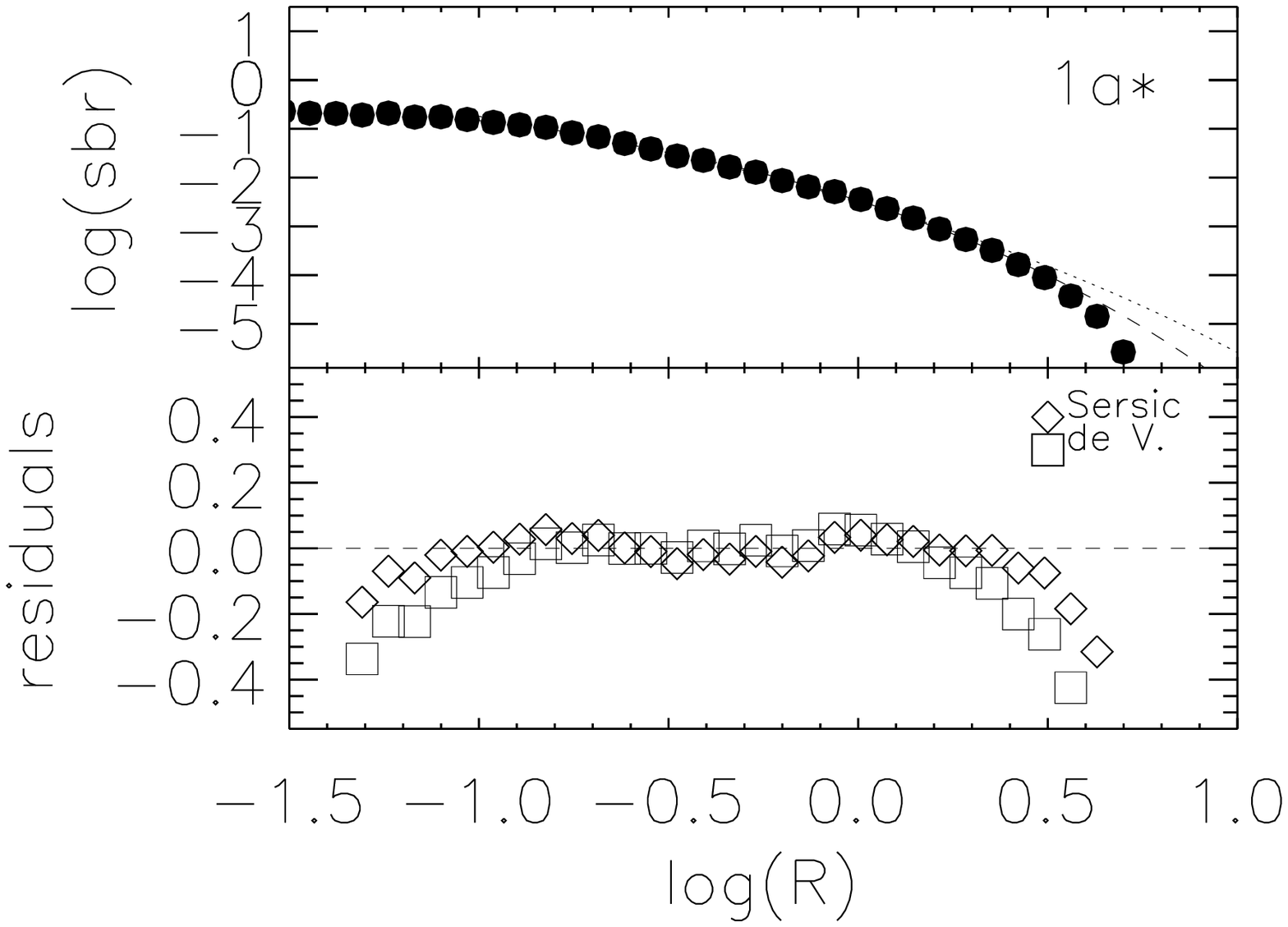}
\hspace{0.cm}
\includegraphics[width=4.5cm]{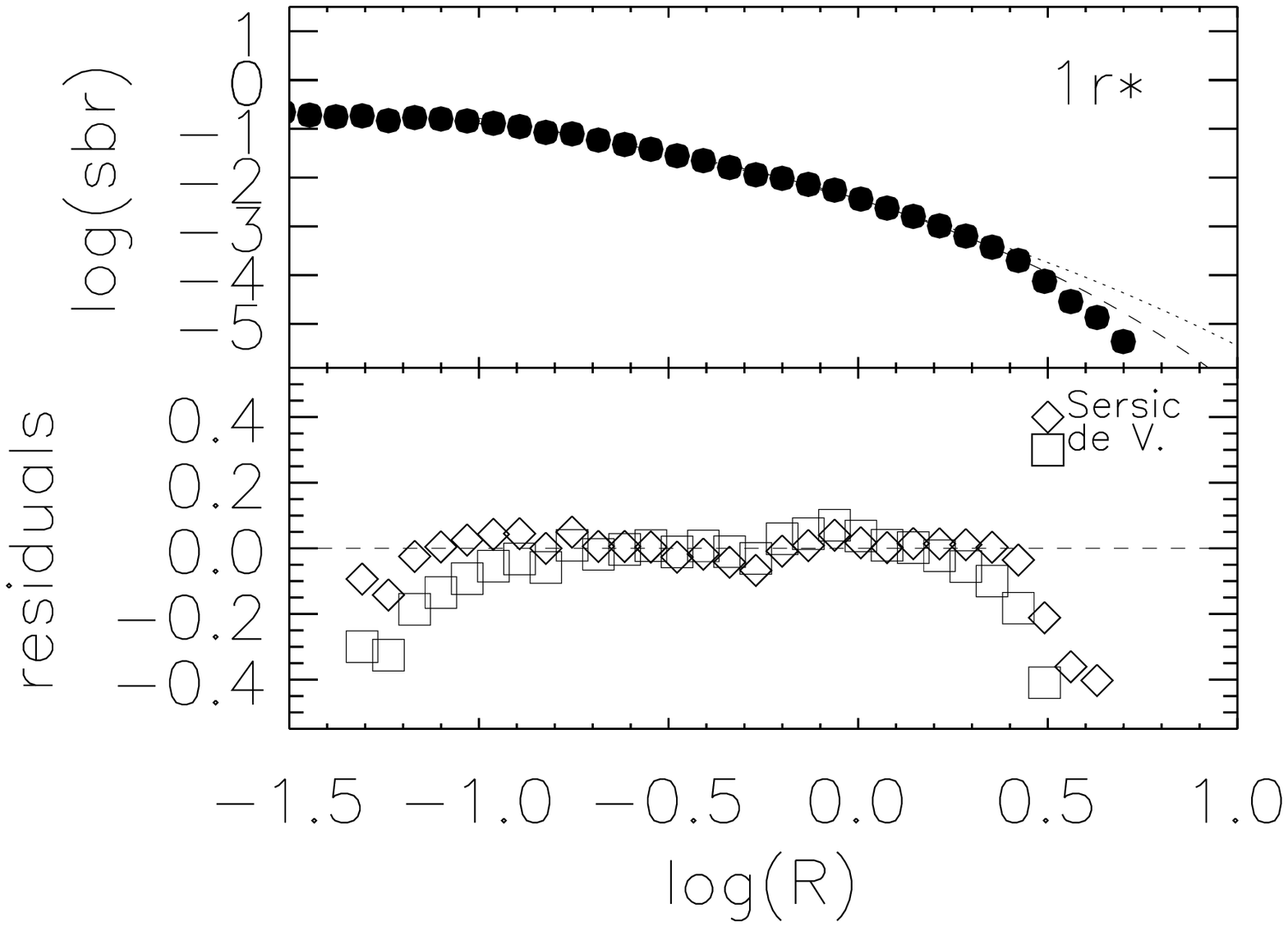}
\vspace{0.cm}
\includegraphics[width=4.5cm]{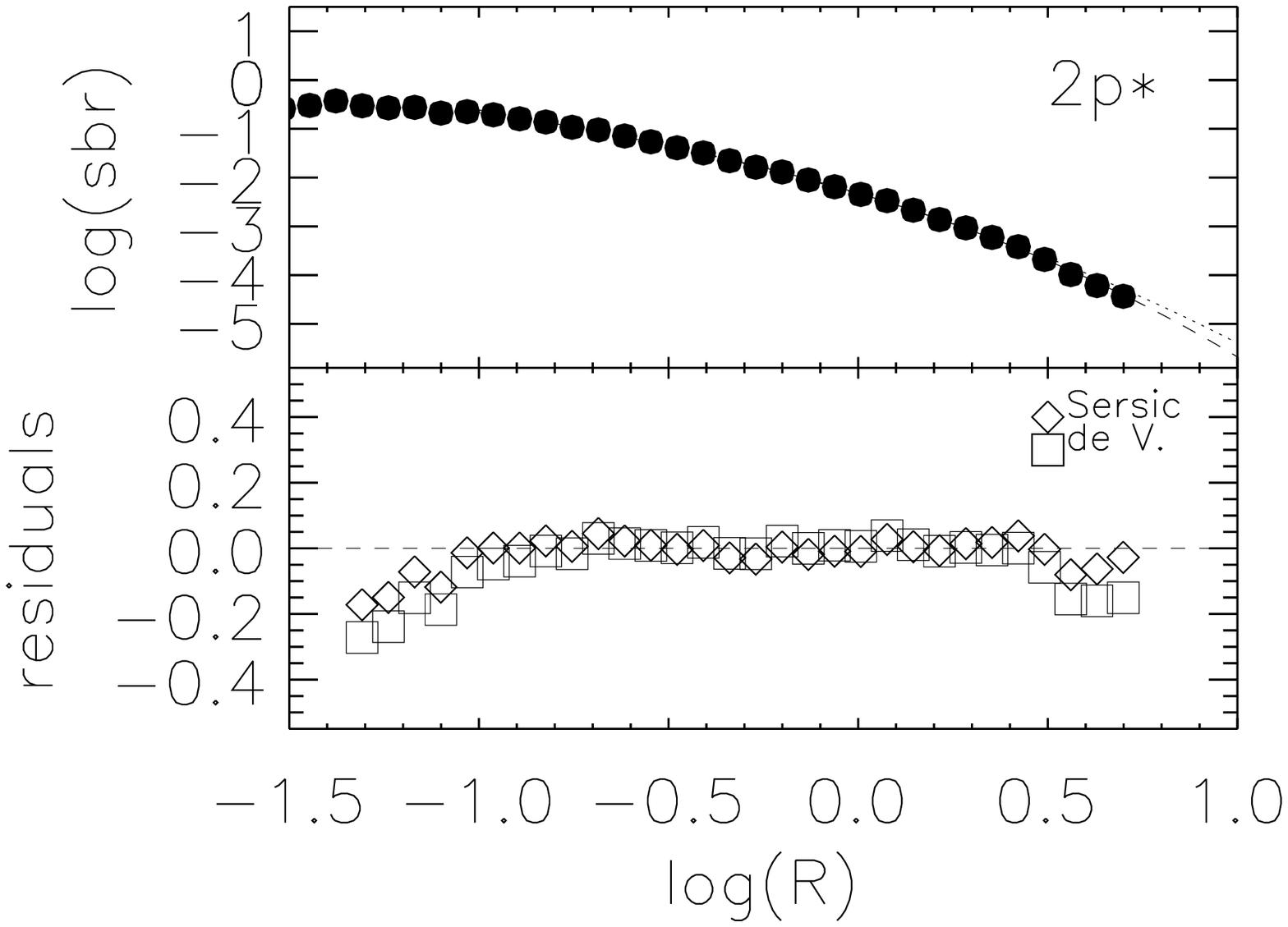}
\hspace{0.cm}
\includegraphics[width=4.5cm]{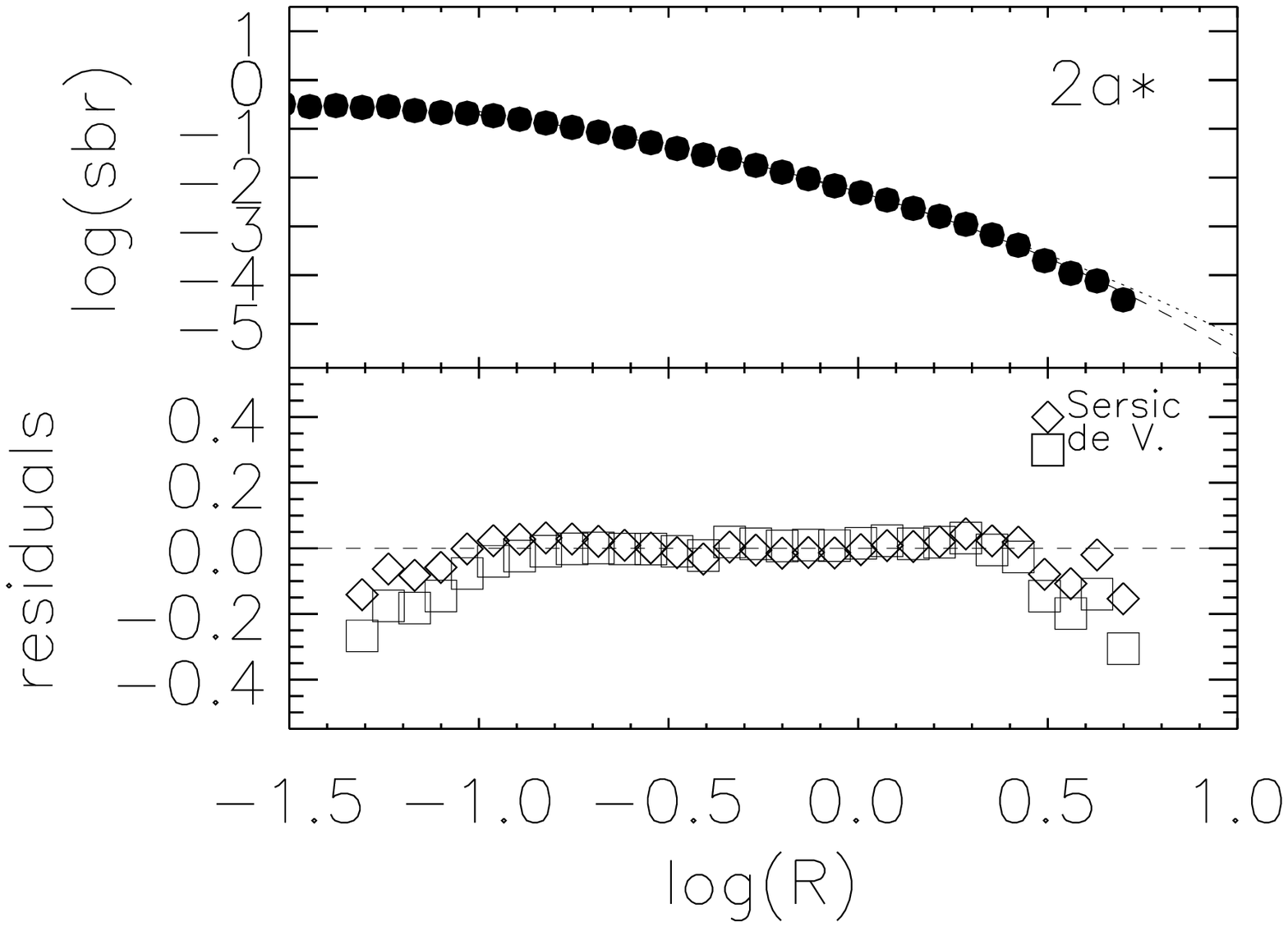}
\hspace{0.cm}
\includegraphics[width=4.5cm]{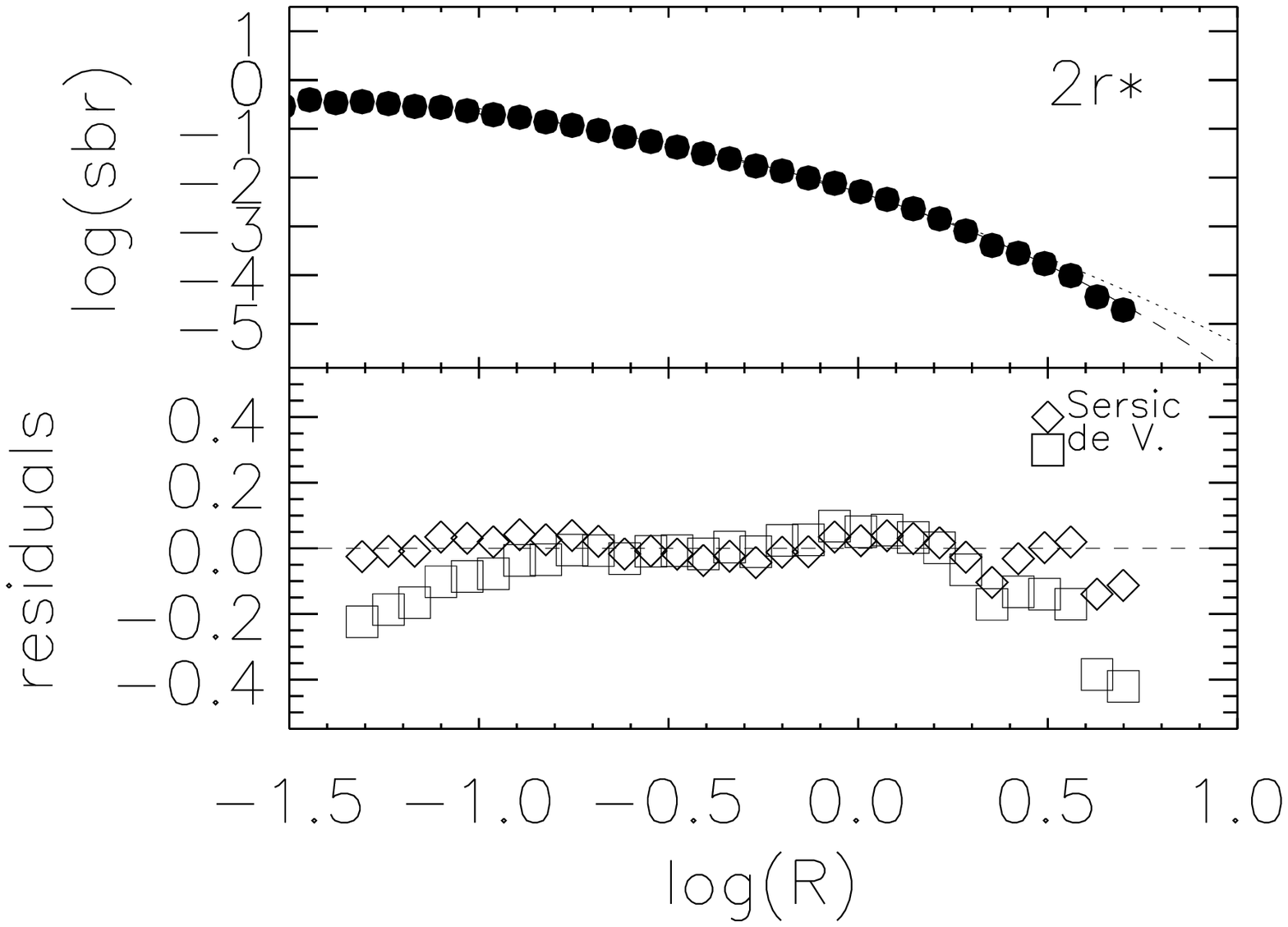}
\vspace{0.cm}
\includegraphics[width=4.5cm]{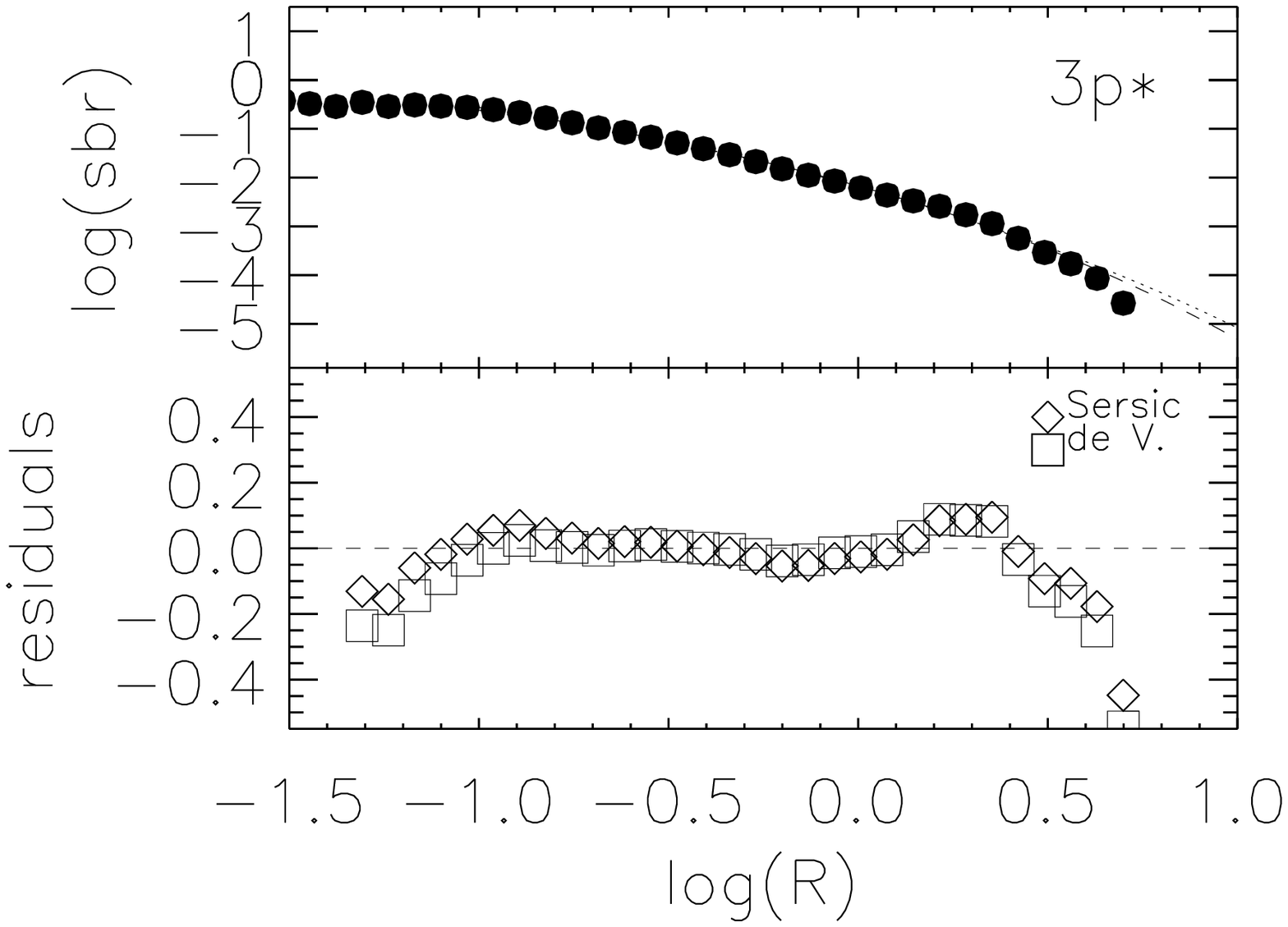}
\hspace{0.cm}
\includegraphics[width=4.5cm]{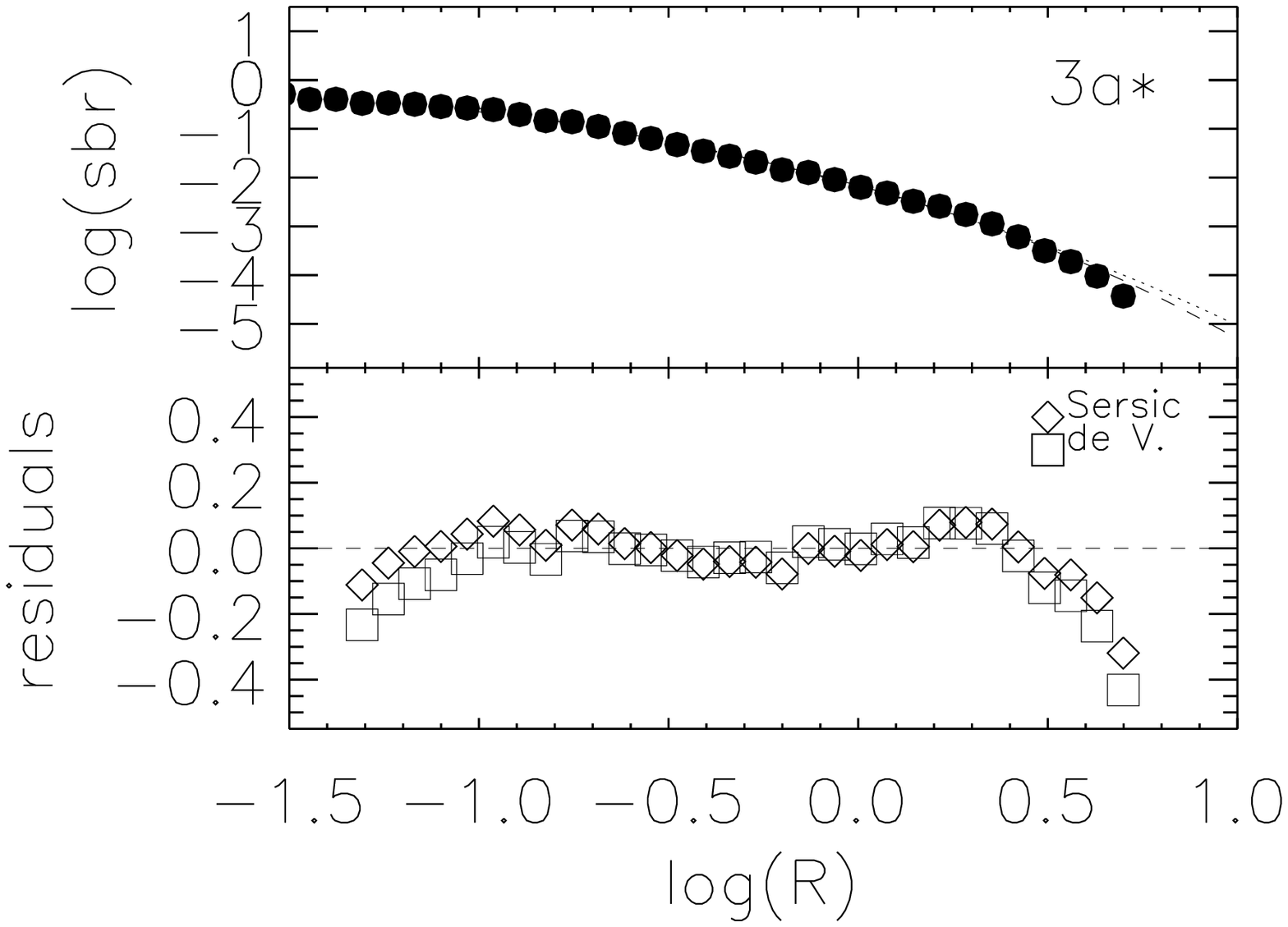}
\hspace{0.cm}
\includegraphics[width=4.5cm]{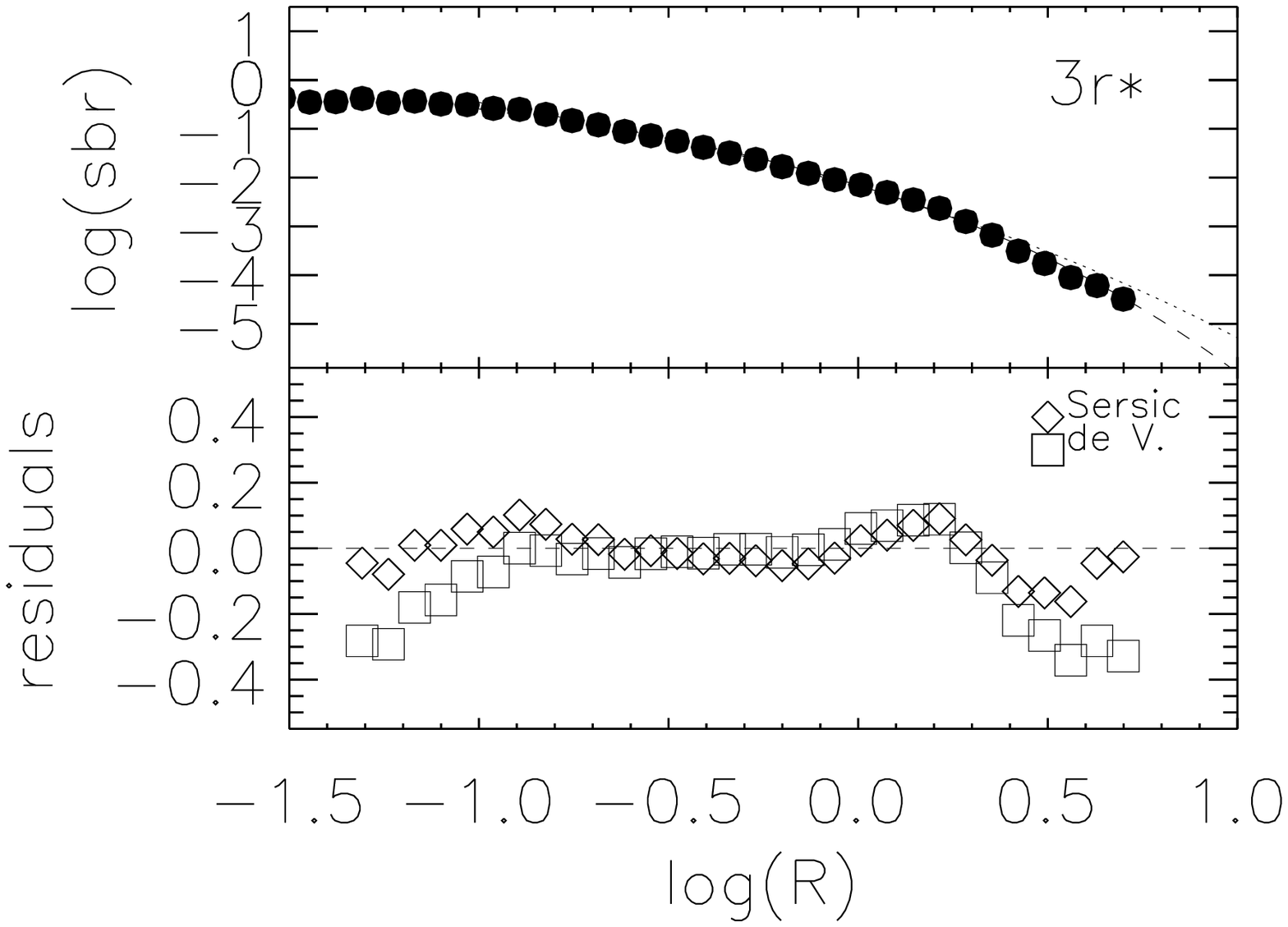}
\vspace{0.cm}
\caption{Surface density profiles for the nine dh models. For each model, the top panel shows the surface density profile (dots), the best-fitting \R14\ model (dotted line), and the best-fitting S\'ersic model (dashed line). The bottom panel shows the residuals from the two fits.  \label{sbrdh}}
\end{figure*}

We derive surface density profiles from ellipse fits to the isodensity contours of the merger remnants, as viewed from a point of view parallel to the initial orbital angular momentum. Surface density profiles for the nine dbh models are shown in Figure \ref{sbrdbh}, while those of the dh models are given in Figure \ref{sbrdh}.  

These profiles are fitted using a least-square method to a de Vaucoleurs (1958) \R14\ profile:

\begin{equation}
I = I_{\rm e} \exp{\{ -7.61[(R/R_{\rm e})^{1/4}-1] \} },
\end{equation}
 and also to a S\'ersic (1968) \Rn\ profile:

\begin{equation}
I = I_{\rm e} \exp{\{ (-1.9992 n+0.3271)[(R/R_{\rm e})^{1/n}-1] \} }.
\end{equation} 
where \Re~ is the radius enclosing half the luminous mass, \Ie~ is the luminous surface density at \Re, and $n$ is the profile shape index, with $n=1$ corresponding to an exponential profile, and $n=4$ to the de Vaucouleurs law.
In order to avoid softening effects at small radii, we have only fitted points outside $2\varepsilon$, with $\varepsilon = 0.02$ for dbh models and 0.09 for dh models. Fits extend out to $R=5$, the truncation radius of the initial discs.  Best-fitting parameters for the S\'ersic and de Vaucouleurs fits are given in Table~\ref{tabr14}, together with the \rms\ from each fit. In Figures \ref{sbrdbh} and \ref{sbrdh}, the bottom panels for each of the plots give the residuals from both fits.

The surface density profiles are well approximated by a de Vaucouleurs profile, as found in previous works (Barnes 1992,  Hernquist 1992, 1993).  However, inspection of the \rms\ values (Table~\ref{tabr14}) shows that a better fit is provided with the S\'ersic model, especially for the db models.  This is reassuring, since the S\'ersic model provides a better fit to real ellipticals than the \R14\ profile as well (Caon et al.\ 1993).  Remnants of dbh models show S\'ersic indices $n=3.3$ to $n=8$, while dh models have lower $n$ values, $n=2.4$ to $n=3.2$.  The lower $n$ of the dh models are a result of the lower central densities of the initial models, highlighting that mergers of bulge-less discs cannot account for the central densities of elliptical galaxies (Hernquist 1992).  However, recent results from Trujillo et al. (2004) show that giant elliptical galaxies, brighter than $-18 M_B$, cover a range of values for the Sersic exponent from 2 to 9, which would be in agreement with the range for dh experiments. We note that the S\'ersic indices of dbh models do not reflect the density profiles of the initial bulges, which are modeled as King profiles and therefore approach an exponential rather than a \R14\ profile.  

All the profiles deviate from both the de Vaucouleurs and the S\'ersic laws at radii $R \gtrsim 3$.  The mass involved is small ($\sim$5\% of the initial luminous mass).  Most likely, the deviations reflect the initial disc truncation radius ($R=5$) combined with the fact that the outer parts are not yet fully virialized:  after merging, the systems were let to relax for about 10 time units (i.e. 40 half-light radius crossing times), which is of the order of the remnants' crossing times at a radius $R=3$.  

\begin{table*}
\begin{minipage}{140mm}
\begin{center}
\caption{ Best-fitting parameters for Sersic and de Vaucouleurs fits to the surface density profiles. For each model, the first line gives the best-fit parameters for a Sersic profile, and the second line the parameters for a de Vaucouleurs fit ($n=4$). \label{tabr14}}
\begin{tabular}{cccccccccc}
\hline

{\bf Model} &{\bf $\log($\Ie$)$}& {\bf \Re} & {\bf $n$}& {\bf rms} &{\bf Model} &{\bf $\log(I_{\rm e})$}& {\bf \Re} & {\bf $n$}  & {\bf rms}\\
(1) & (2) & (3) & (4) & (5) & (6) & (7) & (8) & (9) & (10)\\
\hline
\hline
$1p$ & -1.779    & 0.519    &   8.05 & 0.059& $1p*$ & -1.937  &     0.545  &     3.23&0.056\\
	&     -1.700   &   0.505   &    4.00 & 0.084& &    -1.953    &  0.549   &    4.00&0.080\\

$1a$ & -1.538 &    0.407 &      5.54& 0.049 &$1a*$ &-1.802   &    0.491   &    2.65& 0.059\\
	 &   -1.538   &   0.423    &   4.00& 0.051 &&     -1.851  &   0.494  &     4.00 & 0.124\\

$1r$ & -1.492     &  0.398   &     4.34& 0.027& $1r*$ & 1.892  &     0.552    &   2.50& 0.089\\
	&    -1.497   &    0.404   &    4.00 & 0.028 & &    -1.969  &    0.574  &     4.00 & 0.157\\
\hline
$2p$ &-1.691   &  0.601    &  4.88& 0.045& $2p*$ &-1.748   &   0.533   &    3.24& 0.055\\
   & -1.674 &    0.602   &   4.00  & 0.049& &     -1.738  &    0.519    &   4.00& 0.081\\

$2a$ & -1.666   &  0.584   &   5.17&0.050&$2a*$ & -1.826    &   0.593 &     3.08& 0.048\\
   & -1.649 &    0.585    &  4.00&0.048& &   -1.838   &   0.586  &     4.00& 0.081\\

$2r$ & -1.526   &  0.525    &  3.53&0.044&$2r*$ & -1.691  &     0.520   &    2.66& 0.034\\
&  -1.538   &  0.523  &    4.00&0.048&&      -1.682  &    0.496   &    4.00& 0.082\\
\hline
$3p$ & -1.769  &   0.758 &      6.00&0.055&$3p*$ & -1.802   &    0.654  &     3.26& 0.073\\
  &  -1.709    &  0.724   &    4.00&0.060& &     -1.802   &   0.646   &    4.00 & 0.082\\

$3a$ & -1.682 &     0.698  &     4.83&0.046&$3a*$ & -1.838  &     0.689 &      3.14& 0.061\\
  &  -1.666 &     0.697  &    4.00&    0.047&&   -1.851  &    0.686  &     4.00& 0.070\\

$3r$ & -1.481   &    0.583 &      3.34&0.043&$3r*$ & -1.604    &  0.545 &      2.56& 0.068\\
  &  -1.487 &     0.574     &  4.00&0.049&&-1.563   &   0.498   &    4.00& 0.120\\
\hline  
\hline
\end{tabular}
\end{center}
\end{minipage}
\end{table*}

\begin{figure}
\centering
\includegraphics[width=8.2cm]{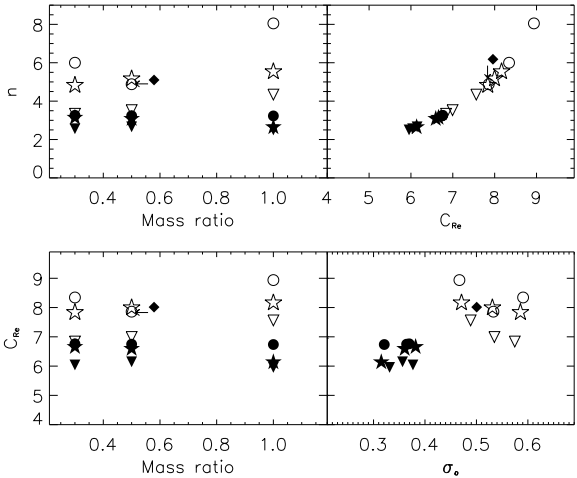}
\caption{Top left: mass ratio versus index $n$ of S\'ersic profile fit. Top right: Trujillo et al (2001) concentration parameter $C_{Re}$ versus $n$. Bottom left: $C_{Re}$ versus mass ratio. Bottom right: $C_{Re}$  versus central velocity dispersion.  Open symbols are dbh models while filled symbols are dh models. Circles are `p' models, stars are `a' models and triangles are `r' models. The diamond indicates model $2aG$.
\label{conc}}
\end{figure}

Perhaps surprisingly, the S\'ersic index $n$ does not depend on the mass ratio of the merged galaxies.  Instead, the final profile shapes are affected by the alignments of the galaxy spins with the orbit.  Prograde and anti-parallel configurations give higher values of $n$ than orbits where both spins are opposite to the orbital angular momentum (see Fig.~\ref{conc}). This is true for both dbh and dh models, 
and probably reflects more extended mass distributions resulting from higher absorption of orbital energy by the galaxy envelopes in prograde mergers.  

Because $N$-body systems are dimensionless, the models cover a small base-line in luminosity (or mass) and are unable to generate a luminosity-S\'ersic index $n$ diagram that reproduces the trend followed by ellipticals (e.g.\ Caon et al.\ 1993).  However, from the lower S\'ersic indices of bulge-less merger remnants, it appears that the S\'ersic index of the remnant is a tracer of the concentration of the precursors.  We use here the Concentration Index $C_{\rm Re}$ introduced by Trujillo, Graham, \& Caon (2001).  $C_{\rm Re}$ is defined as the ratio of the light within $0.3$\Re to the light within \Re, where \Re is the effective radius derived from the S\'ersic fit. $C_{\rm Re}$ is a monotonic function of $n$ (Fig.~\ref{conc}, top-right panel).  The concentration index $C_{\rm Re}$ is distinctly higher for dbh models than for dh models. Clearly, the presence of the bulge in the progenitors leads to a higher central concentration in the light distribution of the remnants. Higher central concentrations are associated to higher central velocity dispersions (Fig,~\ref{conc}, bottom-right panel), as found in real  ellipticals (Graham, Trujillo, \& Caon 2001).

\subsection{Boxiness-diskiness}
\label{Sec:BoxinessDiskiness}

We calculate the boxiness-diskiness of the isodensity contours of the merger remnants by fitting the density residuals $\delta(\phi)$ along the best-fitting ellipses to a Fourier series (Carter 1978),

\begin{equation}
        \delta(\phi) = \delta + \sum a_n cos (n\phi) + \sum b_n sin (n\phi),
\end{equation}
where $\phi$ is the azimuthal angle sweeping the ellipse, $\delta(\phi)$ is the residual intensity along the best-fitting ellipse, $\delta$ is the mean residual intensity and $n=1,2,...$. For well behaved isophotes, $\delta$, $a_n$ and $b_n$ should be small.  Isophotes are called disky when $a_4$ is positive, and boxy when $a_4$ is negative (e.g.\ Bender 1988).   

\begin{figure}
\centering
\includegraphics[width=4cm]{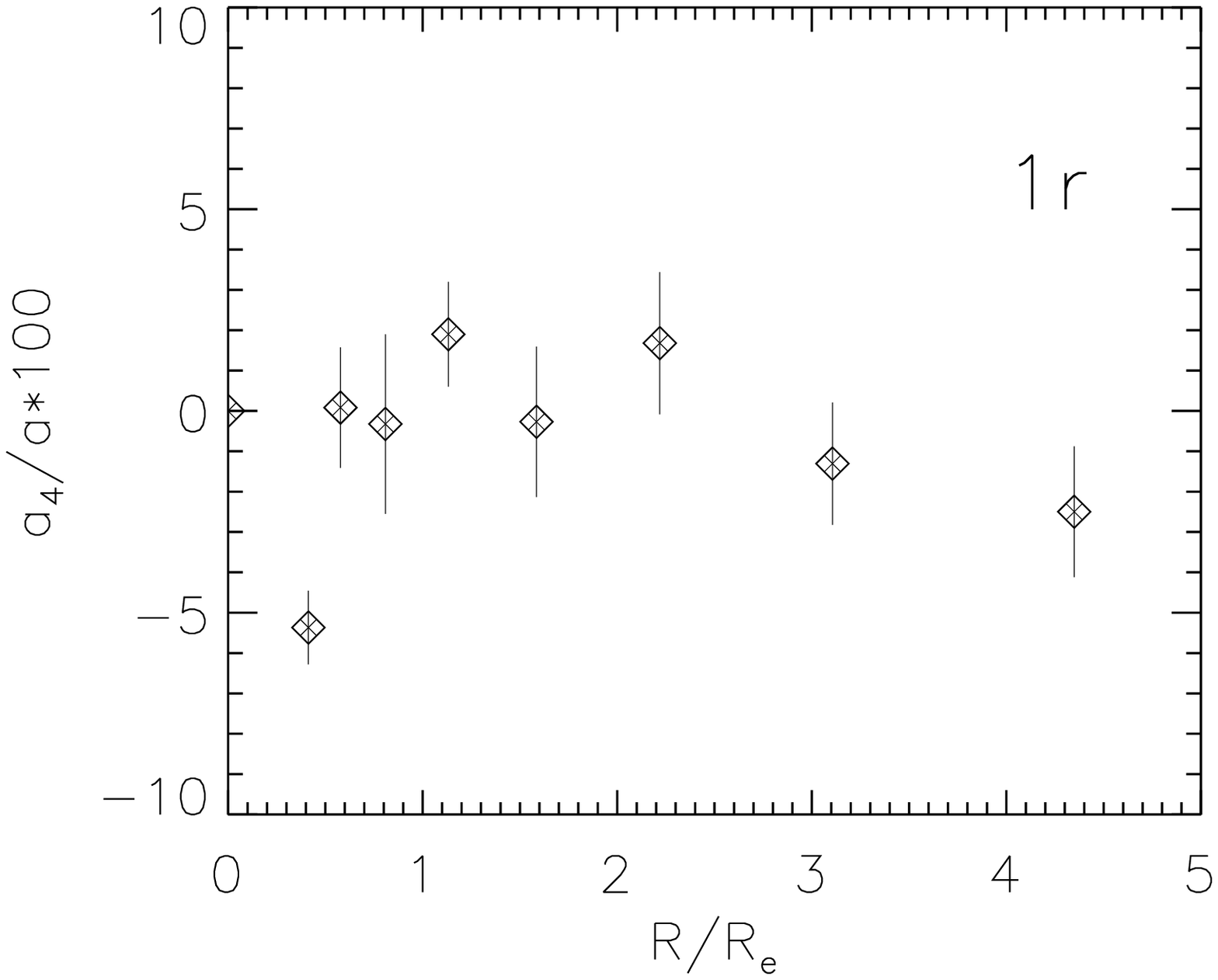}
\hspace{0.cm}
\includegraphics[width=4cm]{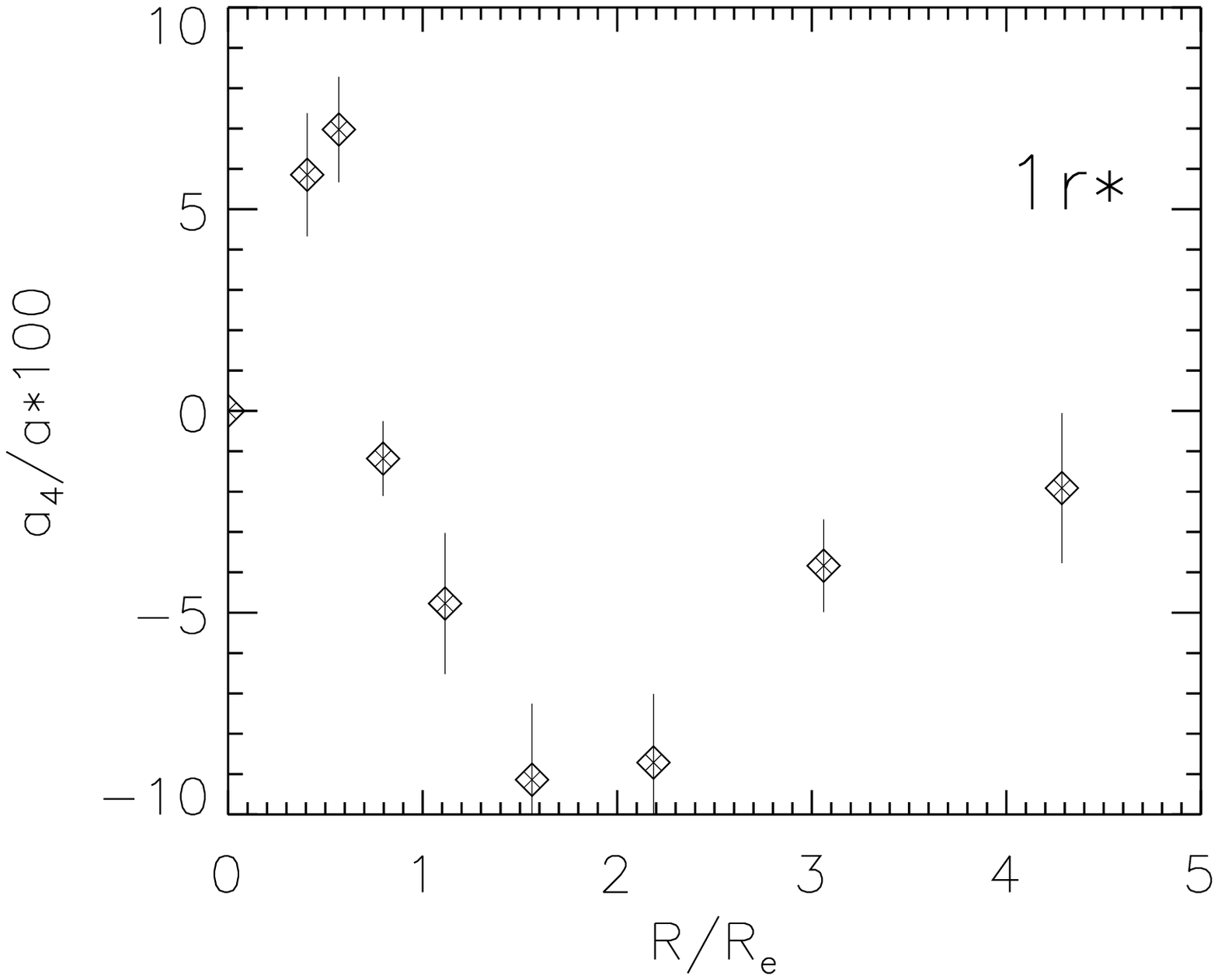}
\vspace{0.cm}
\includegraphics[width=4cm]{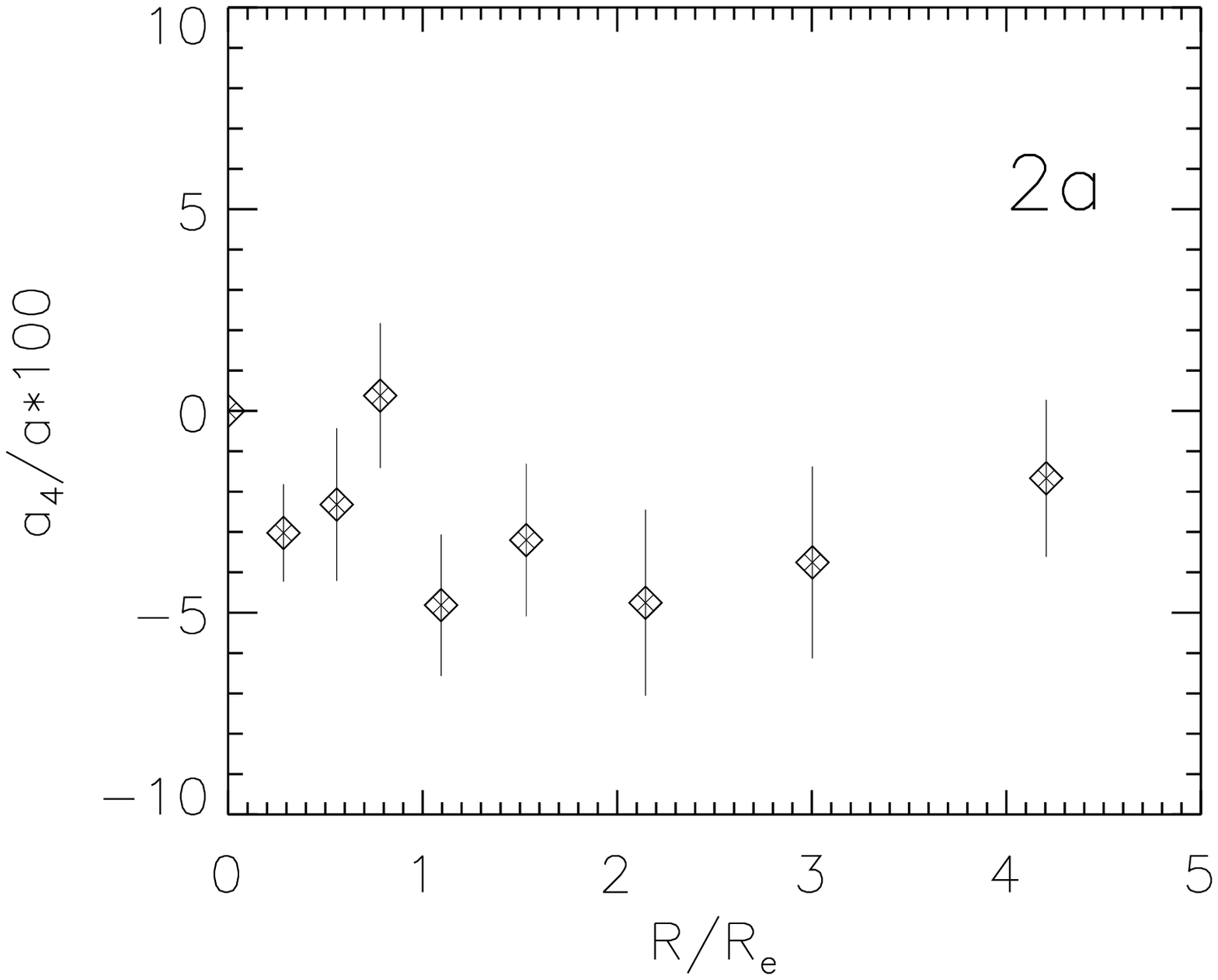}
\hspace{0.cm}
\includegraphics[width=4cm]{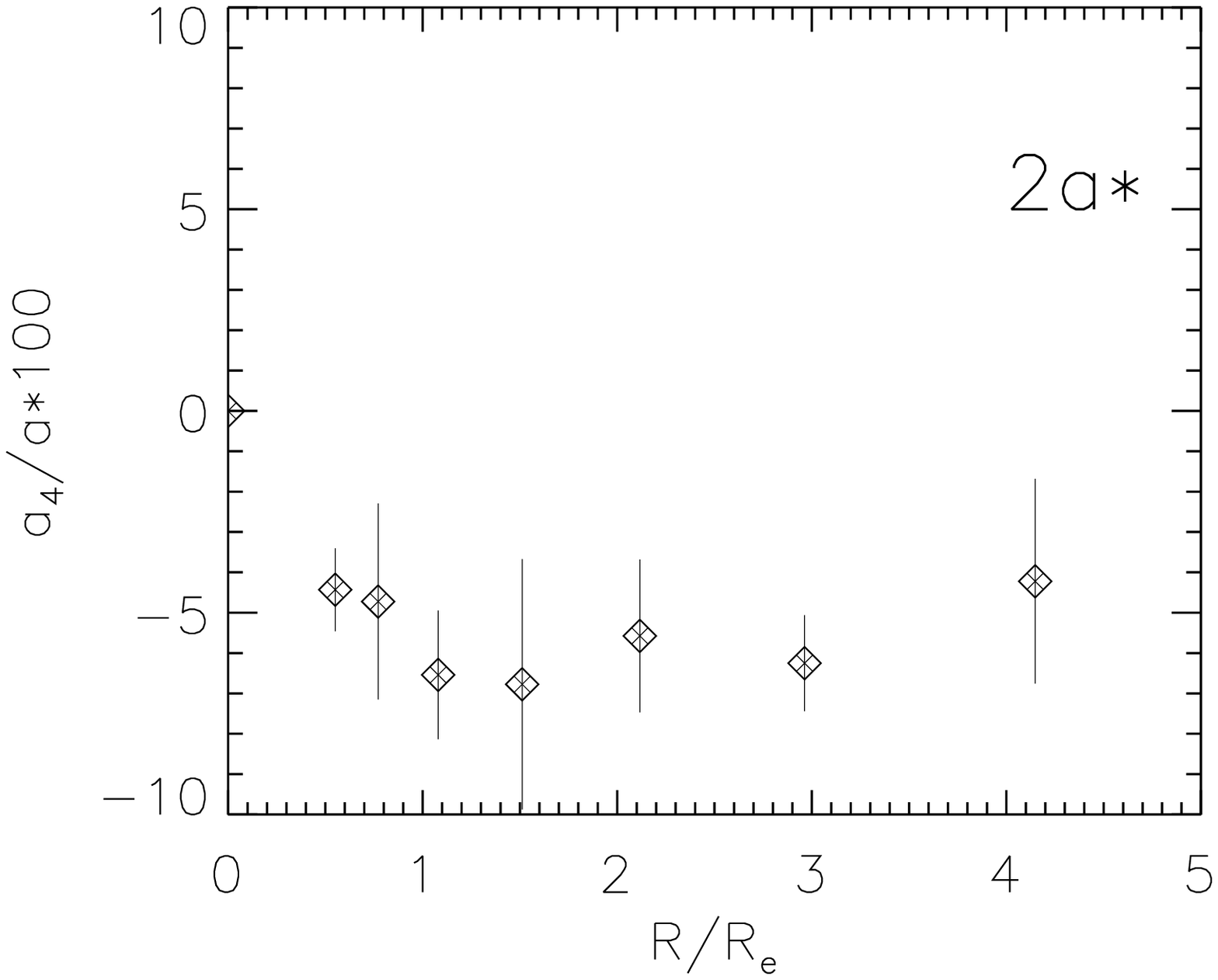}
\vspace{0.cm}
\includegraphics[width=4cm]{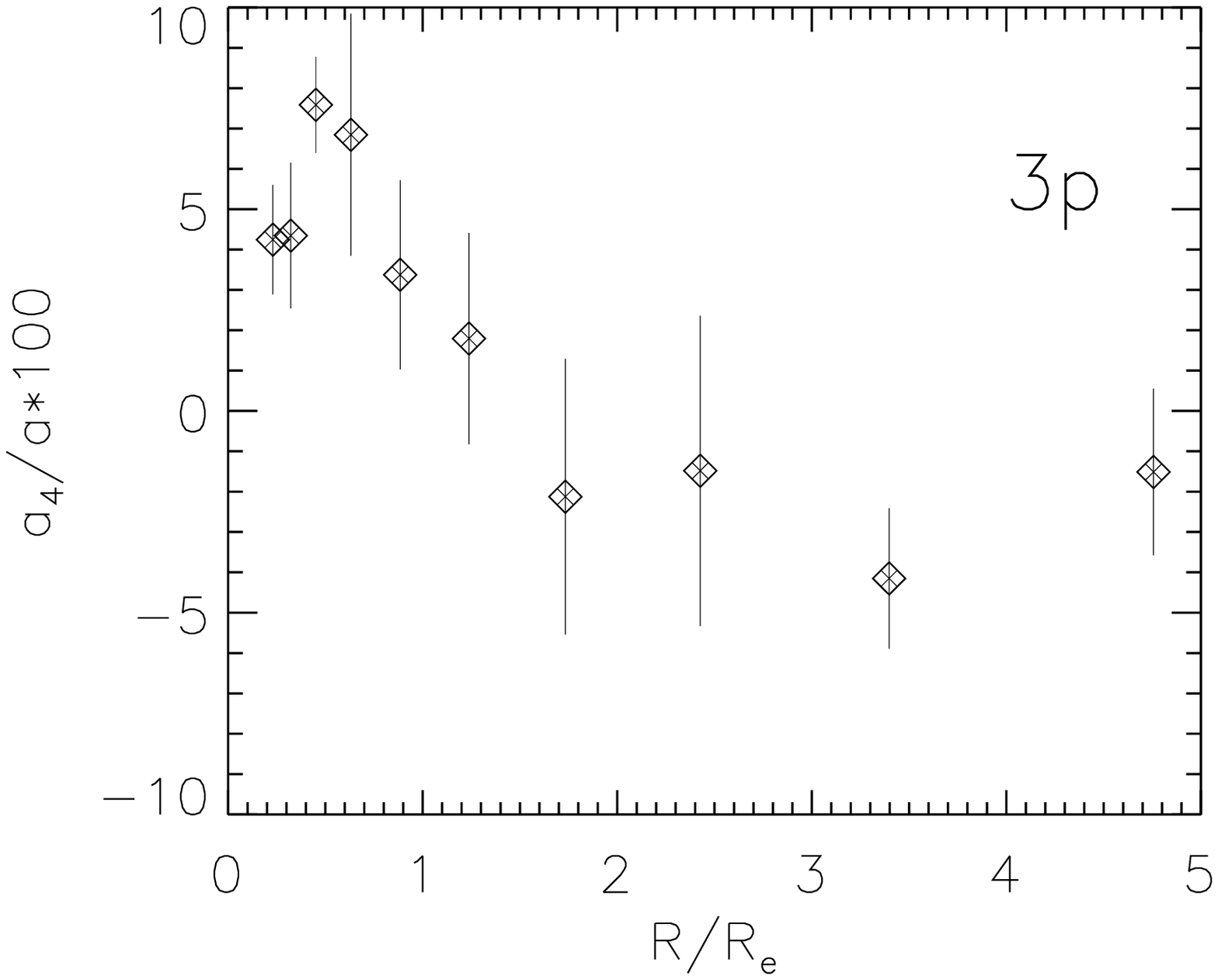}
\hspace{0.cm}
\includegraphics[width=4cm]{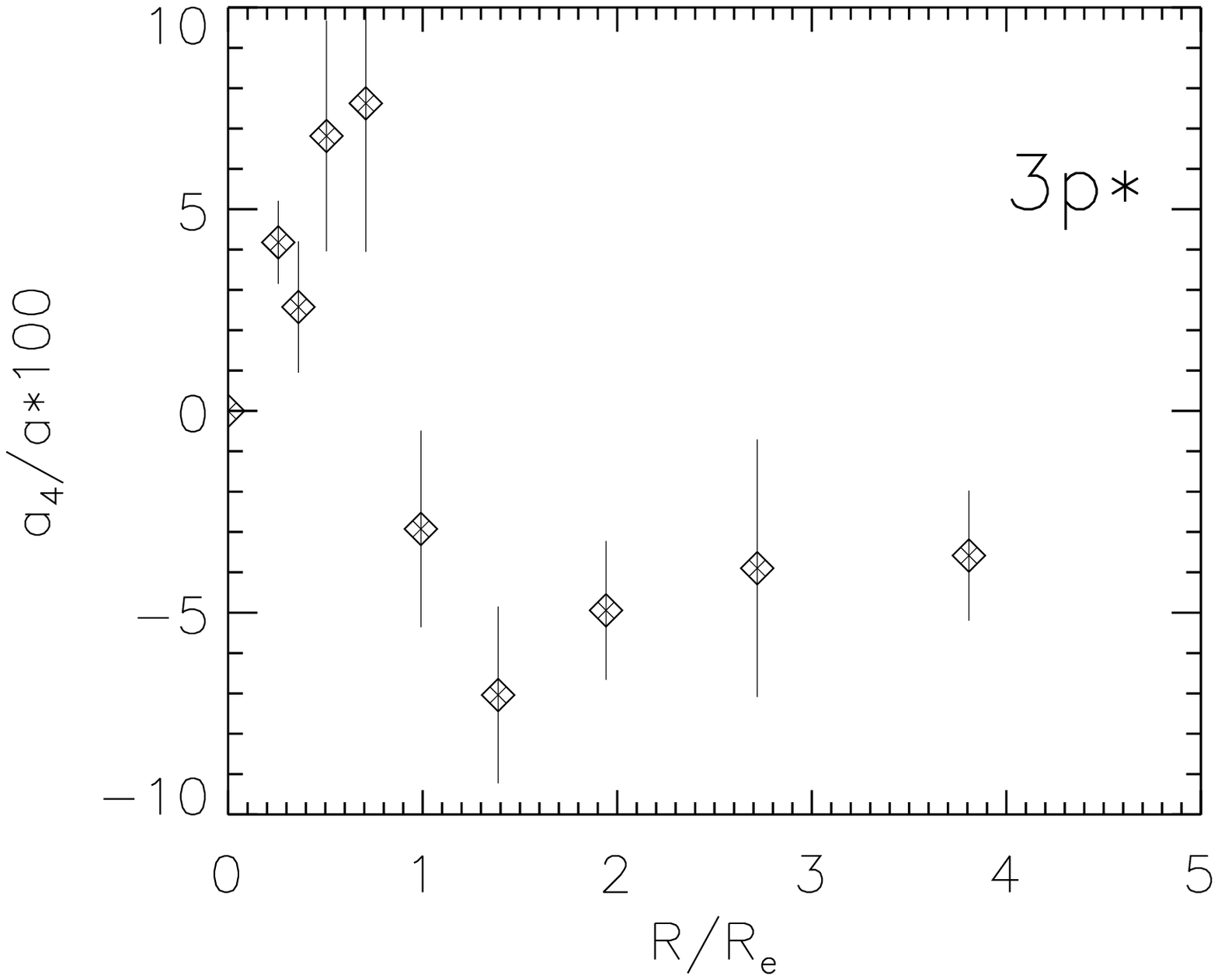}
\vspace{0.cm}
\includegraphics[width=4cm]{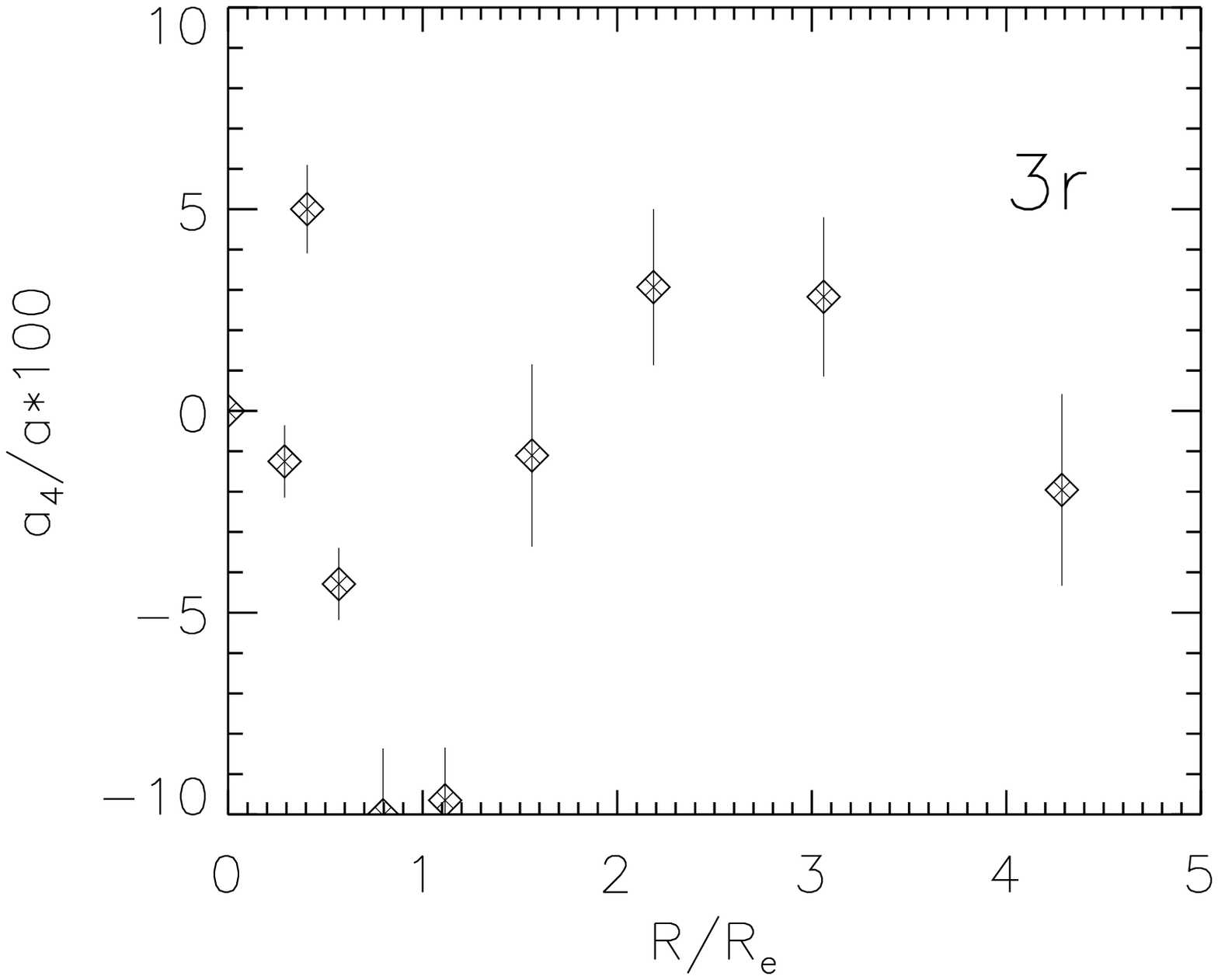}
\hspace{0.cm}
\includegraphics[width=4cm]{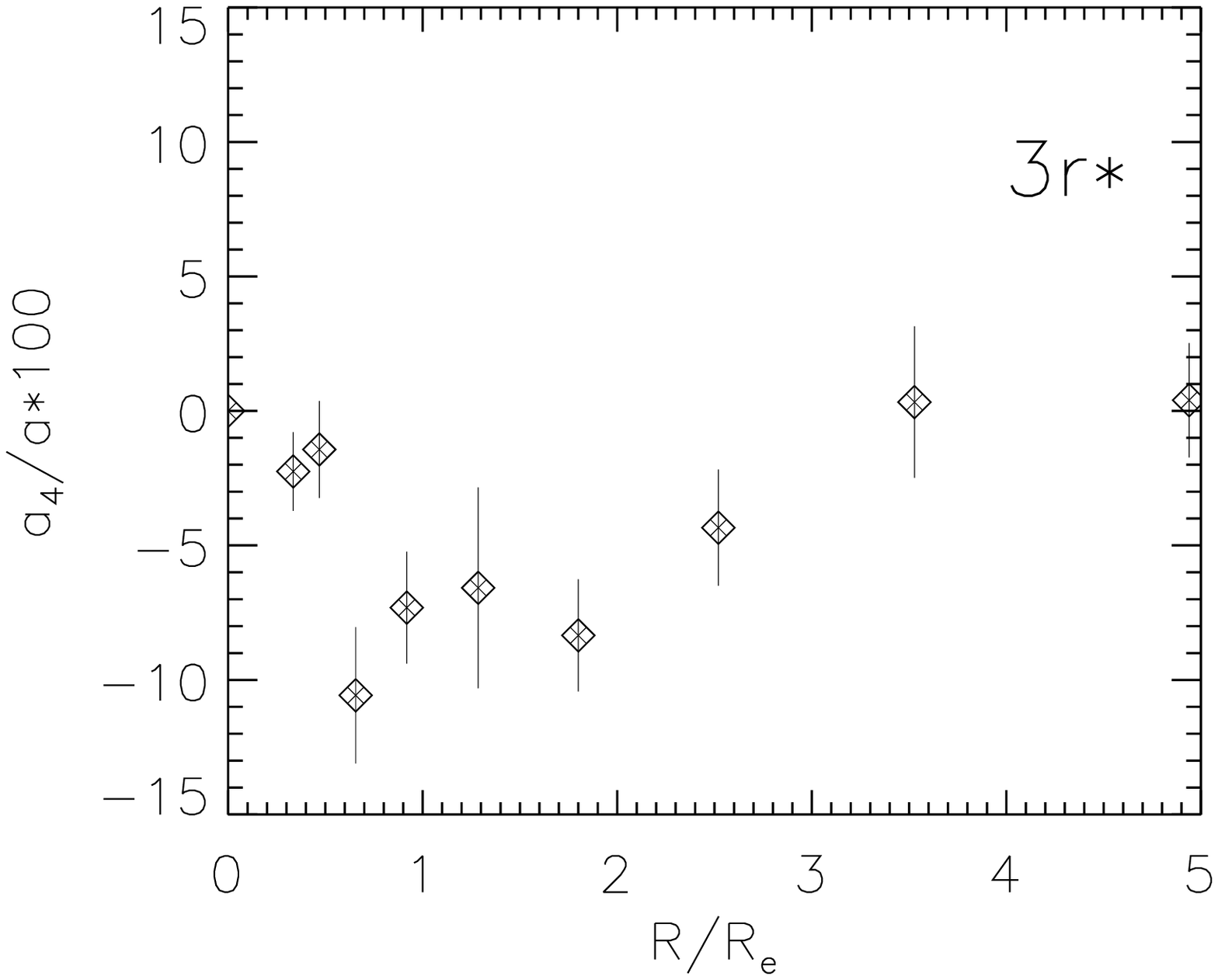}
\vspace{0.cm}
\caption{Boxiness-diskiness, as measured by $100 a_4/a$, vs. $R/R_{\rm e}$ for several models. The point of view is parallel to the intermediate axis at the half-light radius of the remnant. \label{boxidmulti}}
\end{figure}

We show profiles of the $a_4/a \time 100$ parameter for eight of our models in Figure \ref{boxidmulti}.  The view point is parallel to the intermediate axis of the inertia ellipsoid, and the average of 60 snapshots, closely spaced in time, has been taken to increase the signal-to-noise.  
Amplitudes of the boxiness/diskiness deviations are a few percent, but show peaks reaching 10\%, which is of concern as $a_4$ deviations in ellipticals are typically $<3$\% (Bender 1988).  However, a number of patterns are apparent in Figure~\ref{boxidmulti}, which suggest that we are not just measuring noise.  We find that the isophotal properties of dbh and dh remnants are quite different.  dbh models gradually change from boxy ($1r$) to disky ($3p$) as the mass ratio of the merging galaxies increases.  Naab et al.\ (1999) reported the same trend, which prompted these authors to conjecture that intermediate-luminosity ellipticals might be remnants of mergers of unequal disc galaxies. Naab \& Burkert (2003) also conclude that a trend from boxy to disky can be observed with the mass ratio.  Whether such association is real is currently unclear;  Naab \& Burkert (2001) rejected it on the basis of the line-of-sight velocity distribution properties of the remnants.  

\begin{figure}
\centering
\includegraphics[width=4cm]{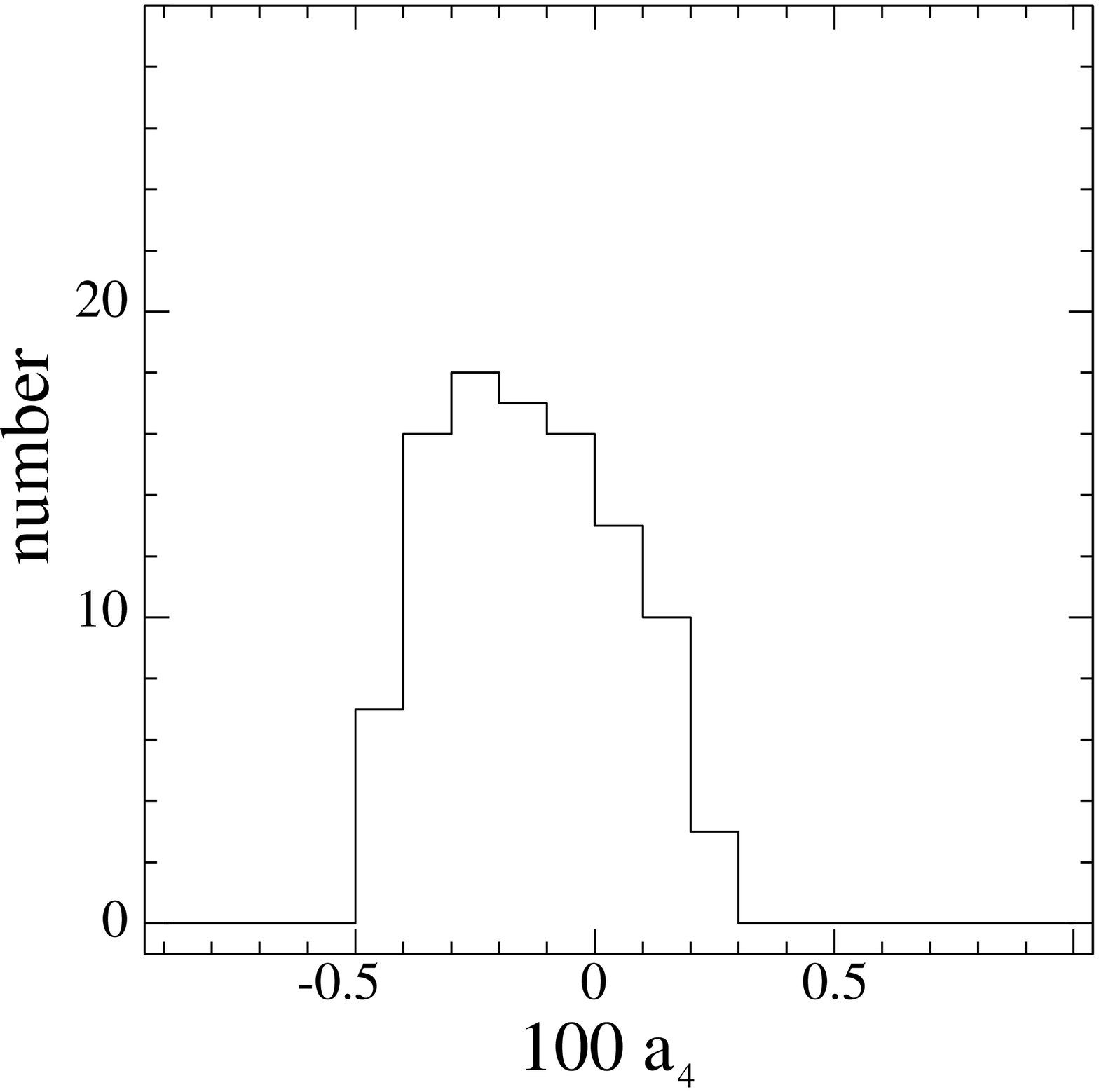}
\hspace{0.cm}
\includegraphics[width=4cm]{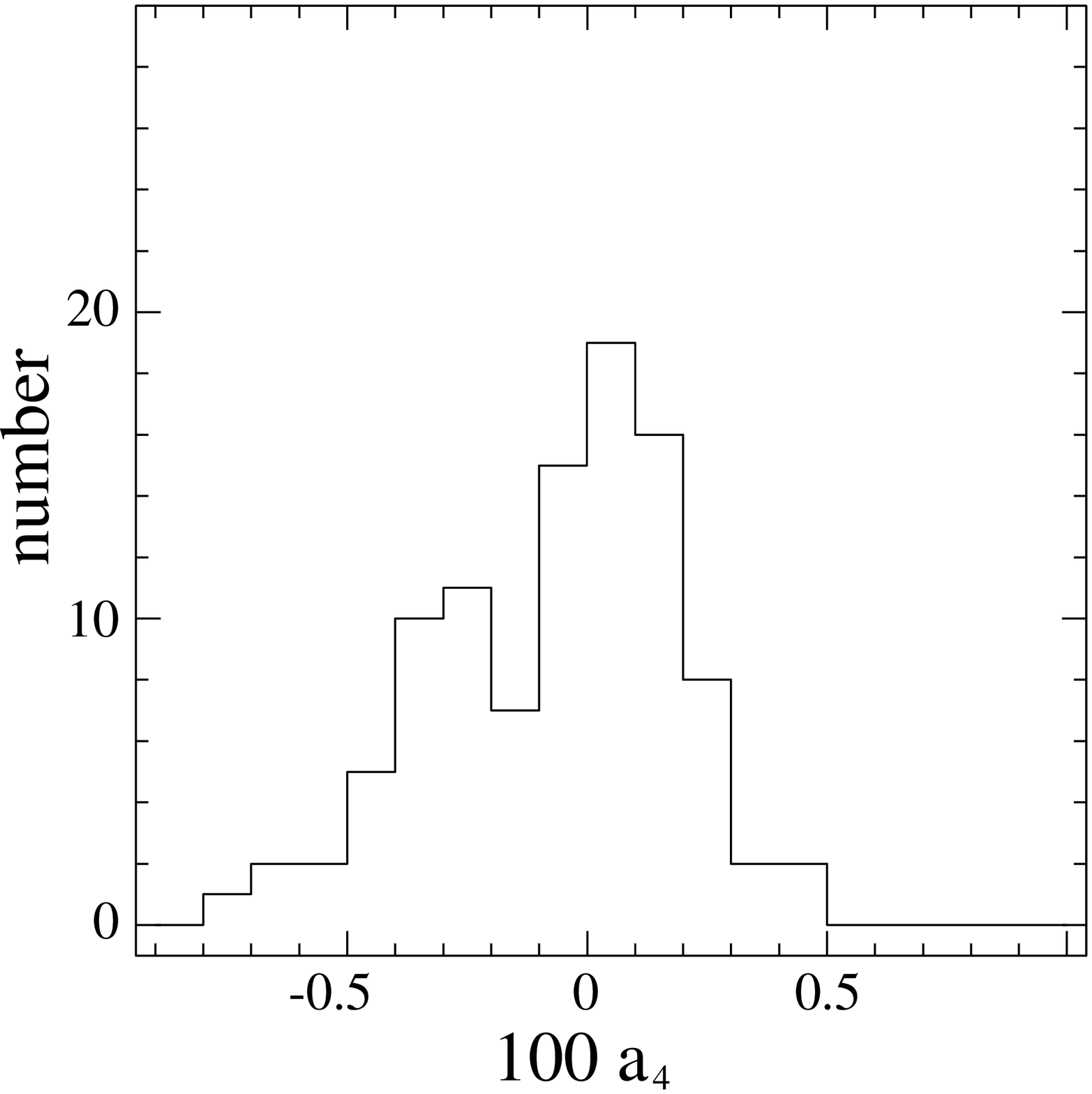}
\caption{Histograms of boxiness distribution for model $3r$. We have calculated the mean $a_4$ parameter inside $1.5$\Re (left) and $3$\Re  (right) of this model as measured from one hundred points of view (randomly chosen) for 60 snapshots .\label{boxid}}
\end{figure}

Models without bulges show an altogether new and different pattern, as their main bodies are strongly boxy.  Such isophotal profiles are not found in elliptical galaxies, and, together with their strong triaxiality (\S~\ref{Sec:AxisRatios}) represent another failure of collisionless, bulge-less galaxy mergers to yield objects resembling real ellipticals. 

Boxiness-diskiness has been claimed to be point-of-view dependent (Fasano 1991, Stiavelli et al.\ 1991, Ryden 1992, Governato et al.\ 1993, Heyl et al.\ 1994, Gonz\'alez-Garc\'{\i}a 2003, Naab \& Burkert 2003). We have studied the point-of-view dependence of the $a_4$ parameter of model $3r$ by plotting histograms of the mean values of $a_4$ inside $1.5$\Re~ and $3$\Re~ for one hundred random viewpoints (Fig.~\ref{boxid}).  A time average over 60 snapshots was taken to improve the signal-to-noise ratio. Overall, the $a_4$ parameter is found to vary from positive to negative, in agreement with previous works, although the distribution is often skewed (Fig.~\ref{boxid}, left panel) indicating a larger probability to show a distinct isophotal signature (boxy in the plotted example). Clearly, the distribution of $a_4$ depends on the radial range of the measurement (cf.\ left vs.\ right panels in Fig.~\ref{boxid}), highlighting that comparison to observations needs to be performed over similar radial ranges.

\subsection{Rotation curves}

\begin{figure*}
\centering
\includegraphics[width=8.5cm]{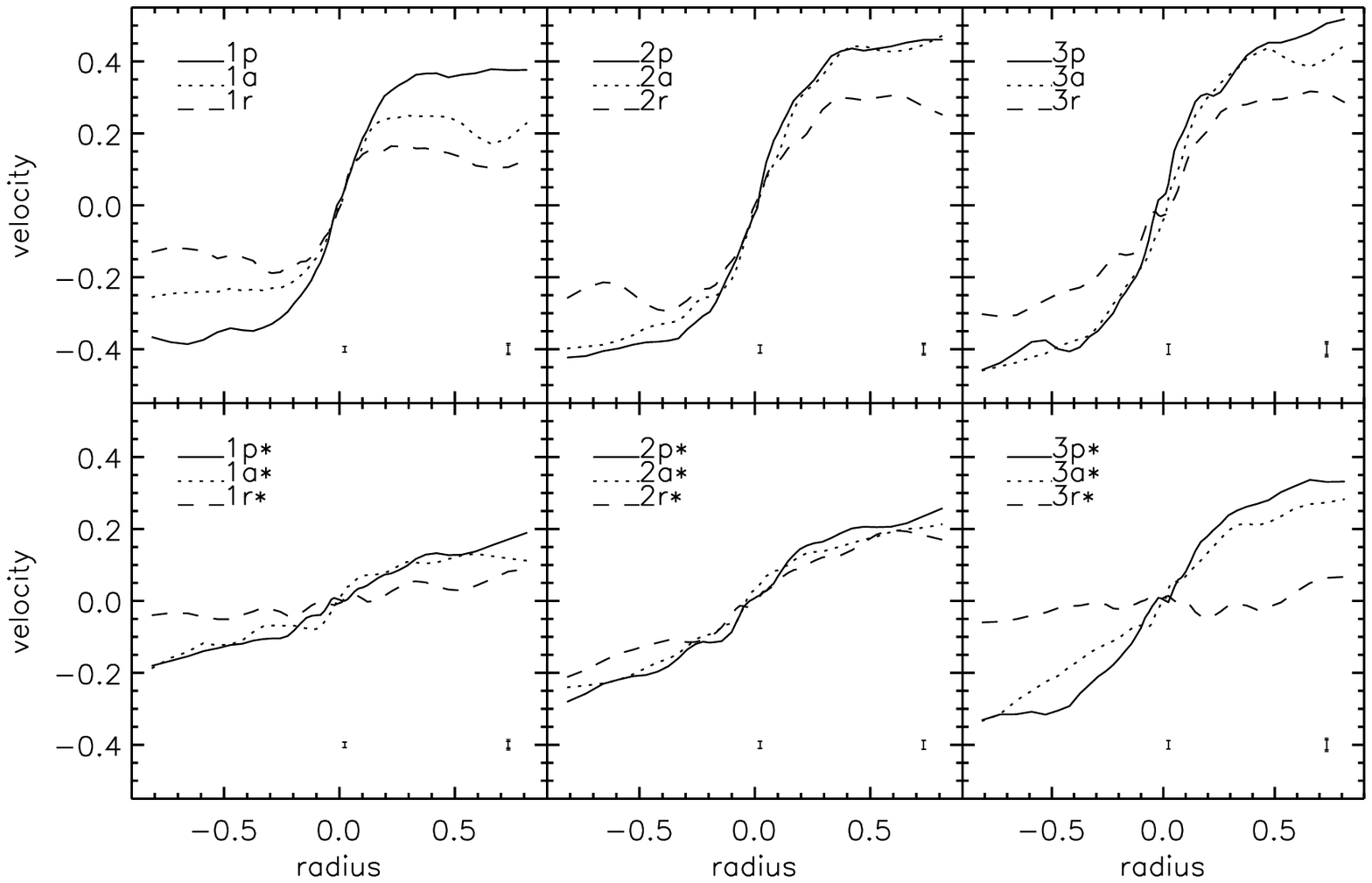}
\hspace{0.cm}
\includegraphics[width=8.5cm]{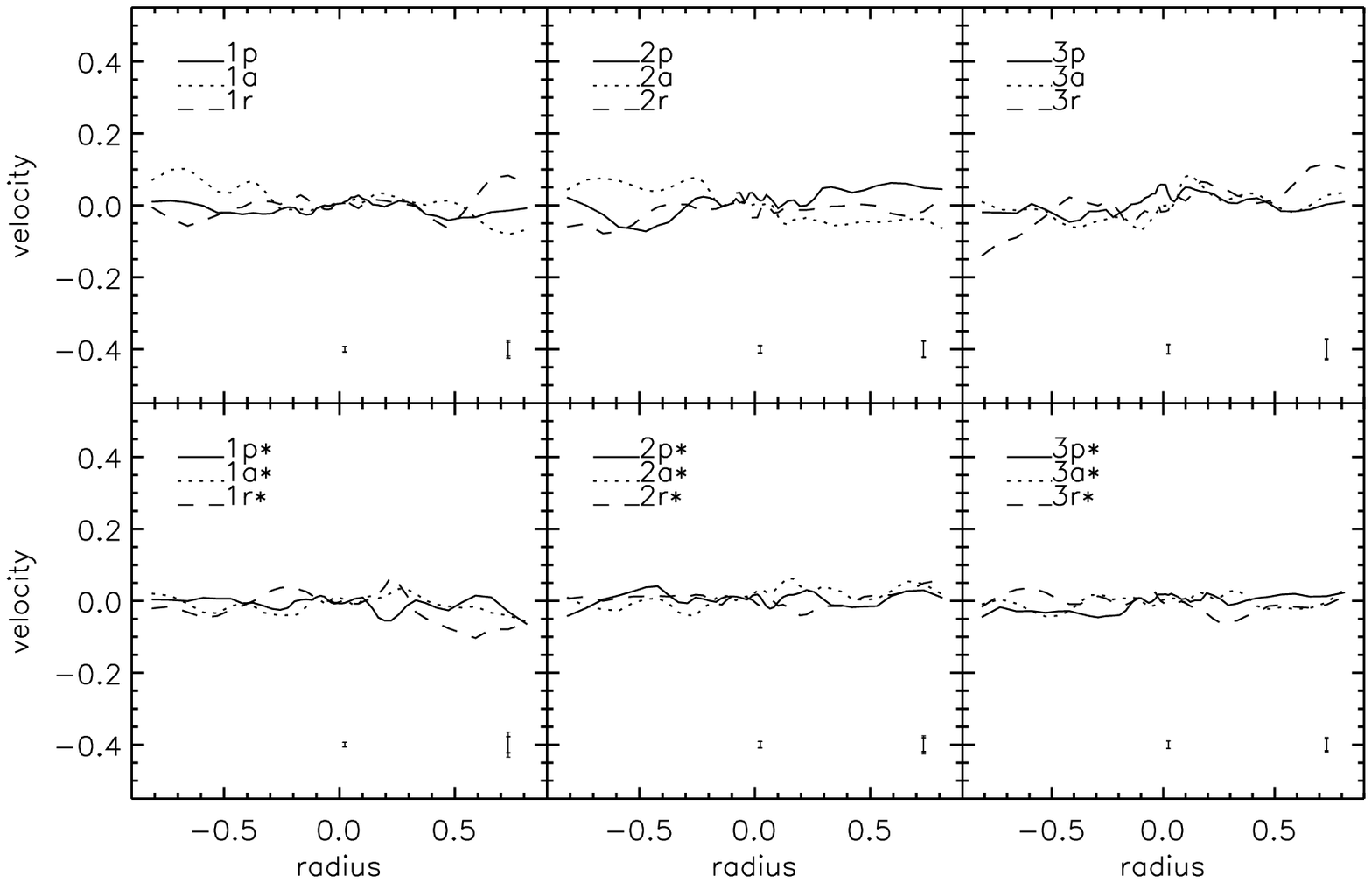}
\caption{The left panels show the rotation curves of the luminous matter for the remnants of the 18 simulations, as measured by placing a viewing slit of width of 0.3 length units along the major axis.  Models are viewed from the positive $y$-axis. The right panels show the rotation curves along the minor axis from the same point of view.  The vertical bars indicate representative velocity error bars, computed using Poisson statistics.   \label{rotd}}
\end{figure*}

Whether mergers produce rotation curves with the amplitudes observed in elliptical galaxies is a key test on the merger hypothesis.   We show major and minor axis rotation curves for all the merger remnants in Figure~\ref{rotd}.  Rotation curves were derived by looking at the models from the $y$-axis and using a viewing slit of length of 2 units and a width of 0.3 units; such rotation curves extend up to $2$\Re~ in all cases.  All the remnants show rotation along the major axis and nearly zero minor-axis rotation, i.e.\ they rotate like oblate ellipsoids.  In particular, this is true for most of our 1:1 mergers, which are commonly assumed to lead to kinematic misalignment (e.g.\ Heyl et al.\ 1996, Naab \& Burkert 2003).  Of the 1:1 mergers, only model $1a$ shows a mild misalignment.   A slight nuclear minor axis rotation is also seen in model $3r$.  

\begin{figure}
\centering
\includegraphics[width=8.cm]{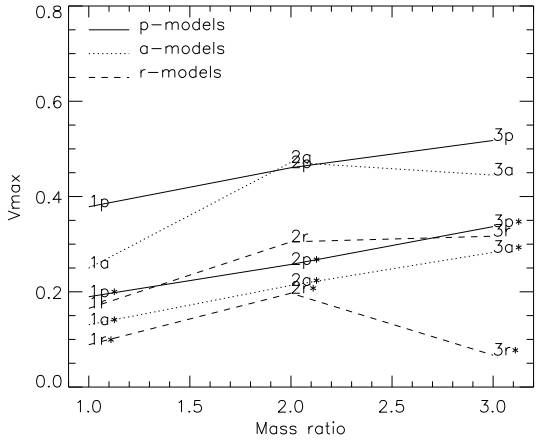}
\caption{Maximum rotation velocity, $V_{\rm max}$, vs. merger mass ratio.\label{vmmr}}
\end{figure}

Two factors affect the rotation amplitude.  First is the spin alignment with the orbit: as expected, $p$ models show the highest rotation.
Second is the presence of a bulge in the progenitors: dbh remnants show both a higher rotation amplitude than their corresponding dh models, and a distinct rotation profile shape, given by a rapid inner rise and an outer plateau.  In contrast, the rotation curves of dh remnants show a gentle, uniform rise to the outermost measured point.  Clearly, the bulge components, which remain self-gravitating during most of the merger, deliver their orbital angular momentum further in than bulge-less galaxies.   These results are summarized in Figure~\ref{vmmr}, which plots $V_{\rm max}$ for all the models vs.\ the mass ratio.  
Model $3r*$ is the only model showing no rotation; this 3:1 retrograde merger of two dh galaxies shows nearly zero rotation, on both the major and minor axes, and a mild nuclear counter-rotation.  Its configuration is similar to those studied by Balcells \& Gonz\'alez (1998) to produce counterrotating cores from mergers of disc galaxies.  

\begin{figure}
\centering
\includegraphics[width=8.5cm]{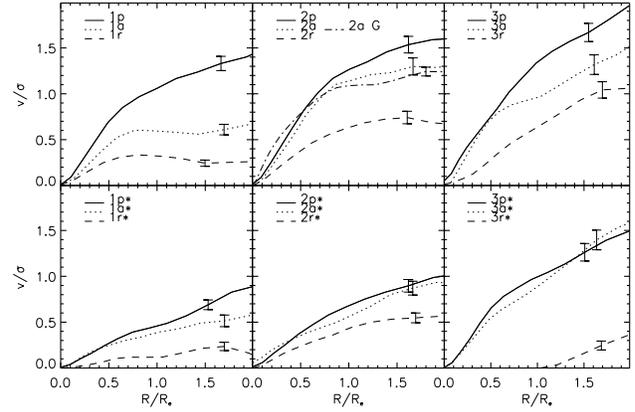}
\caption{$V(R)/\sigma(R)$ vs. radius $R/$\Re. Notation as in Figure \ref{rotd}.  {\it Top panels}:  dbh mergers. {\it Bottom panels}: dh mergers. Error bars for one of the  highest $R/$\Re are given, computed using Poisson statistics.  \label{rotd2}}
\end{figure}

We obtain the degree of rotational support by normalizing the rotation curves by the velocity dispersion.  In Figure \ref{rotd2}, we plot $V(R)/\sigma(R)$ vs. $R/$\Re~ for each model, where $R$ is the projected radius.   As before, these plots are produced for a point of view parallel to the y-axis. A mean has been done for both sides of the rotation curve.
The patterns outlined above for $V(R)$ are reproduced here, with $V(R)/\sigma(R)$ being higher for dbh remnants and for direct configurations.  Additionally, Figure~\ref{rotd2} shows that the amplitude of $V(R)/\sigma(R)$ is higher for higher mass ratios (the dependency of the rotation with mass ratio could not be derived from Figure~\ref{rotd} as the total mass varies, from 2 to 4, between 1:1 and 3:1 models).  This effect mostly results from the lower velocity dispersions in the outer parts of 2:1 and 3:1 remnants as compared to the 1:1 remnants, i.e.\ lower-mass companions produce less heating in the outer parts of the larger disc, and its cold dynamics is partly retained in the remnant.  

The variety of rotation amplitudes, for given mass ratio and spin orientation, emphasizes that the rotation of merger remnants arises from a combination of retained initial spin and redistribution of orbital angular momentum (Bendo \& Barnes 2000).  Therefore, ellipticals of merger origin should show a wide range of rotation amplitudes.   Figure~\ref{rotd2} may be directly compared to observations.  While giant ellipticals have long been known to rotate slowly, with $V_{\rm max}/\sigma_0 << 1$ (Illingworth 1977),  recent, deep spectroscopy of intermediate-luminosity ellipticals shows high rotation, with $V_{\rm max}/\sigma$ in the range of 1--3 (Rix et al.\ 1999; $V_{\rm max}/\sigma=3.8$ for an S0 galaxy in their sample).  These values have been compared to $N$-body 3:1 merger models by Bendo \& Barnes (2000) and by Cretton et al.\ (2001), with conflicting conclusions.  At $R = 2$\Re, our 3:1 mergers yield at most $V_{\rm max}/\sigma=2$ (model $3p$), hence we concur with Cretton et al.\ that 3:1 mergers do not yield the highest $V_{\rm max}/\sigma$ values observed in intermediate-luminosity ellipticals.  We note however that, out to $R\sim $\Re, our model $3p$ matches the fastest-rotating ellipticals.  The discrepancy sets in at larger radii: although $V(R)/\sigma(R)$ continues to rise to the limit of our measurement, its amplitude at $1<R/$\Re$<2$ is lower than observations.  The rotation curve $\sqrt{G\,M_{\rm r}/r}$ is halo-dominated at these radii;  more massive, extended haloes might yield faster rotation in the merger remnants.  Alternatively, the collisional formation of an outer gaseous disc, with ensuing star formation, could lie at the origin of the high outer $V_{\rm max}/\sigma$ values observed by Rix et al.

\subsection{Rotation support vs pressure support}\label{sec:rota} 

\begin{figure}
\centering
\includegraphics[width=8cm]{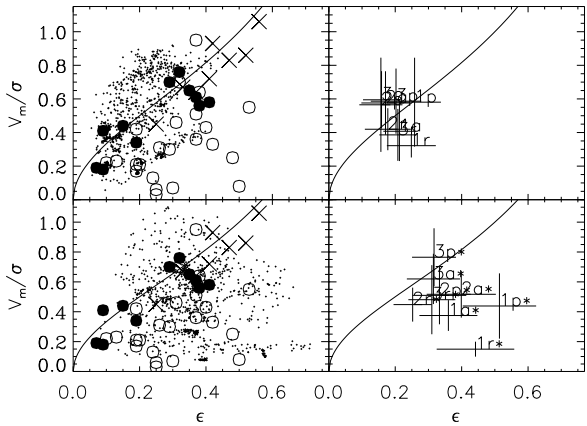}
\caption{Model rotation, expressed as $V_{\rm m}/\sigma_0$ vs.\ the projected ellipticity. {\it Top panels}:  dbh mergers. {\it Bottom panels}: dh mergers.  Measurements are made for one hundred points of view for each model, and are plotted with small dots in the left panels. The mean for each model is plotted with the model name in the right panels. Observations from Davies et al.\ (1983) are shown in the left panels; open circles are bright ellipticals, filled circles are low luminosity ellipticals and crosses are bulges. \label{evsed}}
\end{figure}

Whether ellipsoids owe their flattening to rotation or to anisotropy of the velocity dispersion tensor can be studied with the $V_{\rm max}/\sigma_0$ vs $\epsilon$ diagram (Fig.~\ref{evsed}).  Oblate ellipsoids flattened by rotation show a well defined dependency of $V_{\rm max}/\sigma_0$ with the ellipticity, given by the solid line in Figure~\ref{evsed} (Binney 1978). In this diagram, giant elliptical galaxies fall well below the oblate line (Illingworth, 1977), while low-luminosity ellipticals and bulges scatter along the oblate line (Davies et al.\ 1983; Kormendy \& Illingworth 1982, 1983).  The location of our merger remnants in the $V_{\rm max}/\sigma_0$ vs $\epsilon$ diagram has been determined for 100 random view-points for each model.  The results are shown in Figure~\ref{evsed} (left panels), while the mean position for each model is plotted in Figure~\ref{evsed} (right panels).  

The mean for bulge dominated models lies close to the isotropic oblate rotator line.
dbh models do keep a high amount of rotation while being fairly round ($\epsilon = 0.2$). The spin configuration plays a role, as direct orbits yield higher rotation support than antiparallel or retrograde orbits.   The mass ratio also plays a role since equal mass mergers are less rotationally supported than non-equal mass mergers, as noted by Naab et al.\ (1999) and Naab \& Burkert (2003). Overall, the ranges in $V/\sigma$ and in ellipticity are similar to those of intermediate-luminosity ellipticals.  

Remnants of bulge-less systems show a larger scatter in the $V_{\rm max}/\sigma_0$ vs $\epsilon$ diagram. Remnants of unequal mass galaxies lie close to the oblate line, but equal-mass mergers lead to highly-flattened, slowly rotating systems that are pressure-supported.  As discussed in \S~\ref{Sec:AxisRatios}, these systems are strongly triaxial as a result of transient bars during the merger evolution.  The overall distribution in the $V_{\rm max}/\sigma_0$ vs $\epsilon$ diagram is different both from that of intermediate-luminosity ellipticals and  that of giant ellipticals, providing another line of evidence that mergers of bulge-less galaxies do not yield objects resembling elliptical galaxies.  

\subsection{Anisotropy}

We have calculated the anisotropy parameter $\beta$ as a function of radius for our models. 
This quantity is defined as:

\begin{equation}
	\beta = 1-\frac{\sigma_{\rm t}^2}{2 \sigma_{\rm r}^2},
\end{equation}
where $\sigma_{\rm r}$ is the radial and $\sigma_{\rm t}$ the tangential velocity dispersion:

\begin{equation}
	\sigma_{\rm t}^2=\sigma_{\rm \theta}^2 + \sigma_{\rm \phi}^2 .
\end{equation}

\begin{figure}
\centering
\includegraphics[width=8.cm]{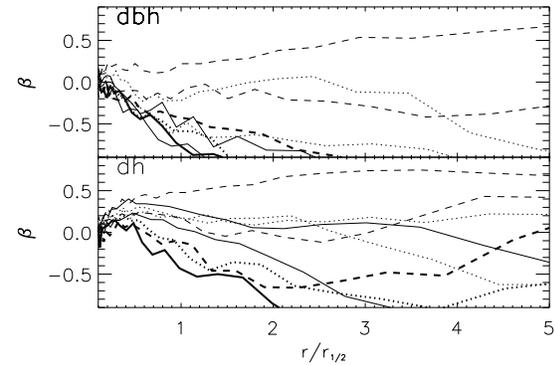}
\caption{ Anisotropy profiles $\beta$ vs.\ radius. {\it Top panel}: dbh models.  {\it Bottom panel}: dh models. 'p' models are depicted with solid lines, 'a' models with dotted lines, and 'r' models with dashed lines. The thickness of the lines gives the mass ratio; the thinnest lines are for 1:1 mass ratios and the thickest for 3:1. \label{anisodisc}}
\end{figure}

Radial anisotropy yields $\beta>0$, while $\beta<0$ corresponds to tangential anisotropy.  Profiles of $\beta$ vs. radius are shown in Figure~\ref{anisodisc}. The top panel shows the dbh models, while dh models are shown in the bottom panel.  Clearly, most remnants show tangential anisotropy.  This is most pronounced in dbh remnants, which have tangential anisotropy at all radii.  Anisotropy in strongest for unequal-mass remnants, an indication that these systems retain a dynamical memory of the disc kinematics of the larger precursor disc.  The only models deviating from tangential anisotropy are model $1a$, which is nearly isotropic over the entire radial range, and model $1r$ which is radially anisotropic. This is likely a result of the cancellation of spin and orbital angular momenta which leads to dominantly radial motions in the remnant.    

Remnants of dh models show more diversity in their anisotropy characteristics. All models show some degree of radial anisotropy inside the half mass radius. 
Such pattern is related to bar formation during the merging stages, as mentioned in \S~\ref{Sec:AxisRatios}.  In the outer parts ($r/r_{1/2} > 1$), similarly to dbh remnants, most dh systems are tangentially anisotropic; some are isotropic, while spin cancellation leads model $1r*$ to show radial anisotropy.  

The tangential anisotropy characteristics are at odds with common lore that elliptical galaxies should be radially anisotropic at large radii (e.g., Bertin, Saglia, \& Stiavelli 1992, Bertin \& Stiavelli 1993).  This belief is rooted  in general properties of violent relaxation.  Under strictly violent relaxation conditions, stars lose their dynamical memory and owe their final dynamics to the energy exchanges with the time-varying global modes of the potential.  Because in mergers 'all the action' occurs at the centre of the system, stars at large radii must have acquired their energy near the centre and henceforth must be on dominantly radial orbits.  The tangential anisotropy observed here is at odds with this framework, and emphasizes that merger relaxation is far from being completely violent (Zurek, Quinn, \& Salmon 1988).  Owing to the preservation of the energy and angular momentum ordering during the merger, the outer parts of the remnants are comprised of stars in the outer parts of the discs, the tangential anisotropy being a relic of the disc kinematics of the larger galaxy.  Such preservation is maximal in dbh mergers, due to the stabilizing effect of the bulge's potential.   In our models, radial anisotropy is restricted to two situations:  when spins and orbital angular momentum cancel out in retrograde, equal-mass encounters, or in the central parts of dh remnants, as a result of transient bars triggered by the tidal field; in neither case is the radial anisotropy related to violent relaxation phenomena.  

Winsall \& Freeman (1993) do not find evidence for anisotropy in giant ellipticals. Bender, Saglia \& Gerhard (1994) study 44 elliptical galaxies fitting Gauss-Hermite polynomials to the LOSVD for each of them. The coefficient of the four order polynomial is related to the tangential anisotropy. Bender et al.\ do not find a clear trend in anisotropy, but find evidence that more rotation means more positive $h_4$ (negative values would mean radial anisotropy), although the conclusion is not firm.

\section{Effects of particle number}\label{sec:cluster}

As described in Section~\ref{Sec:SimulationDetails}, model $2a$ was replicated at high resolution (5 times more particles).  Here we compare the high-resolution model ($2aG$) and its low-resolution counterpart ($2a$) over the set of properties analysed in previous sections;  resolution limitations could in principle affect all of them.  The high-resolution model should be more accurate, therefore the differences between models $2a$ and $2aG$ provide a quantitative estimate of the errors in the models due to low particle number.  

Of particular concern was the ability to generate shells in low-$N$ models due to discreteness effects (Hernquist \& Spergel 1992).  
Like all dbh models, model $2a$ had been shown to have small, if any, shells (Figure~\ref{Fig:Shellsdbh}), which led to the claim that mergers of galaxies with central bulges do not form prominent shells (Section~\ref{Sec:Shells}). For run $2aG$, sharp structures in the projected particle distribution are similar to those of model $2a$.  In phase space, where shells are generally easier to identify, the distribution for model $2aG$ is quite similar to that of model $2a$ shown in Figure~\ref{Fig:shellsvre}.  In particular, only the clumps at low binding energy that show up in Figure~\ref{Fig:shellsvre} are visible in the corresponding diagram for model $2aG$.  This similarity suggests that shells are indeed faint or non-existent for mergers of galaxies harboring central bulges.  

Variations in structural parameters are also small.  Axis ratios are smaller by 9\% at the 30\% mass radius, and by less for outer radii; see Figure~\ref{axidi}, where model $2aG$ is displayed with a diamond.  The lower mass ratios may reflect a stronger inward transport of orbital angular momentum as the nuclei of the high-resolution models manage to get closer to each other before merging together.   The surface brightness profile of model $2aG$ is nearly identical to that of model $2a$.   The effective radius for model $2aG$  is 0.5\% larger than that from run $2a$.  The S\'ersic exponent ($n_{2aG} = 4.82$) is just 7\% lower, while typical uncertainties for the S\'ersic exponent are of order 10\%; and the concentration parameter is 2.5\% lower; see Figure~\ref{conc}, there model $2aG$ is shown as a diamond. 

The rotation profile from model $2aG$ is shown in Figure~\ref{rotd2}, top middle panel, together with that from model $2a$.  
The two curves are very similar. Differences arise in the inner parts, where particle resolution is more important. In the high resolution case, the $V(R)/\sigma(R)$ values are slightly higher at small radii, pointing perhaps to a more efficient transfer of the angular momentum to the inner parts. This may affect our results for other runs, pointing to an even steeper rise of the rotation curve for the inner parts, which would bring our models to a closer agreement with observations (see Section~\ref{sec:rota}).

In conclusion, most properties of the simulations are unchanged with the increase of resolution, which suggests that particle discreteness effects are not blurring our principal conclusions. Some minor differences are observed, specially in the inner parts where particle resolution is better.

\section{Discussion}
 
Our models show that the remnants of collisionless mergers of disc galaxies have markedly different properties depending on whether the precursors harbor central bulges.   Models with bulges (dbh) show higher central densities, oblate shapes, oblate-rotator kinematic support, and a mixture of disky and boxy isophotal shapes.  Bulge-less models (dh) show shallow central densities, strongly prolate or triaxial figure shapes, sub-oblate kinematic support, and boxy isophotes.   In addition, dbh merger remnants show prominent tidal tails and no shells, while dh remnants do not have tails but show shells.  

We trace the origin of these differences to the fact that the tidal field distorts bulge-less discs into transient bars (Hernquist 1992).  These remove angular momentum from the luminous bodies (Hernquist \& Weinberg 1992), resulting in low rotation, and generate prolate figures which, upon merging, yield prolate or triaxial remnants.  The nearly radial orbit makeup of tidally-distorted dh models may explain the formation of phase-wrapped shells.  
In contrast, central bulges help stabilize the discs against bar distortion (Hernquist 1993), which allows the  discs to retain part of their spin throughout the merger decay.  This yields remnants with near oblate shapes and oblate-like kinematic support (Barnes 1992, 1999; Bendo \& Barnes 2000; Cretton et al. 2001).  It also allows for more efficient spin-orbit coupling which is at the origin of tidal tail formation in prograde encounters (Toomre \& Toomre 1972, Barnes 1992).

Besides the well known difficulty in reproducing the central densities of elliptical galaxies (\S~\ref{Sec:Introduction}), dh merger models show additional properties which make them unlikely precursors of today's ellipticals:  they yield figure shapes that are too strongly triaxial, and their remnants show too little kinematic support.  One could argue that their level of rotational support, their triaxial figure shapes and their boxy isophotes, make them akin to giant ellipticals; however, bulge-less galaxies are significantly less massive than galaxies with massive bulges (e.g.\ Trujillo et al.\ 2001; Balcells, Graham, \& Peletier 2004), and their luminosities are lower than those of giant ellipticals by factors of 10-100, hence an evolutionary link between dh mergers and anisotropic giant ellipticals cannot be established.  

Merger models with central bulges compare better with real ellipticals, specifically with intermediate-luminosity ellipticals.  In agreement with Hernquist (1993), our dbh models yield realistic central densities.  Further, figure shapes are oblate, and are supported by rotation.  The level of rotational support, out to $\sim$~\Re, compares well with even the fastest-rotating ellipticals.  Note that, because matching the rotation requires disc kinematics in the precursors, and matching the central densities requires a bulge, precursors are required to harbor both a disc and a bulge if, as claimed by currently popular galaxy formation models, ellipticals formed by mergers.  

A number of shortcomings are present in dbh collisionless merger models.  As noted above, the models cannot reproduce the rapid rotation in the outer parts of intermediate-luminosity ellipticals.   Also, the way dbh models reproduce the isophotal properties of ellipticals is not fully satisfactory.  We concur with previous works (Naab et al.\ 1999; Bendo \& Barnes 2000) that equal-mass mergers yield boxy isophotes, while 3:1 mergers yield diskiness.  However, not all initial orbit configurations of our 3:1 mergers produce disky isophotes; the values of the diskiness parameter $a_4/a$ are higher than generally observed;  and, a particular system can be seen as either boxy or disky depending on the point of view.   More generally, diskiness is a property of intermediate-luminosity ellipticals, and boxiness of giant ellipticals (Bender 1988; Rix \& White 1990), a relation our models cannot explain given that gravitational N-body models are scale-free. The role of post-merger gas infall needs to be taken into account to explain the systematics of boxiness-diskiness with galaxy luminosity (khochfar \& Burkert 2004). 

On the whole, our models are more successful in reproducing oblate-rotator, intermediate-luminosity ellipticals than giant ellipticals, in general agreement with Naab \& Burkert (2003).  This may be surprising given that triaxiality and low rotation are properties generally expected from collisionless merger dynamics.  Pre-existing discs are a problem for the formation of slowly-rotating, triaxial ellipticals, as the precursors retain too much of their spin and yield remnants that rotate too fast.  Therefore, galaxies lacking or with less massive discs are potential merger precursors of giant ellipticals, as in the elliptical-elliptical merger models of Gonz\'alez-Garc\'\i a \& van Albada (2005a,b).  These may be seen as the final stages of the sequence of mergers galaxies are believed to follow (Khochfar \& Burkert 2003, van Dokkum \& Ellis 2003, Bell et al. 2004).

\section{acknowledgments}
We thank K. Kuijken and J. Dubinski for making their galaxy generating code available, L. Hernquist for making his version of the TREECODE available to us, and T.S. van Albada for useful comments and stimulating discussions.

\end{document}